\documentclass[a4paper,11pt]{article}
\pdfoutput=1

\usepackage{jheppub} 
\usepackage{amsmath,amssymb,amscd}
\usepackage{graphicx}
\usepackage{subcaption}
\usepackage[percent]{overpic}
\usepackage{color}
\usepackage{multirow}
\usepackage{bbm}

\def\be{\begin{equation}}
\def\ee{\end{equation}}

\newcommand{\CS}{\mathcal{S}}

\newcommand{\CE}{\mathcal{E}}

\newcommand{\CG}{\mathcal{G}}

\newcommand{\CN}{\mathcal{N}}

\newcommand{\CK}{\mathcal{K}}

\newcommand{\CO}{\mathcal{O}}
\newcommand{\CW}{\mathcal{W}}
\newcommand{\CZ}{\mathcal{Z}}

\newcommand{\tSigma}{{\widetilde{\Sigma}}}

\newcommand{\fOmega}{\overline{\underline{\Omega}}}

\newcommand{\IC}{\mathbb{C}}
\newcommand{\IP}{\mathbb{P}}
\newcommand{\IZ}{\mathbb{Z}}

\newcommand{\IR}{\mathbb{R}}

\newcommand{\Li}{\mathrm{Li}}

\newcommand{\one}{\mathbbm{1}}
\newcommand{\cN}{\mathcal{N}}
\newcommand{\cW}{\mathcal{W}}

\newcommand{\Tr}{\mathrm{Tr}}

\renewcommand{\(}{\left(}
\renewcommand{\)}{\right)}

\title{Exponential BPS graphs and D brane counting on toric Calabi-Yau threefolds: Part I}

\author[a]{Sibasish Banerjee}
\author[b,c]{Pietro Longhi}
\author[d]{Mauricio Romo}

\affiliation[a] {Faculty of Physics, University of Warsaw, ul.\ Pasteura 5, 02-093 Warsaw, Poland}

\affiliation[b]{
Institute for Theoretical Physics, ETH Zurich, 8093, Zurich, Switzerland
}
 
\affiliation[c]{Mathematical Sciences Research Institute, 17 Gauss Way, Berkeley, CA 94720, USA}

\affiliation[d]{Yau Mathematical Sciences Center, Tsinghua University, Beijing, 100084, China}

\emailAdd{sbanerjee@fuw.edu.pl, longhip@phys.ethz.ch, mromoj@tsinghua.edu.cn}

\abstract{We study BPS spectra of D-branes on local Calabi-Yau threefolds $\mathcal{O}(-p)\oplus\mathcal{O}(p-2)\to \mathbb{P}^1$ with $p=0,1$, corresponding to $\mathbb{C}^3/\mathbb{Z}_{2}$ and the resolved conifold.
Nonabelianization for exponential networks is applied to compute directly unframed BPS indices counting states with D2 and D0 brane charges. 
Known results on these BPS spectra are correctly reproduced by computing new types of BPS invariants of 3d-5d BPS states, encoded by nonabelianization, through their wall-crossing. 
We also develop the notion of exponential BPS graphs for the simplest toric examples, and show that they encode both the quiver and the potential associated to the Calabi-Yau via geometric engineering.
}

\begin{document}

\maketitle
\flushbottom

\section{Introduction}

In this paper we continue developing our geometric approach for counting BPS states in five dimensional gauge theories with eight supercharges compactified on a circle of finite radius. 
This problem has direct relevance to BPS counting of Calabi-Yau compactification of M-theory, which further relates to questions in enumerative geometry \cite{2003math.....12059M,2004math......6092M,2008arXiv0809.3976M,Dijkgraaf:2006um}. 
Important progress on wall-crossing, both in mathematics and in string and gauge theories, over the past years has made headway in addressing these questions and opened up several new ones
\cite{Kontsevich:2008fj,Joyce:2008pc,Gaiotto:2008cd,Gaiotto:2009hg,Denef:2007vg,Alim:2011ae,Alim:2011kw,Manschot:2010qz,Manschot:2012rx,Alexandrov:2008gh,Jafferis:2008uf,Szendroi:2007nu,Morrison:2011rz}. 

Continuing from our previous work \cite{Banerjee:2018syt}, we focus on BPS spectra of 5d gauge theories engineered from M-theory on a Calabi-Yau threefold $X$. 
Quite generally, a very fruitful approach to analyzing nonperturbative aspects of supersymmetric gauge theories is to introduce supersymmetric defects of different types \cite{Alday:2009fs, Gaiotto:2009fs, Gaiotto:2012rg}. 
For our purposes, we consider a codimension-two defect in the 
5d gauge theory, obtained by wrapping an M5-brane on $L\times S^1 \times \IR^2$, where $L$ is a special Lagrangian in $X$. 
The effective worldvolume description of the M5 brane is a 3d $\CN=2$ theory on $S^1 \times\IR^2$,  its field content and coupling are determined by the geometry of $L$. 
This theory $T[L]$ couples to the 5d bulk theory, in a way that is determined by the embedding of $L$ in $X$. 
Part of the  BPS spectrum of $T[L]$ consists of vortices, which can be studied effectively with topological strings techniques \cite{Dimofte:2010tz}. 
In our work we focus on a different kind of BPS states of $T[L]$, corresponding to field configurations that reduce to kinks upon shrinking the $S^1$. 
In the following, we explain how with the aid of such a system, one can compute different invariants 
of the Calabi-Yau threefolds in brief, following \cite{Banerjee:2018syt}.

The coupling between $T[L]$ and the ambient 5d theory consists in the identification of some of the global symmetries of $T[L]$ with global or gauge symmetries of the 5d theory. 
Through this mechanism, the BPS spectrum of the coupled system features interactions between 3d and 5d BPS spectra as well as a whole new sector of hybrid 3d-5d states.\footnote{This is an uplift of the 2d-4d BPS sector first considered in \cite{Gaiotto:2011tf}.} 
This sector is very rich, containing both information about the pure 3d states, and the pure 5d states, in addition to the hybrid 3d-5d states. 
In our previous work \cite{Banerjee:2018syt}, we had developed a systematic framework for analyzing this 3d-5d spectra. 
With that under control, we proceeded to the analysis 
of pure 5d states, which was another main outcome from that work. 
The idea is to formulate the problem geometrically, in terms of exponential networks \cite{Eager:2016yxd}.
Our approach is similar in spirit to the construction of spectral networks in the context of class $\mathcal{S}$ theories, which capture the spectrum and the wall-crossing behavior of hybrid 2d-4d states \cite{Gaiotto:2012rg}. 
There are however some important novelties, stemming from logarithmic ambiguities of the superpotential, which are not present in class $\CS$ theories. As a result the construction, and the types of results, differ significantly from those seen in spectral networks.

Physically speaking, this approach consists of studying moduli spaces of configurations for the 3d defect theory engineered by $L$, by varying certain couplings of Fayet-Ilioupoulos type.
Geometrically, the Lagrangian $L$ has topology $\IR^2 \times S^1$ and therefore has a moduli space of deformations of real dimension $b_1(L) = 1$ \cite{Mclean96deformationsof}. We study exponential networks over the complexification of this moduli space induced by the rpesence of a flat $U(1)$ connection on $L$.
In this way, the physics of 3d-5d systems is naturally related to the setup of open-string mirror symmetry, and the vacuum manifold of the 3d-5d theory coincides with the mirror curve of $L$ fibered over the K\"ahler moduli space of $X$.
In our analysis we consider $L$ to be a toric brane, as a consequence the moduli space of $L$ coincides with the standard mirror curve \cite{Aganagic:2000gs, Aganagic:2001nx}. 
However our approach extends automatically to a much larger class of Lagrangian branes, in particular exponential networks can be defined for any mirror geometry associaetd to $(L,X)$ \cite{Aganagic:2013jpa}.

In this work we carry forward the program started in \cite{Banerjee:2018syt} principally in two ways. 
The first one is by studying exponential networks for toric Calabi-Yau geometries with  compact two-cycles (our previous work focussed entirely on applications to $\IC^3$). We study in detail the resolved conifold and $\mathcal{O}(0) \oplus \mathcal{O}(-2) \rightarrow \IP^1$. 
For these geometries we compute the spectrum of 3d-5d BPS spectra and show how the spectrum of 5d BPS states is recovered from the former, from topological jumps of the exponential network. 
As a check, our results on the 5d BPS spectrum are compared with results from topological strings and the Gromov-Witten/Donaldson-Thomas correspondence using wall-crossing arguments.
These results provide checks on our framework, in particular confirming that 3d-5d wall-crossing provides a new powerful and systematic tool for computations in enumerative geometry of toric Calabi-Yau threefolds.
We hasten to stress that our framework also applies to toric threefolds with compact four-cycles, and interesting preliminary results on the BPS spectrum were obtained in \cite{Eager:2016yxd}. 
In this regard, our approach to computing BPS states appears to capture a broader range of the spectrum than existing techniques based on topological strings.

The second main result of this work is the introduction of exponential BPS graphs. 
We consider exponential networks for values of the (toric) K\"ahler moduli where central charges of all BPS states are either aligned or anti-aligned.
This is the uplift of the definition of the ``Roman locus'' of \cite{Gabella:2017hpz}, where the BPS graph emerges from the exponential network at the critical phase.
We show that exponential BPS graphs encode BPS quivers with potential encoding the D-brane spectrum on the Calabi-Yau.
The connection to BPS quivers had been anticipated in \cite{Eager:2016yxd}, who observed an interesting relation between single BPS states as stable quiver representations. 
We take a different approach, by obtaining directly the quiver with potential from a single distinguished exponential network: the BPS graph.
This object ties together in a neat fashion the different facets of geometry and quiver representation theory in the description of BPS spectra.
An important application of the original BPS graphs is that they encode wall-crossing invariants as shown in \cite{Longhi:2016wtv} (the motivic spectrum generator, in physics).
Although we do not explore this construction in this paper, the physical principles underlying this statement carry over directly to 3d-5d systems, and therefore we fully expect that exponential BPS graphs should likewise encode motivic spectrum generators for BPS states of toric CY threefolds.
This observation is especially important for effective applications to enumerative geometry, since the number of BPS states can quickly grow out of control. By contrast, BPS graphs are relatively simple objects, and the spectrum generator they encode would neatly capture all possible BPS spectra of a model.

\subsection{Implications for problems in enumerative geometry}
The results of this and our previous work \cite{Banerjee:2018syt} may be of interest to mathematicians working on enumerative geometry. One of the purposes of this work is to showcase applications of a framework developed in our previous work, to computations of enumerative invariants of local toric Calabi-Yau threefolds. Let us attempt to summarize the main concepts, referring interested readers to our previous paper for a more extensive mathematical introduction to our work.

BPS states of 5d gauge theories are related to generalized Donaldson-Thomas (DT) invariants.
In String Theory this connection arises as follows. The 5d BPS states that we study descend from M2 and M5 branes wrapping cycles in a toric Calabi-Yau threefold $X$.
Their count is related by a circle compactification to the spectrum of D6-D4-D2-D0 branes in type IIA string theory on $X$. Generalized DT invariants arise as protected indices counting these states.
Although in this and our previous work we focus for simplicity on models without compact four-cycles, our techniques directly apply to models with these features as well. 
This is clear from physical considerations, and a proof of concept can be found in \cite{Eager:2016yxd}.

A main feature of our approach is its novelty from a mathematical viewpoint. The way we arrive at the result is inspired by physics, in particular by works of Gaiotto, Moore and Neitzke \cite{Gaiotto:2012rg}.
A central role is played by exponential networks previously introduced by Eager, Selmani and Walcher \cite{Eager:2016yxd} (inspired by \cite{Klemm:1996bj}), and by a new class of ``3d-5d BPS states'' that we introduce and study. 
A mathematical definition of the latter appears to be lacking, but  we hope that our results, which show how they encode information about generalized DT invariants, may supply motivation for a rigorous study of 3d-5d BPS spectra.
Exponential networks probe the geometry of the mirror curve of $X$. The role of mirror symmetry arises naturally in the gauge theory setup, since the mirror curve is identified with the moduli space of vacua of a certain 3d theory that probes the dynamics of M theory on $X$.
Another important feature of our approach is that it is both systematic and algorithmic.
In particular, it is able to capture BPS states whose charges are associated to compact four-cycles of $X$, unlike techniques based on topological vertex.

Another salient novelty introduced in this work is the concept of ``exponential BPS graph''. 
These are (comparatively) simple  ribbon graphs embedded in a Riemann surface (in this paper, it is just $\IC^*$), whose topology is expected to encode the whole spectrum of generalized DT invariants in degree zero (that is, with zero unit of D6 charge). 
We argue that these graphs in fact encode quivers with potential, whose representation theory indeed is known to reproduce the BPS spectrum. We provide explicit examples for the resolved conifold and for $\CO(0)\oplus\CO(-2)\to \IP^1$.
The existence of these objects is subject to certain conditions on the central charges of BPS states, and it is an interesting open question whether these conditions are always satisfied at some points on the moduli spaces of generic toric threefolds.
Assuming existence, the exponential BPS graph  bridges geometric and algebraic descriptions of BPS spectra, and provides a powerful tool to computing the spectrum. It was shown in \cite{Longhi:2016wtv} that similar BPS graphs encode Kontsevich-Soibelman invariants of wall-crossing, and this statement is expected to carry over to our setting.

\subsection*{Organization of the paper}
The paper is organized as follows. In the section \ref{sec:review}, we review various aspects related to the exponential networks for toric Calabi-Yau threefolds. We discuss briefly the M-theory perspective of our works.
We review the relation between various invariants that are related to the computation through the exponential networks. We also 
describe the novelty of the codimension two defects in 5d theories. Then we give a concise recollection of our works on the exponential networks in \cite{Banerjee:2018syt}. Finally, we end this section with a discussion of a subtle point: 
the criteria for the appropriate choice of framing of the mirror curve for plotting the exponential networks. In section \ref{sec:conifold}, we use our techniques to compute various interesting information for the resolved conifold geometry. 
We obtain the BPS counting for this geometry. We discuss various features of D-brane monodromies. 
Finally, we obtain a superpotential for the related BPS quiver for this geometry. Then in the section \ref{sec:C3modZ2}, we follow 
the structure. We derive the BPS invariants for the geometry $\CO(-2) \oplus \CO(0) \rightarrow \IP^1$. Then we analyze them from the points of view of enumerative geometry, and derive the BPS quiver and its superpotential using 
our techniques. 
Finally, the appendices contain some of the materials to support our computations in the main 
body of this paper.

\section*{Acknowledgements}

We thank Brice Bastian, Thomas Bridgeland, Richard Eager, Dmitry Galakhov,  Stefan Hohenegger, Saebyeok Jeong, Sheldon Katz, Can Kozcaz, Jan Manschot, Ivan Smith, Piotr Sulkowski, Joerg Teschner, Johannes Walcher for discussions.
SB acknowledges the support from the ERC starting grant no.\ 335739 ``Quantum fields and knot homologies'' under the European Union's seventh framework programme. 
SB acknowledges the hospitality of Universit\'e de Montpellier where a part of this work was finished. 
PL thanks the Kavli IPMU, the Institut de Physique Nucl\'eaire de Lyon, Trinity College Dublin, and the Simons Center for Geometry and Physics for hospitality during completion of this work.
MR acknowledges hospitality from University of Warsaw, Fields Institute and Heidelberg University.
The work of PL was supported by NCCR SwissMAP, funded by the Swiss National Science Foundation.
The work of PL was also supported by the National Science Foundation under Grant No. DMS-1440140 while the author was in residence at the Mathematical Sciences Research Institute in Berkeley, California, during the Fall 2019 semester.

\section{Exponential networks and BPS counting}\label{sec:review}

This section is aimed towards providing background materials supporting our construction of 
exponential networks both from mathematical and physical viewpoints. Another purpose of this section is to bridge 
our current results to our previous work in \cite{Banerjee:2018syt}.

\subsection{Five dimensional gauge theories from M-theory} 

Our present work focuses on the BPS spectra of five dimensional gauge theories engineered by M-theory on a toric Calabi-Yau threefold $X$. The goal is to develop a systematic 
framework for analyzing the spectrum of BPS instanton-dyons and magnetic monopole strings. A fruitful approach to study them is to introduce various supersymmetric defects as probes. 
We consider codimesion two defects engineered by an M5 brane on $L \times S^1 \times \IR^2$, where $L$ is a special Lagrangian submanifold of $X$. The M5 brane defect engineers 
a 3d $\cN=2$ theory on $S^1 \times \IR^2$ whose field contents and couplings are determined by the geometry of $L$.This theory $T[L]$ further couples to the five dimensional theory in the 
bulk encoding how $L$ gets embedded in $X$. Defects of this type are well-known for embedding topological string theory in M-theory \cite{Dimofte:2010tz}, in which case they establish a connection 
between vortex partition functions and open topological string amplitudes. However, as in the previous paper \cite{Banerjee:2018syt}, in our current work we study a different kind of BPS states of $T[L]$
that correspond to the field configurations which reduce to BPS kinks when the circle shrinks. 

The coupling of $T[L]$ to the 5d theory includes an interaction between the 3d and the former, leading to a hybrid 3d-5d theory. This sector is sensitive to the stability conditions both of 3d and 5d spectra, as well as to 
the stability condition specific to 3d-5d boundstates. 
In our previous paper, we had developed the systematics for studying these 3d-5d states. With these results at our disposal, we showed how to get the information 
about the 5d BPS spectrum. Both of these constructions were inspired by the seminal ideas of \cite{Gaiotto:2011tf} that involves studying entire families of configurations for the defect engineered by $L$. 
From the perspective of the 3d-5d theory, this is nothing but varying certain couplings of the 3d theory. However, for the geometric side, this is equivalent to the study of the moduli space of $L$. The deformation 
moduli of $L$ is one real dimensional if $b_1(L) = 1$. We restrict ourselves to the case of a single M5 brane wrapping a Lagrangian $L$ of topology $S^1 \times \IR^2$. Thus the space of deformations plus the flat $U(1)$
bundle moduli can be described in the form of a curve.

\subsection{3d-5d systems} 

In this section, we focus on the codimension two defects in 5d gauge theories. In the previous section, we have described how to engineer them from M-theory standpoints. 
For simplicity, we resfrict to the case when $L$ is a toric brane, even though our construction applies for a larger class of branes who mirror geometries are known. The advantage of 
doing this is that, now $T[L]$ can be presented as a $U(1)$ gauge theory with a finite number of charged chiral multiplets coupled to the bulk and background fields. Since they couple to the 3d chiral
multiplet as twisted mass, the latter can be integrated out to obtain an effective 3d-5d system. 

For illustration purposes, assume that the bulk 5d gauge theory has gauge group $SU(N)$. Namely, we consider a circle uplift of the 2d $\cN = (2,2)$ $U(1)$ GLSM to a 3d $\cN = 2$ gauge theory with a charged chiral multiplet 
transforming in the fundamental representation of $SU(N)$. We further turn on a minimal coupling for the 3d chirals to 5d vectormultiplet fields restricted to the defect. The 3d theory may additionally feature 
a Chern-Simons term. This is the description of the 3d theory coupled to the 5d theory. The quantum moduli space of vacua is expected to be captured by the mirror 
curve of the toric brane \cite{Aganagic:2000gs}. The 3d Chern-Simons term is directly related to the choice of framing. 

However, invoking mirror symmetry to study the low energy dynamics of 3d-5d theory is rather optional. It is possible to directly rely on the quantum field theory following the same strategy as \cite{Gaiotto:2013sma}
for 2d-4d systems. For a big class of 5d theories, there are matrix model descriptions. The difference however, is that these matrix models are unitary, contrary to hermitean (as for matrix model descriptions of certain 4d $\CN=2$ theories). For example \cite{Klemm:2008yu} studies 
one such matrix model associated to 5d $\cN=1$ gauge theory. 
The quantum-corrected effective superpotential $\widetilde{\cW}(\sigma,t,u)$ of $T[L]$ is then computed from, the (primitive of the) the resolvent of such matrix models.
Minimization of $\widetilde{W}$ yields the Seiberg-Witten curve for the 5d bulk theory \cite{Nekrasov:1996cz}. 

Recent studies of 3d-5d systems corroborate these expectations  \cite{Ashok:2017lko,Ashok:2017bld}. 
We will make a working assumption that these basic features extend to toric branes of generic toric Calabi-Yau threefolds. 
An indication for the validity of this assumption is that it is possible to view these 3d-5d systems as KK uplifts of 2d-4d systems. Even though it looks quite involved from the point of view of 2d-4d theory, one could nevertheless appeal to the logic used in the study of the latter. Further  evidence for this picture is also provided by mirror symmetry, whose prediction identifies the moduli space of $L$ with the mirror curve, which is again coincides with the Seiberg-Witten curve of the 5d theory.

\subsection{Spectral networks and exponential networks} \label{exp-net}

In this section, we will briefly shed light on the field-theoretic interpretation of exponential networks. 
In this paper, we extend the B-model version of BPS counting for the local Calabi-Yau three folds following our earlier work 
\cite{Banerjee:2018syt} and the work done in \cite{Eager:2016yxd}. We briefly digress to recall some key ideas from spectral networks \cite{Gaiotto:2012rg} before proceeding on to  exponential networks. 

There are two pieces of data that define both spectral and exponential networks: the ``geometric'' data and the ``combinatorial'' (or ``soliton'') data. 
The geometric data arises from the BPS equations for solitonic kinks in a 2d theory in the case of spectral network, for a review of this that is suitable to our purposes see \cite{Longhi:2016bte}.
The combinatorial data encodes counts of BPS kinks computed by the CFIV index \cite{Cecotti:1992qh}. In essence, the physics underlying spectral networks \cite{Gaiotto:2011tf} can be viewed as an extension to 2d-4d systems of the framework of $tt^*$ geometry \cite{Cecotti:1991me,Cecotti:1992rm}.
The situation is morally similar, although technically quite more involved, for exponential networks \cite{Banerjee:2018syt}, whose physical interpretation may be viewed as an extension to 3d-5d systems of 3d $tt^*$ geometry \cite{Cecotti:2010fi,Cecotti:2013mba}.

\subsubsection{Spectral networks}
Let us first recall some aspects of the standard spectral networks and how it fits into the 2d-4d systems and $tt^*$ geometry. The BPS spectrum of $\CN=2$ supersymmetric gauge theories in 4d can be repackaged
into a spectral curve. The basic idea is to embed the gauge theory into a higher dimensional one \cite{Gaiotto:2009hg, Klemm:1996bj, Witten:1997sc, Gaiotto:2009we}. Then the spectral curve becomes a part of the 
spacetime. Then one interprets the BPS particles in 4d as extended object calibrated by the Seiberg-Witten differential associated with the spectral curve. 
BPS spectra of a large class of such theories (class $\mathcal{S}$) can be studied thanks to the pioneering work of \cite{Gaiotto:2012rg}.

Theories of class $\mathcal{S}$ are defined as partially twisted dimensional reduction of the 6d, (2,0) ADE theories, on certain punctured Riemann surfaces. Each of such theories is completely characterized by the
choice of the ADE algebra $\mathfrak{g}$, a punctured Riemann surface $C$ and certain defect data on the punctures. Class $\mathcal{S}$ theories of the $A$ type can be embedded naturally into $M$ theory. This can be viewed
as the low-energy dynamics of the worldvolume theory of rank $(\mathfrak{g})$ M5 branes on $\IR^{1,3} \times C \subset \IR^{1,3} \times T^*C \times \IR^3$. At a generic point on the Coulomb branch in the infrared, the 
stack of M5 branes merges into a single fivebrane wrapped on $\IR^{1,3}\times\Sigma$, where $\Sigma \rightarrow C$ is the spectral cover 
\be
\{\lambda : {\mathrm det} (\phi-\lambda I) = 0\} \subset T^* C
\ee 
where $\phi$ is the $\mathfrak{g}$-valued one form parametrizing the Coulomb branch. 

For the purposes of illustration, let us restrict to the case $\mathfrak{g} =\mathfrak{su}(k)$. In 6d the excitations like the strings arise as the boundaries of the M2 branes ending on the stack of M5 branes. Reducing them 
dimensionally on $C$ gives rise to 4d particles, provided that they extend along a path in $C$. Such paths can be labeled locally by a pair $i,j \in \{1,...,k\}$ of integers, with respect to a local choice of the trivialization, if and only if 
the following BPS condition holds 
\be
M = \int \left| \lambda_{(ij)} \right| \ge \left| \int \lambda_{(ij)} \right| = |Z|,
\ee
where $\lambda_{(ij)} = \lambda_i -\lambda_j$, where $\lambda_i$ is the Liouville formon $T^*C$ restricted to the $i^{\mathrm th}$ sheet of the spectral curve. 
This condition is satisfied only if ${\mathrm Im} \(e^{-i\vartheta} \lambda_{(ij)}\) = 0$ defining the trajectories on $C$ with labels $i,j$. Hence this defines the geometric BPS equation. 

Spectral networks are in fact the evolution of such BPS trajectories, with boundary conditions as follows. The trajectories of type $ij$ can end on a branch point where $\lambda_i -\lambda_j = 0$ or 
on a junction where trajectories of types $jk$ and $ki$ meet, all with the same phase $\vartheta$. At special values of the phase $\vartheta$, generalized saddle connections may appear, formed by trajectories with opposite tangent vectors. Nodal points of such finite webs may include both branch points and junctions. However they always lift to a closed homology 
cycle on the spectral curve to closed cycles, whose homology classes are charges of stable  4d BPS particles.

To obtain the BPS indices from the finite webs, one takes into account that the finite webs might exist as continuous families and the spectral networks produce some critical representatives. The generic ones do not necessarily
pass through the branch points, but still they are calibrated by the one form locally and they still satisfy the junction conditions. The deformation modes of these finite webs realize the zero modes of the BPS particles in four dimensions
\cite{Mikhailov:1997jv}. A direct way would be to quantize all such zero modes. However, practically this is a rather difficult task, especially when there are BPS states with higher spins. An alternative route was advocated in 
\cite{Gaiotto:2012rg}. The crux of the strategy is to consider BPS states in presence of various BPS line and surface defects. The consistency with the wall-crossing behavior of such particles bound to BPS defects then leads to 
certain constraints from which one can solve the BPS spectra. This is the upshot of how the other part of the data, namely the combinatorial data are calculated. 

A beautiful interpretation of this exists in terms of $tt^*$ geometry. The curve $C$ is identified with the parameter space of UV couplings of a ``canonical'' surface defect. The finite webs with an open endpoint $z$ 
are then identified with particles bound to a surface defect $S_z$. Line defects on the other hand are engineered by an infinitely heavy M2 brane which has its boundary along a path $\wp$ on $C$. Viewing a line defect as 
an interface between a surface defect and itself (with a basepoint $z$ in $\wp$), the spectrum of the ``framed'' BPS states can be computed by counting the intersections with open BPS strings of the network. 

Physical arguments dictate that the partition function of framed BPS states $F(\wp,\vartheta)$ must only depend on the homotopy of the path $\wp$, since this is a UV observable. Then $F(\wp,\vartheta$ can be viewed as the 
holonomy of a flat connection over $C$ and this in turn gets identified with Lax connection of the $tt^*$ geometry. In fact, $\wp$ is a path on the moduli space of the 2d theories, deformed by the coordinates on $C$. Crossing the network
with this path, the partition function $F(\wp,\vartheta)$ gets corrected by the 2d-4d states corresponding to the open webs. This is indeed closely related to the IR expansion of the CFIV index \cite{Cecotti:1992qh} where this index is also 
intimately tied with the $tt^*$ connection. The coefficients of the IR expansion gives the 2d-4d soliton degeneracies, the data on the spectral network. This data is completely fixed because  homotopy invariance under deformations of $\wp$ translates into the constraint that framed 2d-4d BPS states bound to the supersymmetric interface modeled by $\wp$ remain invariant. This imposes constraints on the 2d-4d states encoded by the soliton data of the network, completely fixing them. 

BPS degeneracies (more precisely, BPS indices) of 4d BPS particles are computed at $\vartheta$ where the network becomes degenerate, leading to finite webs. 
This degeneration leads to a jump in the network topology and hence a jump of the the soliton data. 
These jumps are interpreted as the statement that 2d-4d states mix with 4d particles, leading to changes in the 2d-4d spectrum. 
Comparing the network data at a phase before $\vartheta^-$ and after $\vartheta^+$  the jump, one can obtain 
information about the phase space for such boundstates or decays. This allows us to compute the 4d states that induce these jumps.

\subsubsection{Exponential networks}

We will now review the basic geometric features of the definition of exponential networks following \cite{Eager:2016yxd, Banerjee:2018syt}. 
The basic setup is similar to that of spectral networks, with a certain covering of Riemann surfaces and a family of trajectories associated to this covering.
The covering surface is the mirror curve describing moduli of a  Lagrangian brane $L$ on a toric Calabi-Yau threefold $X$. 
The quantum-corrected moduli spaces of the brane is described by a mirror geometry $Y$ of the form $ \{uv = H(x,y)\} \subset \IC^2 \times (\IC^*)^2$.
The mirror curve is defined by $H=0$, as a curve $\Sigma$ within $ (\IC^*)^2$. 
Stable A-branes on the mirror Calabi-Yau $Y$ are supported on Lagrangian submanifolds $S$, which we take to be compact, subject to the constraint that the holomorphic top form $\Omega$ has constant phase $\alpha$ along them $\Omega|_S = e^{i\alpha} |\Omega|$.
By an observation of \cite{Klemm:1996bj},  A-branes on $Y$ can be studied via their projection down to the mirror curve $\Sigma$. 
The holomorphic top form $\Omega$ descends, upon integration over fibers of the projection $\IC^2\times(\IC^*)^2 \rightarrow (\IC^*)^2 $,  to the holomorphic differential $\lambda = \log y \, d\log x$ that calibrates one-cycles on $\Sigma$.

Quantum corrections to the moduli space of the A-brane introduce an ambiguity that is absent in the classical picture on the toric side \cite{Aganagic:2001nx}.
Classically speaking, the position of a toric $A$-brane is specified by a point on the toric diagram. In local coordinates, where a vertex sits at the intersection of divisors $z_i=0$ for $i=1,2,3$, the position is specified by $|z_2|^2 - |z_1|^2 = 0$ and $r \sim |z_3|^2 - |z_1|^2$ where $r\in \IR$ is the position modulus. 
This classical picture approximates the quantum one if the brane sits far away from the vertex, where disk instanton corrections are exponentially suppressed. 
But in the quantum regime, and especially closer to vertices, disk instanton corrections become important and introduce a \emph{framing} ambiguity $r \sim |z_3|^2 - |z_1|^2 + f(|z_2|^2 - |z_1|^2 )$, labeled by $f\in \IZ$.
Descending to the mirror curve, different choices of framing are related by a change of variables taking $x\mapsto x y^f$. These correspond to different presentations of $\Sigma$ as a covering map of, for example, $\IC^*_x$. 
Different choices of framing will give rise to different projections of calibrated cycles on $\Sigma$ down to $\IC^*_x$, resulting in different kinds of exponential network geometries. 
These differences notwithstanding, the BPS spectrum of the model (5d BPS states)) is expected to be independent of framing, therefore one is free to make the most convenient choice for this particular purpose.

On the other hand, there is another type of BPS spectrum that exponential networks compute, which does depend on framing.
The spectrum of so-called 3d-5d states, the ``soliton data'' of the network however does change with framing.
In our previous work \cite{Banerjee:2018syt} we introduced BPS counting for exponential networks 
by developing a combinatorial structure associated to the original set-up of \cite{Eager:2016yxd}. 
By analogy with spectral networks, and due to physical motivations, we called this the \emph{nonabelianization map}. (Various features of this construction will be reviewed in section \ref{nonab}.) 
This map counts 3d-5d BPS states, and these are sensitive to a choice of framing, since they depend on the geometry of the network.
In  \cite{Banerjee:2018syt} we use data of 3d-5d states and their wall-crossing with 5d BPS states, to obtain the the correct counting of BPS states for the simplest toric Calabi-Yau 
threefold, namely $\IC^3$. 
In this paper we go further, and apply this construction to BPS counting in more interesting geometries. 
This approach complements, from a rather different viewpoint, the study initiated in \cite{Eager:2016yxd} which focussed on the interpretation of degenerate exponential networks (finite webs) as stable representations of BPS quivers.

\subsection{3d $tt^*$ geometry and physical motivation for nonabelianization}

The physical foundations for our construction of the nonabelianization map in \cite{Banerjee:2018syt} are provided by 3d $tt^*$ geometry \cite{Cecotti:2013mba}.
The latter arises in the study of moduli spaces of vacua, and related solitons, of a 3d $\CN=2$ theory  $T[L]$ engineered by an M5 brane wrapping the toric brane, and producing a 3d defect in the 5d theory.
Exponential networks compute BPS states in this 3d-5d system.
We consider both the 5d and 3d theories compactified on a circle of finite radius $R$, and view 
each as a lower-dimensional theory of the Fourier modes along the circle.
For the 3d theory, this consists of a 2d $\CN= (2,2)$ model.

This is the setup of 3d $tt^*$ geometry, which arises indeed as an uplift of the original 2d $tt^*$ geometry to 3d \cite{Cecotti:1991me, Cecotti:2013mba}. 
In our previous work, studied this geometry in detail for the case of a 3d $\CN=2$ theory engineered by the toric brane in  $\IC^3$, and provided an alternative method of counting the 3d solitons \cite{Banerjee:2018syt}. 
While in this paper we will not attempt to solve the $tt^*$ equations for the geometries that we consider, it is important to keep in mind that the physics of exponential networks can be traced back to this framework.
Here we collect key facts about this story, highlighting those that motivate certain features of the nonabelianization map. 

The theory $T[L]$ can often be modeled by a 3d GLSM with $U(1)$ gauge group.
A novelty in three dimensions, compared to two, is that vectormultiplet scalars get complexified with periodic imaginary part $Y_i \sim Y_i + 2\pi i$, if the theory is placed on a circle. 
The low energy dynamics, and the spectrum of BPS states, are governed by the effective (twisted) superpotential $\widetilde{\CW}$.
For example let us consider a $U(1)$ GLSM with a multiplet of $N$ chirals with unit charge. 
Masses  $m_i$ of the chirals get contributions from bulk 5d vectormultiplet scalars, through 3d-5d couplings, which acquire VEVs on the Coulomb branch. 
Taking into account one-loop effects, they contribute to the superpotential as follows
\be
\widetilde{\CW} = XY + \sum_i \Li_2 \(e^{-2\pi R (Y+m_i)}\) + \frac{\kappa}{2} Y^2
\ee
where $Y$ is the complexified fieldstrength scalar, $X$ is the complexified Fayet-Iliopoulos coupling and $\kappa$ is an effective 3d Chern-Simons level. 
The vacuum manifold takes the form of an algebraic curve $F(x,y) = 0$ defined by $x=e^X, y=e^Y$ (we are omitting numerical factors like $2\pi R$, see \cite{Banerjee:2018syt} for details).

For generic values of the coupling $x$, the vacuum manifold consists of a discrete set of massive vacua $\{y_i(x)\}$.
We then study the spectrum of BPS states arising as field configurations interpolating between two such vacua, labeled by  $i$ and $j$,  and obeying BPS field  equations (these are also known as \emph{kinky vortices}). 
In the 3d setup on $S^1\times \IR^2$, the circle corresponds to one of the spatial directions. 
The other spatial direction is noncompact and $y$ approaches constant vacuum values at $\pm \infty$ at spatial infinity. 
Central charges of these solitons are fixed not just by the pair $ij$ of vacua, but also by the relative homology class of the path traced out in $Y$-space by the field configuration. 
Part of this is due to coupling to the bulk theory, and the corresponding shifts of central charge correspond to the 5d central charges. This phenomenon is familiar from 2d-4d systems \cite{Gaiotto:2011tf}
However the 3d setting features an additional source of monodromy for central charges, owing to the multi-valuedness $Y$ on $\IC^*_y$. 
For this reason it is convenient to pass to the universal covering (the $Y$-plane), introducing the index $N \in \IZ$ and labeling the vacua by $\vert i, N\rangle$.
This introduces a $\IZ$-worth orbit of images of the vacuum $y_i(x)$, corresponding to logarithmic shifts of $Y$. 
Soliton central charges are in fact sensitive to these shifts. 

With these conventions, vacua interpolated by solitons are labeled by additional integers such as $(i,N)$ and $(j,M)$. 
These extra labels also appear in the 3d uplift of 2d $tt^*$ geometry \cite{Cecotti:2013mba} (for example the metric acquires enhanced indices such as $g_{(i,N),(\bar{j},\bar{M})}$).
It can be argued, based on physical \cite{Cecotti:2013mba} or geometric  \cite{Banerjee:2018syt} considerations, that solitons always come in towers of effectively identical copies.
More precisely, physical properties of solitons only depend on the difference of logarithmic branches of the two vacua $M-N$.\footnote{It is thanks to this shift symmetry that the 3d $tt^*$ metric can be recast in the form $g_{(i,N),(\bar{j},{M})} = g_{i\bar{j}}(M-N)$ and that the discrete labels can be eventually traded for a continuous modulus via Bloch-Fourier transform.
For more details see   \cite[Section 4.4]{Banerjee:2018syt}. }
This observation is crucial for consistency of the whole picture, since the geometry of walls in the exponential network are labeled by a single integer $(ij,n)$ which coincides precisely with $n=M-N$.
With this technical detail taken into account, we the propose that nonabelianization based on exponential networks computes the connection arising in the Lax formulation of 3d $tt^*$ equations. Its Stokes data is determined by solitons like the ones we consider, whose counting has a well-defined meaning in physics: the 3d uplift of the CFIV index \cite{Cecotti:1992qh, Cecotti:1992rm}.

We conclude by writing down explicit forms of the superpotentials for the branes considered in this paper.  
First we consider the toric brane in the resolved conifold. 
The 5d theory can be viewed a $U(1)$ gauge theory having two phases, birationally related by the Atiyah flop. 
The D-term mirror equation is given by 
$y_1y_2 = e^{-t} y_3y_4$. 
Then one gets the mirror curve $H_f(x,y) = 1 + xy^f + y + e^{-t} xy^{f+1}$, where the framing $f$ has its origin in the 3d Chern-Simons coupling. 
This gives the following superpotential 
\be
\widetilde{\CW}_f = (X-i\pi) Y + \frac{f}{2} Y^2 + \Li_2(-e^{Y}) - \Li_2(-Qe^Y). 
\ee
For the other example, i.e. $\CO(0)\oplus \CO(-2) \rightarrow \IP^1$, the superpotential is instead 
\be
\widetilde{\CW}_f = (X- i\pi) Y +\frac{f}{2} Y^2 + \Li_2(-e^Y) + \Li_2(- T e^Y)
\ee
corresponding to the curve $1+(1+T)y + Ty^2 + xy^f = 0$.\footnote{Here one can rescale $y \rightarrow y/(1+T)$ and $x \rightarrow (1+T)^f x$, reproducing the GLSM curve through  \eqref{eq:TandQrel}. }
The $tt^*$ analysis for these models are indeed challenging and rewarding works. However this is beyond the scope of the 
current work and we postpone it to a future investigation.

\subsection{Nonabelianization for exponential networks} {\label{nonab}}

In this section, we discuss the construction of exponential networks more elaborately. This is essentially a review of section 3 of \cite{Banerjee:2018syt} and thus we will be rather brief, just 
recalling the salient features along with the general strategy. As was mentioned previously, there are two sets of data relevant for this: geometric and combinatorial. Let us describe them in the following.

\subsubsection{Geometric data}

Our starting point is the mirror curve $\Sigma$ which is an algebraic curve in $\IC^*_x \times \IC^*_y$, endowed with a natural projection map $\pi : \Sigma \rightarrow C$, where we denote the $x$-plane 
$\IC^*_x$ by $C$. This projection map a $K:1$ and it presents $\Sigma$ as a ramified covering of $C$. We make a genericity assumption that all the branch points are of the square root types. This indeed is the 
case for the examples that we shall discuss in this paper. The choice of square root branch cuts for $\pi$ and labeling each of the sheets away from the branch points provide a trivialization. Different sheets will be indicated 
by $y_i(x)$ where $i=1,...,K$. 

Given $\vartheta \in \IR/2\pi \IZ$, $\CW(\vartheta)$ is a network of trajectories drawn on $C$ related to the covering~$\pi$. In \cite{Banerjee:2018syt}, we dubbed these trajectories as $\CE$-walls, labeled by 
$(ij,n)$ and their shape is dictated by the differential equation 
\be
(\log y_j - \log y_i + 2\pi i n) \frac{d\log x}{dt} \in e^{i\vartheta} \IR^+. 
\ee
There are certain boundary conditions on the $\CE$-walls. If they start at the branch points, we call them primary walls. At the branch point one has $y_i = y_j$ and furthermore $n=0$. Thus primary walls starting from the 
branch points are always of the type $(ij,0)$. One can have secondary walls too, emanating from the intersection of walls $\CE,\CE'$. We call the new walls $\CE''$ generated from the intersection as the secondary walls,
determined by the types of $\CE,\CE'$. 

Recall from \cite{Banerjee:2018syt}, that the most of novelties of exponential networks arise from the presence of the extra integer label $n\in \IZ$. 
To have this label well-defined, one needs to introduce an additional covering 
$\tilde{\pi} : \tilde{\Sigma} \rightarrow\Sigma$ having branching at points $(x,y) \in \Sigma$ such that $y_i(x)$ goes to either zero or infinity. One has to choose the trivialization for the logarithmic branching too. This is achieved
by specifying each of the sheets away from the log-cuts by an integer $N\in \IZ$, away from the cuts. In this way $\tilde{\Sigma}$ can be regarded as an infinite sheeted covering of $C$. Hence above each $x\in C$, there is 
an infinite sheeted covering $(i,N)$ corresponding to the points located on $\tilde{\Sigma}$ located at $(x, \log y_i + 2\pi i N)$. 

Crossing the square root cut, the $\CE$-wall of type $(ij,n)$ gets its square root sheets permuted rendering the logarithmic label unchanged.
That is $i\rightarrow \sigma(i)$ being the permutation, it becomes $(\sigma(i)\sigma(j),n)$. 
On the other hand crossing the logarithmic cut does not change square root branching. However, the logarithmic index jumps. Consider the lift of a wall $p$ to $\Sigma$ as $\pi^{-1}(p) = p^{(j)} - p^{(i)}$ consisting of the preimages 
to $\Sigma$. Now assume that $p^{(i,j)}$ cross log-cuts such that $\log y_{i,j} \rightarrow \log y_{i,j} + 2\pi i \delta n_{i,j}$. Then the net jump is given by $n \rightarrow n + \delta n_j - \delta n_i$.

\subsubsection{Combinatorial data}

Now we come to the combinatorial data (soliton data) carried by the $\CE$-walls. Fix a point $x$ on the wall and consider the affine lattice 
\be
\Gamma_{ij,n} = H_1^{\mathrm rel} (\tilde\Sigma; (i,N), (j, N+n), \IZ).
\ee
This is the relative homology lattice of open paths on $ a \subset\tilde\Sigma$ starting at $(i,N)$ and ending at $(j,N+n)$. This definition was introduced in \cite{Banerjee:2018syt}, motivated by geometric and physical considerations analogous to those in the 2d-4d setting, where similar definitions of (affine) charge lattices first appeared \cite{Gaiotto:2011tf}.
The homology classes correspond to possible central charges for the 3d-5d particles. 
The central charge is given by $Z_a = \frac{1}{2\pi R} \int_a Y(x) d\log x$, where $Y(x) = \log y(x) + 2\pi i N$, with $N$ keeping track of the appropriate logarithmic index of the path $a$ on $\tilde\Sigma$. 
The shape of the $\CE$-wall depends on the integer $n$. However, the soliton data depends on a pair $N,N+n$ the difference of which is $n$. We shall denote
\be
\Gamma_{ij,n} (x) = \bigsqcup_{N\in \IZ} \Gamma_{ij, N, N+n} (x). 
\ee 
The soliton data is an assignment $\mu(a)$ to each of $a \in \Gamma_{ij,n} (x)$. 

The next objective is to construct the ``nonabelianization'' map. This construction leads to the computation of the aforementioned soliton indices. The starting point is the GL(1) flat connection $\nabla^{ab}$ on $\tilde\Sigma$. 
Through the data of the exponential network $\CW$ one constructs a GL(N) flat connection $\nabla^{na}$ as $\Psi_{\CW} : \nabla^{ab} \rightarrow \nabla^{na}$. Given a path $\wp$ on $C$, one can compute the parallel 
transport as 
\be
F(\wp) = P\exp \(\int_{\wp} \nabla^{na}\)
\ee
and for the solitonic paths $a$ on $\tilde\Sigma$, one has 
\be
X_a = P\exp \(\int_a \nabla^{ab}\). 
\ee
The latter variables satisfy the concatenation rules that $X_a X_b = X_{ab}$ only if the ${\mathrm{end}} (a) = {\mathrm{beg}} (b)$ and zero otherwise. 
For both types of connections, flatness ensures that the quantities depend only on the relative homotopies of $\wp,a$.\footnote{Subtleties due tangent framing are not important for this discussion.} 

In \cite{Banerjee:2018syt}, we gave a detailed account of how to express $F(\wp)$ in terms of $X_a$'s. Instead of repeating it here verbatim, let us just point out some of the important features. 
The first striking aspect is the existence of the $(ii,n)$ solitons. Solitons of such type do not exist for standard spectral networks. This led us to modify the detour rules along the $\CE$-walls.
Unlike for the standard spectral networks where they are of the polynomial nature, in this case we had to allow for exponential detours. Indeed, in our computation we had seen that to ensure flatness around a joint 
of two incoming $\CE$-walls of types $(ij,n)$ and $(ji,n)$, we require this detour rule. Another novelty is that, out of the junction an infinite set of trajectories come out. An infinite set of each of $(ij, n + (k-1)(n+m)), (ji, n+(k-1)(n+m)),
(ii, k(n+m))$ and $(jj, k(n+m))$ where $k = 1,2,...$, trajectories come out, completely consistently with flatness. A direct computation also led to the fact that the degeneracies for the walls of types $(ii,k(n+m))$ and $(jj, k(n+m))$ are fractional 
weighted by a factor $1/k$. However, it is extremely gratifying that the same conclusion can also be reached from the $tt^*$ computation \cite{Cecotti:2010fi}. 

One more important point to recall before moving to the computation of the 5d states, is the shift symmetry. This proved crucial for solving flatness equations. Indeed, as we described before, it can be best understood in the light of the 
underlying $tt^*$ geometry. At the level of the soliton degeneracy of the network, it manifests itself in two variants. The first one is an overall log-branching shift
\be
\begin{split}
& (+l) : \Gamma_{ij,N,N+k} (x) \rightarrow \Gamma_{ij,N+l,N+l+k} (x) 
\\ & 
a \mapsto a^{(+l)}
\end{split}
\ee
under which soliton degeneracies (interpreted as 3d uplifts of CFIV indices, as discussed above) are the same for all images: $\mu(a) = \mu(a^{(+l)})$. 

The other one affects $ii$ type solitons as follows 
\be
\begin{split}
& (i\rightarrow j, +l) : \Gamma_{ii,N,N+k} (x) \rightarrow \Gamma_{jj,N+l,N+l+k} (x) 
\\ & 
a \mapsto a^{(i\rightarrow j,+l)}
\end{split}
\ee
such that $\mu(a) = \mu(a^{(i\rightarrow j,+l)})$. 

\subsubsection{$\CK$-wall formula and 5d BPS states}

Until now, we have assumed that $\CW(\vartheta)$ is generic, that is there aren't any degenerate $\CE$-walls. Under this condition, the soliton data is computed combinatorially, and essentially just depends on the global topology of the network $\CW$, rather that its geometry. However, at some critical phases $\vartheta_c$ the network becomes degenerate. Certain $\CE$-walls of opposite types in this case may overlap partially or entirely. 
These form double walls or ``two way streets''. The network $\CW(\vartheta_c)$ admits two natural resolutions at $\vartheta_c^{\pm} = \vartheta_c \pm \epsilon$ \cite{Banerjee:2018syt, Gaiotto:2012rg}.

Since the topologies of the networks $\CW(\vartheta_c^{\pm})$ are different, there must be a jump in the nonabelianization map. This jump is defined as the $\CK$-wall jump. Physically, as was mentioned before, this is a mixing of the 
BPS solitons supported on the codimension-2 defects and the BPS states supported on the bulk theory. 

The way to compute the $\CK$-wall formula is to compare $F(\wp,\vartheta_c^+)$ and $F(\wp,\vartheta_c^-)$. Then read it off from the fact $F(\wp,\vartheta_c^+)= \CK(F(\wp,\vartheta_c^-))$. This exercise is done explicitly in 
 \cite{Banerjee:2018syt}. We just repeat the important points from there. Let's assume that a path $\wp$ crosses a two way street $(ij,n)/(ji,-n)$. The first step is to write down the soliton data in terms of a new type of function 
\be
Q_N (p) = 1 + \sum_{\substack{a \in \Gamma_{ij,N,N+n}(p) \\  \\ b\in\Gamma_{ji,N+n,N}(p)}} \mu(a) \mu(b) X_{{\mathrm{cl}} (ab)},
\ee
where we denoted simply by $\mu(a)$ the soliton degeneracies computed by the network both before and after the critical phase (equality was discussed in \cite{Banerjee:2018syt}).
Shift symmetry furthermore ensures that $Q_N(p) = Q_{N+1}(p)$. Crucially, unlike the generating functions that we have considered so far, the functions $Q_N(p)$ depend on the variables 
$X_\gamma$ which depend on the closed homology class. By a genericity assumption, we restrict $Q_N$ to be a function of a single variable $X_{\gamma_c}$. More precisely, different generating functions $Q_N$ can depend on different 
formal variables associated with the homology classes $\gamma_N$ on $\tilde\Sigma$. However, the periods of the one form $\lambda$ are all equal along those cycles because of the shift symmetry. We take a quotient on the homology lattice by 
ker$(Z)$. $\gamma_c$ is the equivalence class of this quotient.  

To this end, it is crucial to obtain a factorized form for the $Q_N$ as 
\be
Q_N(p)  = \prod_{k=1}^\infty \(1+ X_{k\gamma_c} \)^{\alpha_{k\gamma_c}(p)}, \quad \alpha_\gamma(p) \in \IZ.
\ee
Then from 
\be
F(\wp,\vartheta_c^+) = \CK (F(\wp,\vartheta_c^-)),
\ee
one can find 
\be
\begin{split}
& \CK(X_{\wp^{(i,N)}}) = X_{\wp^{(i,N)}}\prod_{k=1}^\infty \(1+X_{k\gamma_c}\)^{-\alpha_{k\gamma_c}(p)}
\\ & 
\CK(X_{\wp^{(j,N)}}) = X_{\wp^{(j,N)}}\prod_{k=1}^\infty \(1+X_{k\gamma_c}\)^{\alpha_{k\gamma_c}(p)}
\end{split}
\ee
A more useful way to represent this jump for our purpose, is to write it in terms of the lifts of the street $p$ to $\tilde\Sigma$. Writing 
\be
\pi^{-1} (p) = \sum_{N\in \IZ} (p_{(j,N+n)} - p_{(i,N)})
\ee
and recognizing $\langle \pi^{-1}(p), \wp^{(i,N)} \rangle = -1 = -\langle \pi^{-1}(p), \wp^{(j,N)}\rangle$, one has 
\be
\CK(X_a) = X_a \prod_{k=1}^\infty \(1+X_{k\gamma_c}\)^{\alpha_{k\gamma_c}(p)\langle \pi^{-1}(p),a\rangle}.
\ee

For exponential networks, there is another type of two way street, made of $\CE$ walls of types $(ii,k(n+m))$ and $(ii,-k(n+m)$. Even if the $\CK$-wall formula continues to hold through, one can not 
use it to determine $\alpha_\gamma(p)$ in this case, for more details we refer the reader to \cite{Banerjee:2018syt}.

Now consider a one chain $\bf{L}$ on $\tilde\Sigma$ such that the parallel transport along an arbitrary path $\wp$ jumps as 
\be
\CK(X_a) = X_a \prod_{n=1}^\infty \(1+X_{n\gamma_c}\)^{\langle {\bf{L}}(n\gamma_c),a\rangle}.
\ee
Here ${\bf{L}}(n\gamma_c)$ is defined as the canonical lift of the two way streets determined by the soliton data 
\be
{\bf{L}}(n\gamma) : = \sum_{p\in \CW_c} \alpha_\gamma(p) \pi^{-1}(p) = \sum_{N\in \IZ} L_N(\gamma). 
\ee
As it is written, the definition is ambiguous because the $\alpha_\gamma$of the two way streets of type $ii$ remained undetermined. However, the precise values of these $\alpha_\gamma$'s do not 
affect the validity $\CK$-wall formula. Hence, we can fix them arbitrarily. We choose to do in such a way that $L_N(\gamma)$'s lift to closed cycles. Quotient with ker$(Z)$ leads to 
\be
[L_N(\gamma) ] = [L_{N'}(\gamma)].
\ee
Invoking the same logic as \cite{Gaiotto:2012rg,Longhi:2016rjt, Banerjee:2018syt} the BPS deneracies are given by 
\be\label{eq:Omega-formula-review}
\Omega(\gamma) =[L_N(\gamma)]/\gamma.
\ee
We will use this formula to compute the BPS degeneracies for the examples in this paper.

\subsection{Donaldson-Thomas and Gopakumar-Vafa invariants} 

In this section we recall some facts about the DT and GV invariants and the relations among them from both physical and mathematical points of view. Although the GV invariants arise very naturally 
from the physics perspectives, defined by the spin contents of the BPS states of M2-branes wrapping curves in a Calabi-Yau threefold $X$, their mathematical interpretation is still not well-understood, 
in the most general settings. 

The invariants defined in \cite{Gopakumar:1998ii, Gopakumar:1998ki} arise from M-theory compactified on a Calabi-Yau threefold $X$, which engineers a 5d, $\CN=1$ 
theory, which we call $T[X]$. The relevant BPS states in $T[X]$, for the computation in \cite{Gopakumar:1998ii, Gopakumar:1998ki}, corresponds to M2-branes wrapping a 2-cycle $\beta \in H_2(X,\IZ)$. This gives rise to $n_{(j_{L},j_{R})}^{\beta}$ particles in the representation\footnote{Here $j$ denotes the spin of the representation of dimension $2j+1$.} 
\be
\left[\left(\frac{1}{2},0 \right)\oplus 2(0,0)\right]\otimes (j_{L},j_{R})\qquad j_{L,R}\in\frac{1}{2}\mathbb{Z}_{\geq 0} 
\ee
of the little group $\mathrm{Spin}(4)\cong SU(2)_{L}\times SU(2)_{R}$ of $T[X]$. Each of these particles have an extra index $n\in\mathbb{Z}$ (labelling the units of momentum along the M-theory circle) and their central charge is given by
\be
\frac{1}{\lambda}\left(\int_{\beta}\omega(X)+2\pi i n\right)
\ee
where $\omega(X)$ is the complexified K\"ahler class of $X$. The number $n_{j_{L}}^{\beta}$ defined by
\be
n_{j_{L}}^{\beta}=\sum_{j_{R}}(-1)^{2j_{R}}(2j_{R}+1)n_{(j_{L},j_{R})}^{\beta}
\ee
The number $n_{j_{L}}^{\beta}$ is shown to be invariant under small deformations of the complex structure of $X$, however it can jump as one moves in the space of  complex structures, as short multiplets combine to produce long multiplets. The GV invariants are then by simply making a change of basis for the left representation $(j_{L})$. Every irreducible representation of $SU(2)$ can be decomposed as 
\be
\label{basischange}
(j_{L})=\sum_{r=0}^{2j_{L}}\alpha_{r}I_{r}\qquad I_{r}:=[(\frac{1}{2})\oplus (0)]^{\otimes r},I_{0}:=(0)\qquad \alpha_{r}\in \mathbb{Z}
\ee
Then, the GV invariants $n^{\beta}_{r}$ are defined by the formula
\be
\sum_{r}n^{\beta}_{r}I_{r}=\sum_{j_{L}}n^{\beta}_{j_{L}}(j_{L})
\ee

The invariants $n_{r}^{\beta}$ are related to the Gromov-Witten (GW) invariants (at genus $g$) $N_g^\beta$ by the formula \cite{Gopakumar:1998ii, Gopakumar:1998ki}
\be
\label{def:GW-GV}
F(\lambda,Q):=\sum_{\beta>0, g\ge 0} N_g^\beta  \lambda^{2g-2} Q^\beta = \sum_{\beta>0,g\ge 0, k\ge 1} \frac{n_g^\beta}{k}\(2 \sin\(\frac{k\lambda}{2}\)\)^{2g-2} Q^{k\beta}. 
\ee 
here the sum goes over effective curve classes $\beta\neq 0$ and $Q^{\beta}=\exp(-2\pi\int_{\beta}\omega(X))$, 
More intrinsic mathematical definitions for $n_{r}^{\beta}$ have been proposed in \cite{Katz:1999xq, Hosono:2001gf, Maulik:2016rip}, but for us the definition in terms of BPS state counting will suffice. For us it will be particularly important the values of GV invariants $n^{\beta}_{0}$ because we can interpret them as the number of bound states of a D2-brane or M2-brane of charge $\beta$ bounded with any number of D0-branes, this means:
\be
\Omega(\beta+kD0)=n^{\beta}_{0}
\ee 
which is a quantity that will be computed by exponential networks as we will verify in subsequent sections. More generally we will be computing generalized Donaldson-Thomas (DT) invariants 
which arise in physics as BPS indices counting stable D-brane boundstates with fixed D6-D4-D2-D0 charges. 
Mathematically they are roughly Euler characteristics of moduli spaces of semi-stable sheaves, with fixed K-theory classes \cite{Joyce:2008pc, Katz:2004js}.
For the case of bound states of D2 and D0 branes bounded to a single D6 the DT invariants are labelled by their D2 charge $\beta\in H_{2}(X,\mathbb{Z})$ and the number $n$ of D0 branes. The DT partition function is given by 
\be
Z_{DT} = \sum_{n\geq 0,\beta} DT_\beta^n (X) q^n Q^\beta\,
\ee
where $q$ weights the number of D0 branes and $DT_\beta^n (X)$ are the DT invariants. Note that here we allow $\beta$ to be trivial in the sum.

The DT/GW correspondence \cite{2003math.....12059M} implies a relation between the DT and the GV invariants. 
\be
M(-q)^{-\chi(X)}Z_{DT}=e^{F(\lambda,Q)}
\ee
after we have identified $q=-e^{i\lambda}$ and $M(q)=\prod_{k=1}^{\infty}(1-q^{k})^{-k}$ is the McMahon function. The contribution to $Z_{DT}$ from the $\beta=0$ sector, i.e. bound states of a single D6 with an arbitrary number of D0 branes is given by
\be
Z_{DT}^{\beta=0}=M(-q)^{\chi(X)}
\ee
so, we identify\footnote{The sign is induced by the exponents in the factorization of McMahon's function $M(q)=\prod_n (1-q^n)^{-n}$.}
\be
\Omega(nD0)=-\chi(X)
\ee
a quantity that will be also be verified by our computations.

The DT invariants can be related to generalized DT invariants in degree zero (which are the objects we actually compute with exponential networks) by wall-crossing formulae  \cite{Kontsevich:2008fj}. This will be reviewed below in explicit examples, and will serve as a check of our results against known previous ones. Generalized DT invariants are generally very hard to compute from first principles. Our framework provides a new systematic approach to computing them.

\section{$\CO(-1)\oplus\CO(-1)\to\IP^1$}\label{sec:conifold}

In this section we set up the formalism of exponential networks for the resolved conifold following our previous work \cite{Banerjee:2018syt}.
This geometry is among the most well-studied local Calabi-Yau threefolds in the literature. 
Important work on its enumerative invariants include pioneering work by Szendr\"oi \cite{Szendroi:2007nu}  on non-commutative Donaldson-Thomas invariants. These results were interpreted by Jafferis and Moore, in terms of framed wall-crossing of D6-D2-D0 states, in presence of a single noncompact D6 brane \cite{Jafferis:2008uf}.

Here we will start from scratch, and rederive some of these results using the framework of exponential networks. As reviewed above, these compute directly BPS states of a 5d theory engineered by the local Calabi-Yau times $\IR^4\times S^1$. 
The BPS indices are therefore  degree-zero generalized Donaldson-Thomas invariants of D2-D0 boundstates, which are directly related to Gopakumar-Vafa invariants (in our context, the former are ``KK modes'' of the latter).
Since our technique for computing these BPS states differs significantly from the original definition of these invariants, the fact that we find a perfect match provides a highly nontrivial check of nonabelianization for exponential networks.

\subsection{Geometry and its mirror}

Let us start by reviewing some of the relevant geometric background.
The resolved conifold may be defined starting from a $U(1)$ gauged linear sigma model with charges $(+1,+1,-1,-1)$ for coordinates $(w_1,\dots ,w_4)\in \IC^4$. 
Physically one considers the zero-locus of the $D$-term equations
\be
\label{sympl-1-con}
	|w_1|^2 + |w_2|^2 - |w_3|^2 - |w_4|^2 =  r
\ee
and takes a $U(1)$ quotient for gauge-invariance .
Mathematically this corresponds to a symplectic quotient  of $\IC^4$, of which there are really three cases, depending on whether $r<0$, $r=0$ or $r>0$. 
All are birationally equivalent, and in fact $r>0$ and $r<0$ are smoothly connected by turning on a generic $\theta$ term \cite{Witten:1993yc}.
For example taking $r\in \IR^+$, we may consider the locus $\tilde V \subset \IC^4$ with $w_i\neq 0$ and the locus $V_1\subset \IC^4$ where $b_i=0$ but not all $a_i$ vanish. Then the symplectic quotient is $(\tilde V\cup V_1)/\IC^*$.
The mirror curve for this can be obtained by T-duality as defined in Hori-Vafa mirror symmetry \cite{Hori:2000kt}
\be
x_1 x_2 x_3^{-1} x_4^{-1} = Q^{-1}, \quad x_1 + x_2 + x_3 + x_ 4 = 0
\ee
where $|x_i| = e^{-|w_i|^2}$ and $|{Q}| = e^{r}$. Fixing the patch $x_3 = 1$, and setting $x_4 = x, x_1 = y$, one gets the following as the 
mirror curve 
\be
1 + x + y + Q^{-1} xy^{-1} = 0. 
\ee
Changing framing $x \mapsto xy^2$
and rescaling  $x \mapsto x Q$,\footnote{These leave the symplectic form $d\log x\wedge d\log y$ invariant up to holomorphic factors.} one obtains the curve 
\be
\label{mirr-con}
1 + y + xy + Qxy^2  = 0 \quad \subset\quad \IC^*_x \times \IC^*_y \,.
\ee
We will study this form of the mirror curve with exponential networks. 
As we will see presently, the curve (\ref{mirr-con}) satisfies the criteria explained in Appendix \ref{app:framing} regarding the choice of framing.

As we explained in \cite{Banerjee:2018syt}, it is natural from the viewpoint of 5d gauge theory on a circle to view the curve as a ramified covering of $\IC^*_x$.
There are two branches located at 
\be
	y_\pm = 
	\frac{-1-x\pm\sqrt{(x+1)^2-4 Q x}}{2 Q x}\,.
\ee
These branches meet in correspondence of two branch points, whose locations are
\be\label{eq:conifold-branch-points}
	x_\pm = -1+2 Q \pm 2 \sqrt{Q^2-Q}\,.
\ee
Punctures on this curve are defined to be all those points where the curve intersects the lines $x=0,\infty$ or $y=0,\infty$.
There are four such points: 
\be \label{punc1}
\begin{split}
	&\mathfrak{p}_1:\ (x=0,y=\infty)
	\qquad
	\mathfrak{p}_2:\ (x=0,y=-1)
	\\
	&\mathfrak{p}_3:\ (x=\infty,y=0)
	\qquad
	\mathfrak{p}_4: \ (x=\infty,y=-Q^{-1})
\end{split}
\ee

The curve is then a two-fold cover of the $x$-plane with two branch points, i.e. it is a sphere. 
The punctures project to $x=0,\infty$ but at each of these points they live on both branches, making $\Sigma$ a four-punctured sphere.

\begin{figure}[h!]
\begin{center}
\includegraphics[width=0.35\textwidth]{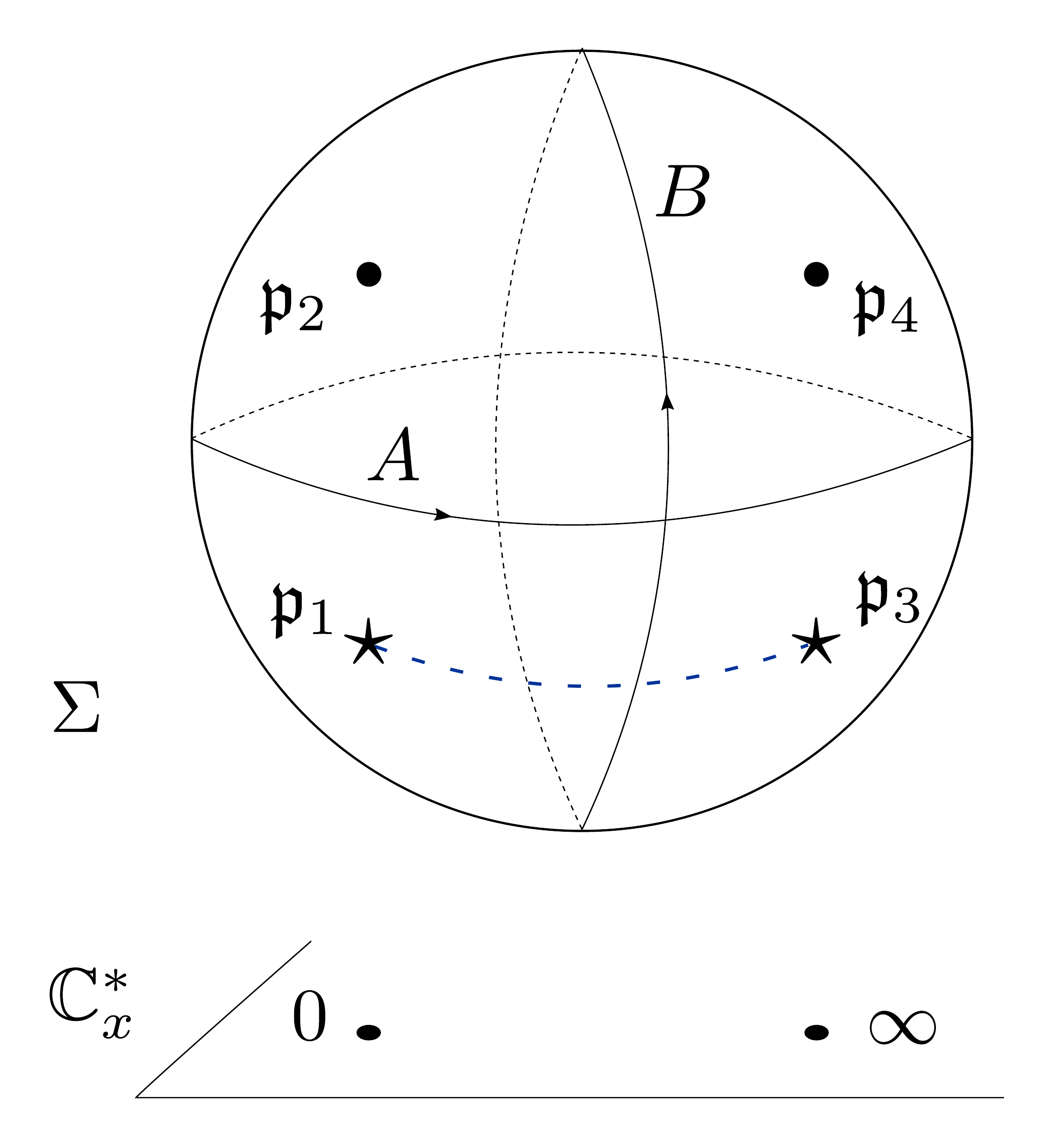}
\caption{Punctures of the conifold curve (\ref{mirr-con}). The dashed line denotes the logarithmic cut for $\lambda$.}
\label{fig:conifold-punctures}
\end{center}
\end{figure}

\subsubsection{Types of punctures}

Although the curve has four punctures, not all of them are on the same footing for certain purposes.
We now briefly discuss the different types of punctures, since this seems to have gathered little attention in previous literature. 

Our criterion for classifying punctures is to study the behavior of the one-form 
\be\label{eq:differential}
	\lambda = \log y \, d\log x
\ee
in correspondence of each.
On the mirror curve, the punctures where $y=0,\infty$ are $\mathfrak{p}_1$ and $\mathfrak{p}_3$, shown in Figure \ref{fig:conifold-punctures}.
Note that 
\be
	\mathfrak{p}_1 = \lim_{x\to 0} y_-(x)\,,
	\qquad
	\mathfrak{p}_3 = \lim_{x\to \infty} y_-(x)\,.
\ee
To compare labeling of sheets $\pm$ at $x=0,\infty$ one needs to choose a trivialization for the covering $\Sigma\to C$. 
Here we have chosen real $Q>1$, then $x_\pm$ are on the positive real axis, and we take the square-root branch cut to run  between them on the real axis. Then sheets at $x=0$ and $x=\infty$ can be properly matched by analytic continuation.\footnote{Continuation must be done away from the real axis, since branch points live there. A convenient choice is to continue sheets along the imaginary axis.}

Therefore there is a logarithmic cut for (\ref{eq:differential}) running between $\mathfrak{p}_1,\mathfrak{p}_3$, its projection to $\IC^*_x$ must not cross the square-root cut connecting sheet $-$ to sheet $+$. 
An example of trivialization satisfying this constraint is shown  below in Figure \ref{fig:conifold-Q-3-2-theta-0}. 
Let us stress at this point that the square-root cut concerns the projection $\Sigma\to\IC^*_x$, while the logarithmic cut concerns the projection $\tSigma\to\Sigma$, a $\IZ$-covering resolving logarithmic ambiguities of (\ref{eq:differential}).

The holomorphic differential (\ref{eq:differential}) has the following asymptotics near each puncture  
\be\label{eq:differential-p1-p2-p3-p4}
\begin{split}
&(\mathfrak{p}_2)
\\
&(\mathfrak{p}_1)
\end{split}
\quad
\begin{split}
&\lambda_+ = \pi i \ d\log x + \dots
\\
& \lambda_- =  -\log x \ d\log x + \dots
\end{split}
\qquad
\qquad
\begin{split}
&(\mathfrak{p}_4)
\\
&(\mathfrak{p}_3)
\end{split}
\quad
\begin{split}
&\lambda_+ = \(\log Q-\pi i\)\ d\log w + \dots
\\ & 
\lambda_- = - \log w\ d\log w +\dots
\end{split}
\ee
where for the two punctures above $x=\infty$, we adopted the local coordinate $w=x^{-1}$.

Above $x=0$ on the positive sheet (puncture $\mathfrak{p}_2$) the differential approaches $\frac{dx}{x}$ and is single valued, but on the negative sheet (puncture $\mathfrak{p}_1$) the differential approaches ${\log x}\frac{dx}{x}$ and is  multivalued.
Likewise above $x=\infty$ we find that puncture $\mathfrak{p}_3$ is of logarithmic type, whereas puncture $\mathfrak{p}_4$ is regular. Throughout the paper we depict regular punctures by $\bullet$ and logarithmic ones by $\star$.

\subsubsection{Homotopy vs homology, periods and flop transition}

Periods of differentials with logarithmic branching require some care, compared to more familiar meromorphic differentials encountered, for example, in Seiberg-Witten theory.
The integration cycles must be specified by \emph{homotopy} classes, instead of homology classes.
Such refined information enters both through the dependence of periods on a choice of basepoint, and through the non-abelian nature of the monodromy group.\footnote{A basic example of these features is the integral representation of the dilogarithm function. See \cite[Section 4.4.1]{Banerjee:2018syt} for a discussion.}
It may happen that periods of $\lambda$ along certain homotopy classes do not depend on the choice of basepoint, in that case one may talk about periods of \emph{homology cycles}. This is the case for cycles corresponding to (mirror) charges of physical states, such as D2 and D0.

\subsubsection*{D2 cycle}

\begin{figure}[h!]
\begin{center}
\includegraphics[width=0.55\textwidth]{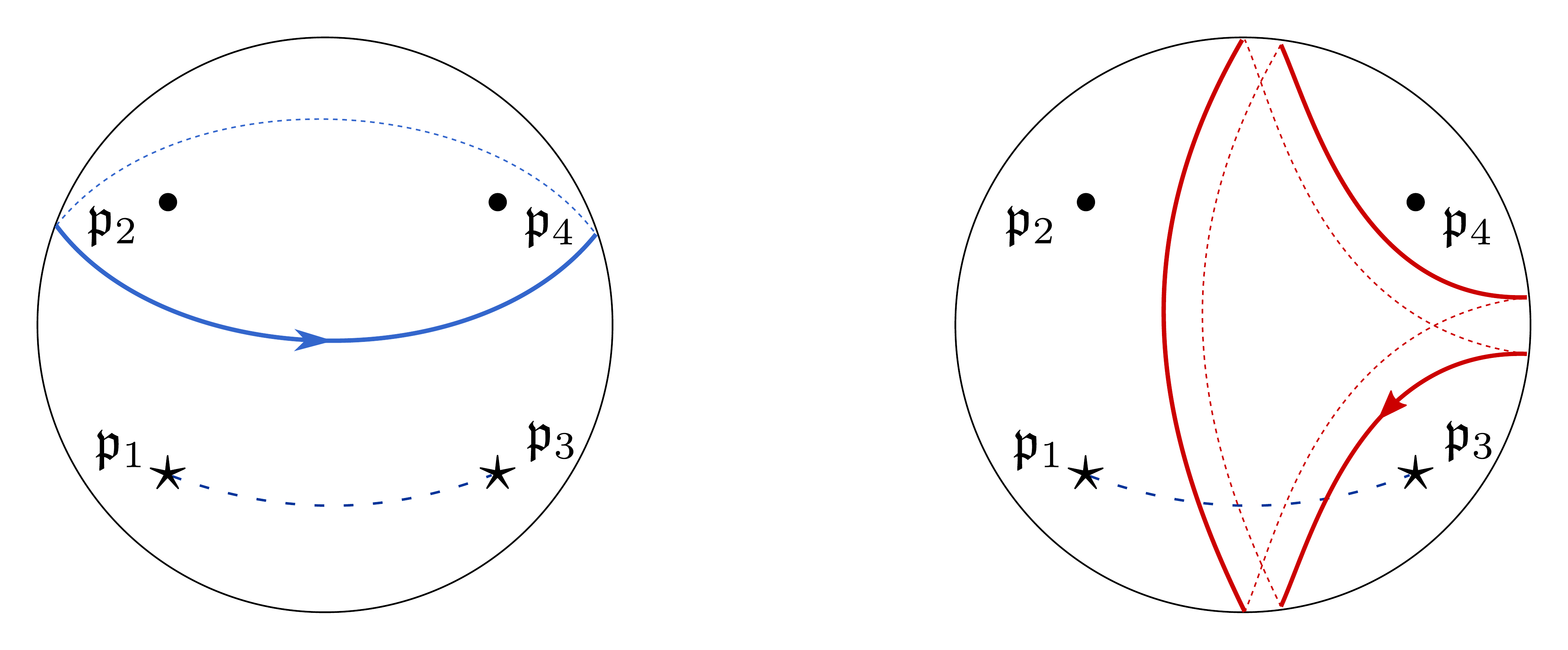}
\caption{Cycles corresponding D2 (blue) and D0 (red) branes on the conifold mirror curve. Solid lines represent parts of the cycles on the front side of the sphere, dashed lines represent parts on the back side. The dashed line connecting two punctures is the logarithmic cut, it lies on the front.}
\label{fig:conifold-cycles}
\end{center}
\end{figure}

Let $p\in \Sigma$ be an arbitrary basepoint, such that it does not coincide with any of the punctures.
Let $C_i \in \pi_1(\Sigma,p)$ be a cycle based at $p$  and encircling  $\mathfrak{p}_i$ counterclockwise.
Periods of $C_1$ and $C_3$ depend on $p$, because these paths cross the logarithmic cut once.
On the other hand, periods of $C_2$ and $C_4$ don't depend on $p$, they are well-defined functions of   homology classes $[C_2]$ and $[C_4]$ .
Noting that $C_1\circ C_3\circ C_2\circ C_4 $ is trivial, it follows that the period of $(C_1\circ C_3)^{-1}$ is the same as that of $C_2\circ C_4$, and therefore is independent of the basepoint.\footnote{It is  easy to find a representative of this homotopy class that does not intersect the logarithmic cut.}
We define $A$ to be the homology class of this distinguished cycle 
\be
	A : = [C_2\circ C_4] = [C_2]+[C_4] \in H_1(\Sigma,\IZ) \,.
\ee
Its period is easily obtained as the sum of residues from (\ref{eq:differential-p1-p2-p3-p4})
\be\label{eq:Z-A}
	Z_A 
	= \frac{1}{2\pi R} \, \int_{C_2\circ C_4} \lambda
	= \frac{i}{R} \log Q \,.
\ee
This is the D2 central charge, hence we take the D2 cycle to be
\be
	\gamma_{\text{D2}} = A\,.
\ee 
as also depicted in Figure \ref{fig:conifold-cycles}.

The flop transition involves continuation from  $r>0$ to $r<0$ in (\ref{sympl-1-con}), or after complexification taking $|Q|>1$ to $|Q|<1$.
It will be useful to note that performing the change of variables\footnote{This is a composition of framing change with coordinate rescaling and inversion.} $x\to 1/(xy^2 Q), y\to yQ$ in (\ref{mirr-con}) gives
\be\label{eq:conifold-curve-flopped}
	1 + y + xy + Q^{-1} x y^2 = 0 \,,
\ee
up to an overall monomial prefactor.
It is thus manifest that $Q\to Q^{-1}$ can be undone by a change of coordinates.
For later purposes, let us stress that $\mathfrak{p}_2$ and $\mathfrak{p}_4$ get exchanged, while $\mathfrak{p}_1$ and $\mathfrak{p}_3$ stay fixed.

\subsubsection*{D0 cycle}

The D0 cycle is more subtle, and shown in Figure \ref{fig:conifold-cycles} (see also \cite[Figure 49]{Eager:2016yxd}).
Since this cycle crosses the logarithmic cut, it is not obvious a priori that its period is independent of the basepoint, and that it descends to a functional of the homology class of the cycle.
To show that, let us consider the limit $Q\to\infty$ in which, as we argue in detail below, the curve degenerates into two three-punctured spheres.
We will focus on the sphere containing $\mathfrak{p}_3$ and $\mathfrak{p}_4$. After degeneration, the third puncture $\mathfrak{p}_B$ arises from pinching the B cycle shown in Figure \ref{fig:conifold-punctures} and is also of logarithmic type.
Switching to a local coordinate $w\sim x^{-1}$ the punctures above infinity map to two punctures above $w=0$, and the third puncture maps to $w=\infty$ (see discussion around (\ref{eq:mirr-conifold-factorized-1})).
The projection of the $\overline{\text{D0}}$ cycle has been studied in detail in our previous paper \cite[Figure 10]{Banerjee:2018syt}. The D0 cycle is obtained by reversing the orientation, and it is straightforward to check that its lift is indeed the one shown in Figure~\ref{fig:conifold-cycles}.

To compute the period of $\overline{\text{D0}}$, we choose a linear framing for the three-punctured sphere, with equation $1-z-y=0$. In this framing $\lambda= \log y\, d\log z$ has logarithmic branching at $z=1,\infty$ and has standard monodromy at $z=0$. 
In fact, the primitive of  $\lambda$ is $\Li_2(z)$, whose branching structure on the $z$-plane is well-known. In the change of framing, $\mathfrak{p}_B$ maps to $z=\infty$, $\mathfrak{p}_3$ to $z=1$ and $\mathfrak{p}_4$ to $z=0$.
Choosing a basepoint $p$ on the $\overline{\text{D0}}$  cycle on the solid line near $\mathfrak{p}_3$ in Figure \ref{fig:conifold-cycles}, we may represent the path as an element of $\pi_1(C_{0,3},p)$
\be
	\gamma_{D0} =  \pi_\infty \circ \pi_0^{-1} \circ \pi_\infty^{-1} \circ \pi_0\,,
\ee
where $\pi_0$ (resp. $\pi_\infty$) is a cointerclockwise path around $z=0$ corresponding to $\mathfrak{p}_4$ (resp. around $z=\infty$ corresponding to $\mathfrak{p}_B$).
The monodromy matrices for $\Li_2(z)$ corresponding to counterclockwise $\pi_0,\pi_\infty$ are (see for example \cite[Section 4]{Banerjee:2018syt})
\be
	M_0 = \left(
\begin{array}{ccc}
 1 & -2 i \pi  & 0 \\
 0 & 1 & 0 \\
 0 & 0 & 1 \\
\end{array}
\right)
	\qquad
	M_\infty = \left(
\begin{array}{ccc}
 1 & 2 i \pi  & 4 \pi ^2 \\
 0 & 1 & -2 i \pi  \\
 0 & 0 & 1 \\
\end{array}
\right)
\ee
acting from the left on $\(\Li_2(z),\log z,1\)^t$.
Since
\be	
	M_{\infty}\cdot M_0^{-1} \cdot M_\infty^{-1} \cdot M_0 = 
	\left(
\begin{array}{ccc}
 1 & 0 & -4 \pi ^2 \\
 0 & 1 & 0 \\
 0 & 0 & 1 \\
\end{array}
\right)
\ee
it follows the the overall shift of $\Li_2(z)$ along $\gamma_{D0}$ is $-4\pi^2$.
This is the shift of the primitive of $\lambda$, and by our normalization (\ref{eq:Z-A}) this implies that 
\be\label{eq:conifold-D0-period}
	Z_{\overline{\text{D0}}} = \frac{1}{2\pi R}\oint_{-\gamma_{D0}}\lambda = -\frac{2\pi}{R}\,.
\ee
Obviously the D0 central charge is obtained by reversing the integration cycle, which gets rid of the sign. 
This confirms that the period is independent of the basepoint, hence $\gamma_{D0}$ is a well-defined homology cycle. (There is also a specular D0 cycle, encircling $\mathfrak{p}_1,\mathfrak{p}_2$ for which a similar analysis applies. It will appear in the study of BPS states below).
Also note that (\ref{eq:conifold-D0-period}) coincides with a unit of KK momentum, as expected for the D0 central charge, and matches precisely with the independent computation in \cite[eq. (4.12)]{Banerjee:2018syt}.

\subsubsection{Distinguished values of $Q$ and curve degeneration}\label{sec:conifold-factorization}

There are three distinguished regions in the moduli space of the curve (\ref{mirr-con}): the region near $Q=1$ and those near $Q=0,\infty$.

\subsubsection*{Conifold point}
$Q=1$ corresponds to the conifold point, where the curve (\ref{eq:conifold-curve-flopped}) factorizes. 
At this point the curve is maximally symmetric (see positions of punctures (\ref{punc1})), and branch points coincide at $x_{+}=x_{-} = 1$ reflecting the fact that cycle $A$ pinches.
The cycle around the two branch points coincides precisely with $A$, since both separate $\{\mathfrak{p}_1,\mathfrak{p}_3\}$ from $\{\mathfrak{p}_2,\mathfrak{p}_4\}$.

\subsubsection*{Degeneration at large $Q$}
Near $Q= \infty$ branch points become infinitely separated 
\be
	x_- \to 0,\qquad x_+\to \infty\,,
\ee
and the curve develops a long, thin tube. 
The two punctures $\{\mathfrak{p}_3,\mathfrak{p}_4\}$  above $x=\infty$ get merged, while those above $x=0$, $\{\mathfrak{p}_1,\mathfrak{p}_2\}$, stay at finite distance from each other. 
In fact rescaling $x\to x/Q$ while taking the limit $Q\to\infty$ corresponds to zooming into $x=0$, and brings the curve (\ref{mirr-con}) into the form
\be\label{eq:mirr-conifold-factorized-1}
	1 + y  + x y^2 = 0
\ee
which is the mirror curve of $\IC^3$ in quadratic framing.
Similarly one can zoom in near the region $x=\infty$. Changing variables $y\to y Q^{-1} , x\to w^{-1} Q$ and taking the limit $Q\to\infty$ gives
\be\label{eq:mirr-conifold-factorized-2}
	\frac{1}{w}\(w + y + y^2\) = 0\,.
\ee
Up to the inessential prefactor, this is the $\IC^3$ curve studied in detail in \cite{Banerjee:2018syt}, it has two punctures above $w=0$ (corresponding to $\{\mathfrak{p}_3,\mathfrak{p}_4\}$) and one puncture above $w=\infty$ (arising from the merger of $\{\mathfrak{p}_1,\mathfrak{p}_2\}$).

\subsubsection*{Degeneration at small $Q$}

\begin{figure}[h!]
\begin{center}
\includegraphics[width=0.45\textwidth]{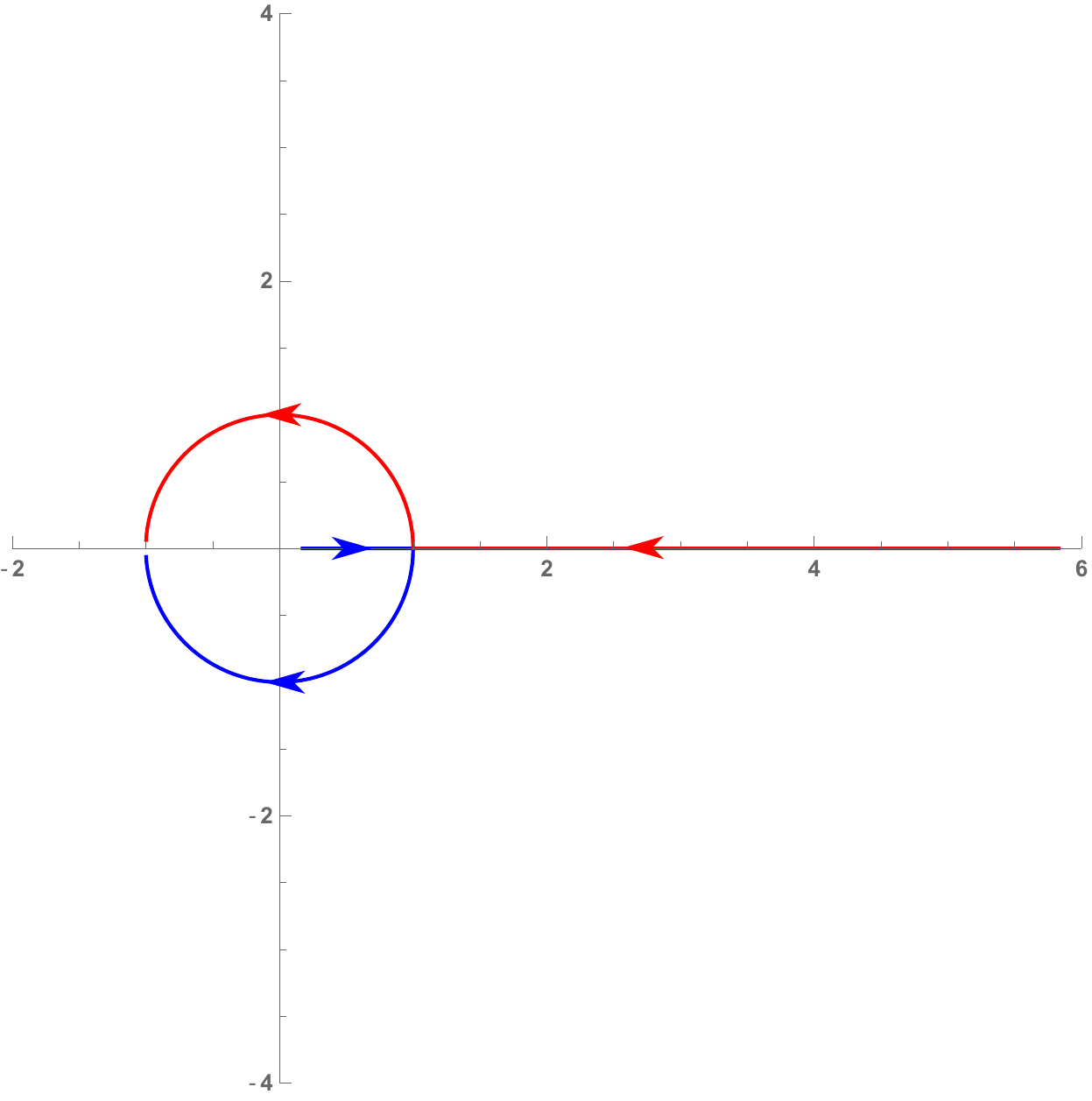}
\caption{Motion of branch points $x_\pm$ with decreasing $Q\to 0$ along the real positive axis.}
\label{fig:conifold-branch-points-motion}
\end{center}
\end{figure}

Finally when $Q\to 0$ the curve also undergoes degeneration, although in a more subtle way.
Let us start with $Q>1$ along the positive real axis, and start decreasing $Q$ all the way to $Q\to 0$ along the real axis.
As long as $Q>1$ the two branch points lie along the positive real $x$-axis, they move towards each other and collide when $Q=1$ at $x_-=x_+ = 1$. 
Decreasing $Q$ further, they separate again and start moving on the unit circle, in opposite directions starting at $x=1$ and ending with a final collision at $x=-1$.
This behavior is shown in Figure \ref{fig:conifold-branch-points-motion}

\begin{figure}[h!]
\begin{center}
\includegraphics[width=0.85\textwidth]{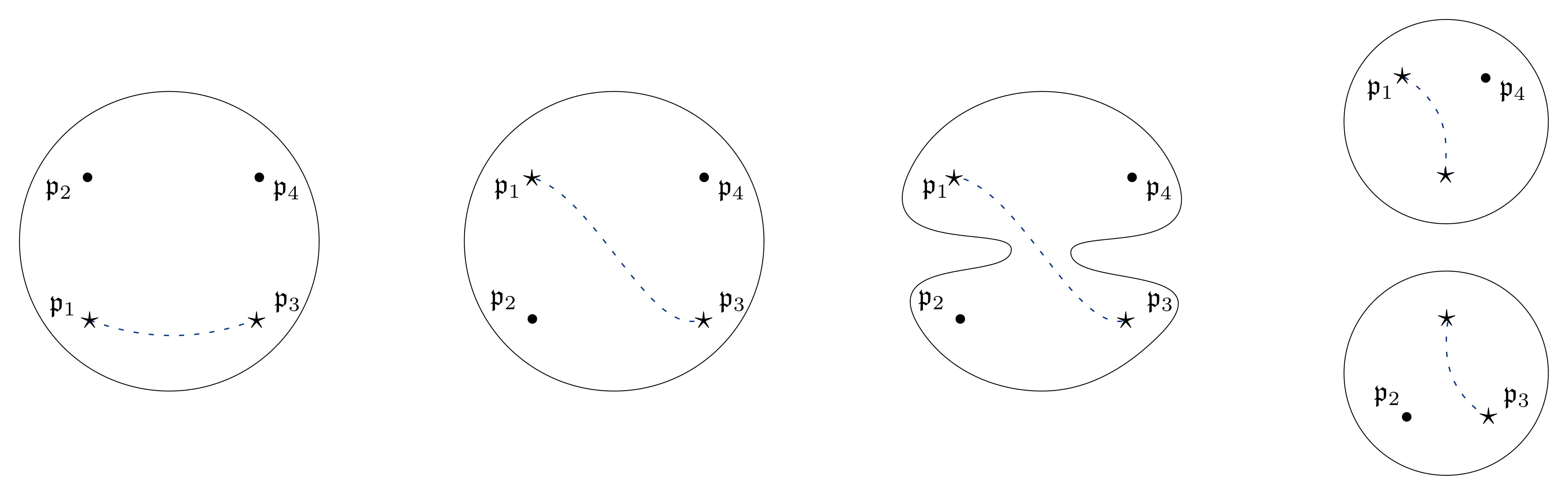}
\caption{Stages in the degeneration of the conifold mirror curve as $Q\to 0$.}
\label{fig:conifold-degeneration-Qeq0}
\end{center}
\end{figure}

As a consequence of this, the branch cut between $x_\pm$, that was originally (at $Q>1$) stretching along the real axis, ends up encircling the puncture at $x=0$.
Sweeping the cut across this point is equivalent to exchanging $\mathfrak{p_1} \leftrightarrow\mathfrak{p}_2$.
Finally, the branch cut becomes contractible and the two branch points simply annihilate, leaving disjoint sheets $y_\pm$, each covering $\IC^*_x$ once. A sketch of this is shown in Figure \ref{fig:conifold-degeneration-Qeq0}.

Let us describe this degeneration in terms of equations.
Taking directly $Q\to0$ (\ref{mirr-con}) gives
\be
	1+y + xy = 0
\ee
this is the mirror curve of $\IC^3$ in linear framing. It is a one-fold covering of the $x$-plane with logarithmic punctures at $x=-1,\infty$. Since the puncture at $x=0$ is regular, it must be $\mathfrak{p}_2$ (the one at infinity being logarithmic, it must be $\mathfrak{p}_3$), therefore this is the sheet corresponding to $y_-$.
Changing instead coordinate $y\to (Q z)^{-1}$ and taking $Q\to 0$ gives
\be
	1+x+x z^{-1} = 0
\ee
which is another curve for $\IC^3$ in linear framing. 
This has logarithmic punctures at $x=-1$  and $x=0$ (corresponding to $\mathfrak{p_1}$) and a regular puncture at $x=\infty$  (corresponding to $\mathfrak{p}_4$), therefore it corresponds to sheet $y_+$.

The two limits $Q\to0$ and $Q\to\infty$ are related by a flop transition, in both cases the period of cycle $A = [C_2\circ C_4]$ grows to infinity. The fact that this happens even when $Q\to0$ is not obvious at first sight, since branch points collide at $x=-1$ and one may be led to conclude (incorrectly) the the cycle is pinching.
This apparent puzzle is explained by the phenomenon illustrated above, which exchanges punctures $\mathfrak{p}_1$ and  $\mathfrak{p}_2$.
Another way to view the flop is by implementing the change of coordinates discussed around (\ref{eq:conifold-curve-flopped}): this would give back the same curve as the one used in the limit $Q\to\infty$, after exchange of $\mathfrak{p}_2$ and $\mathfrak{p}_4$. 
After this change of coordinates we would then find a degeneration into two three-punctured spheres (\ref{eq:mirr-conifold-factorized-1}) and (\ref{eq:mirr-conifold-factorized-2}). However now $\{\mathfrak{p}_1,\mathfrak{p}_4\}$ would be on one side, and $\{\mathfrak{p}_2,\mathfrak{p}_3\}$ on the other, confirming the degeneration depicted in Figure \ref{fig:conifold-degeneration-Qeq0}.

\subsection{BPS states} 

Having examined in detail the geometry of the mirror curve for the conifold, we proceed in this section with the study of the BPS spectrum.
We will plot exponential networks at various phases and detect BPS states of the theory as (generalized) saddles. 
Since we have already shown in (\ref{eq:conifold-curve-flopped}) that the curves for $|Q|>1$ and $|Q|<1$ are related by a simple change of coordinates, it follows that the respective exponential networks are essentially the same. Hence these two regions have the same saddles, and therefore the same BPS spectrum, as should be expected.\footnote{Due to the absence of compact 4-cycles and 6-cycles, the spectrum of D2-D0 states cannot undergo wall-crossing, which would require the presence of mutually non-local BPS states.}
We will therefore focus entirely on the region $|Q|>1$, discussing the behavior of the BPS spectrum both at finite $Q$ and in the degeneration limit $Q\to\infty$.

\subsubsection{Conifold networks for $Q>1$}
\begin{figure}[h!]
\begin{center}
\fbox{\includegraphics[width=0.228\textwidth]{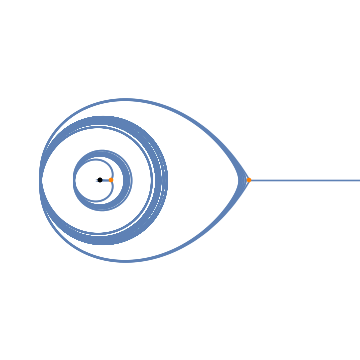}}
\hfill
\fbox{\includegraphics[width=0.228\textwidth]{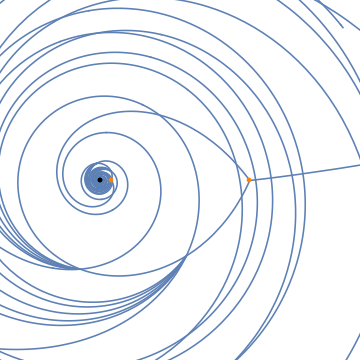}}
\hfill
\fbox{\includegraphics[width=0.228\textwidth]{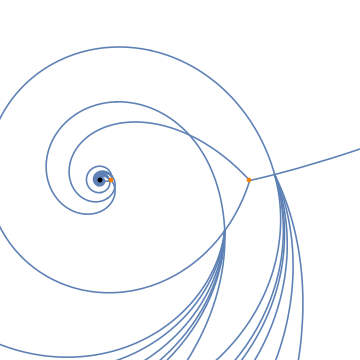}}
\hfill
\fbox{\includegraphics[width=0.228\textwidth]{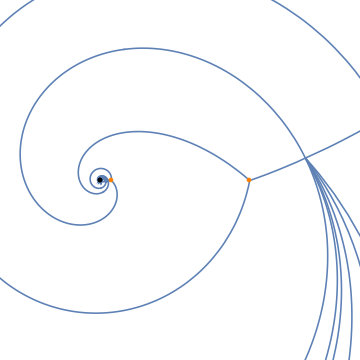}}
\\
\fbox{\includegraphics[width=0.228\textwidth]{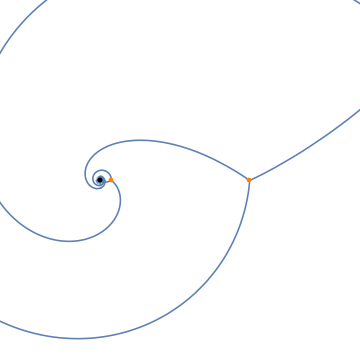}}
\hfill
\fbox{\includegraphics[width=0.228\textwidth]{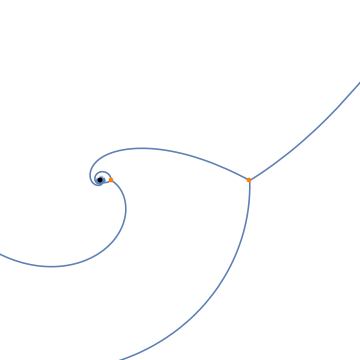}}
\hfill
\fbox{\includegraphics[width=0.228\textwidth]{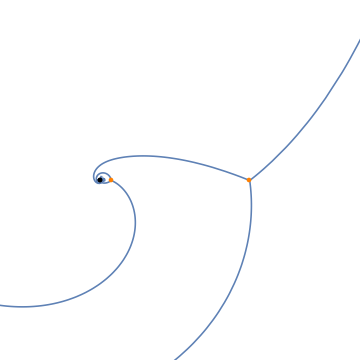}}
\hfill
\fbox{\includegraphics[width=0.228\textwidth]{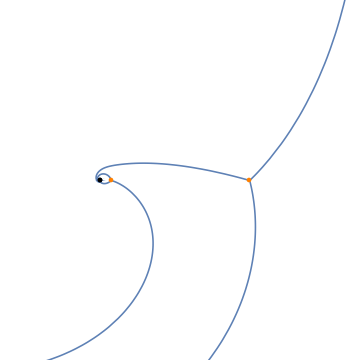}}
\\
\fbox{\includegraphics[width=0.228\textwidth]{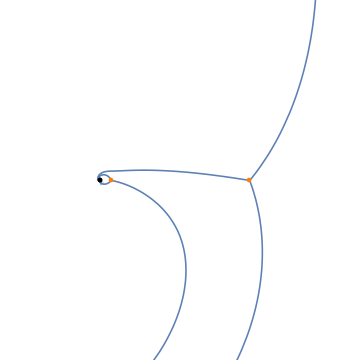}}
\hfill
\fbox{\includegraphics[width=0.228\textwidth]{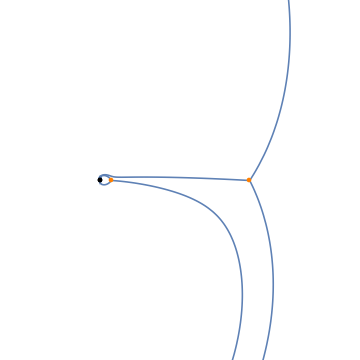}}
\hfill
\fbox{\includegraphics[width=0.228\textwidth]{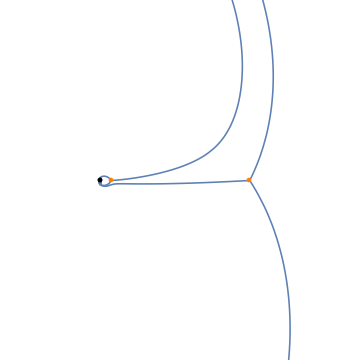}}
\hfill
\fbox{\includegraphics[width=0.228\textwidth]{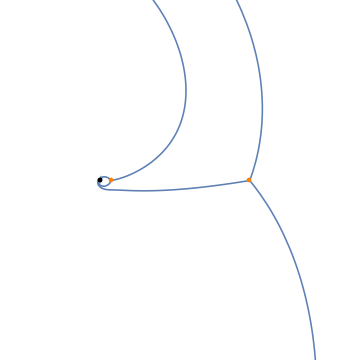}}
\\
\fbox{\includegraphics[width=0.228\textwidth]{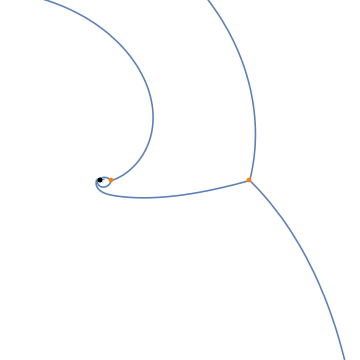}}
\hfill
\fbox{\includegraphics[width=0.228\textwidth]{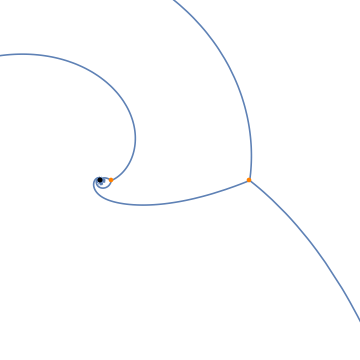}}
\hfill
\fbox{\includegraphics[width=0.228\textwidth]{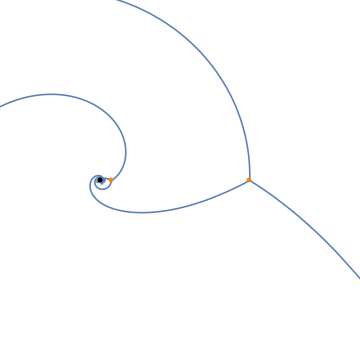}}
\hfill
\fbox{\includegraphics[width=0.228\textwidth]{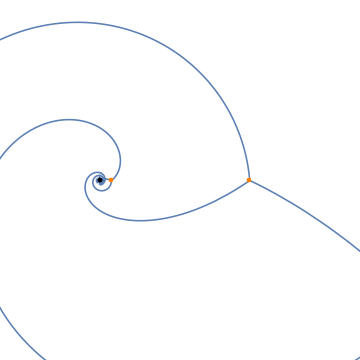}}
\\
\fbox{\includegraphics[width=0.228\textwidth]{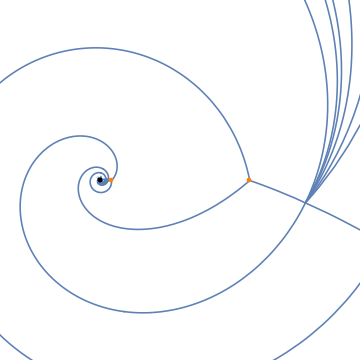}}
\hfill
\fbox{\includegraphics[width=0.228\textwidth]{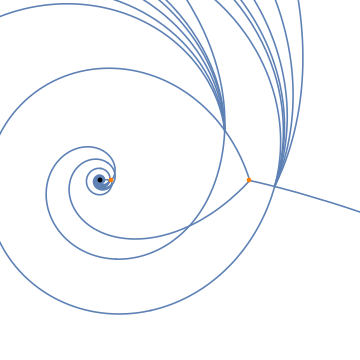}}
\hfill
\fbox{\includegraphics[width=0.228\textwidth]{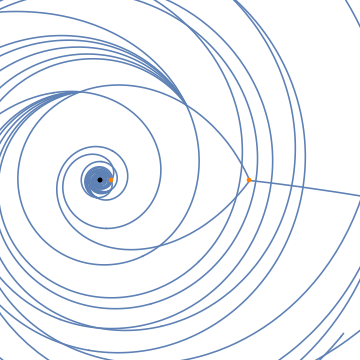}}
\hfill
\fbox{\includegraphics[width=0.228\textwidth]{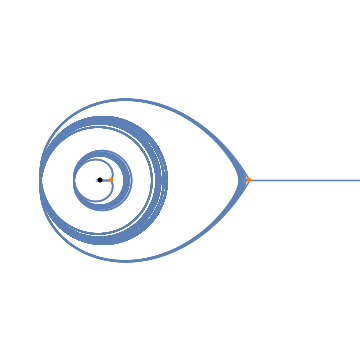}}
\caption{Exponential networks of the conifold for $Q=3/2$ with $\vartheta\in (0,\pi)$}
\label{fig:conifold_Q_3_2}
\end{center}
\end{figure}

Let us begin by choosing $Q$ real and greater than $1$. 
By direct inspection we find that all values of $Q$ in this range are qualitatively the same, in the sense that they have a spectrum of saddles with the same topological types. Moreover, while the phases of saddles depend on $Q$, the topological types of networks in-between these phases are unchanged throughout the region. 
A sample of exponential networks for $Q=3/2$ is shown in Figure \ref{fig:conifold_Q_3_2}.

To look for BPS states we must tune $\vartheta$ to the phases of their central charges, and verify the presence of a saddle.
Central charge of D0 branes and D2 branes were computed above: the former are real whereas the latter are purely imaginary (with the present choice of $Q$).

\subsubsection{Generalized saddles}\label{sec:conifold-saddles}

At $Q=3/2$ we find the following saddles through the evolution depicted in Figure \ref{fig:conifold_Q_3_2}.

\subsubsection*{D0 states}

\begin{figure}[h!]
\begin{center}
\includegraphics[width=0.5\textwidth]{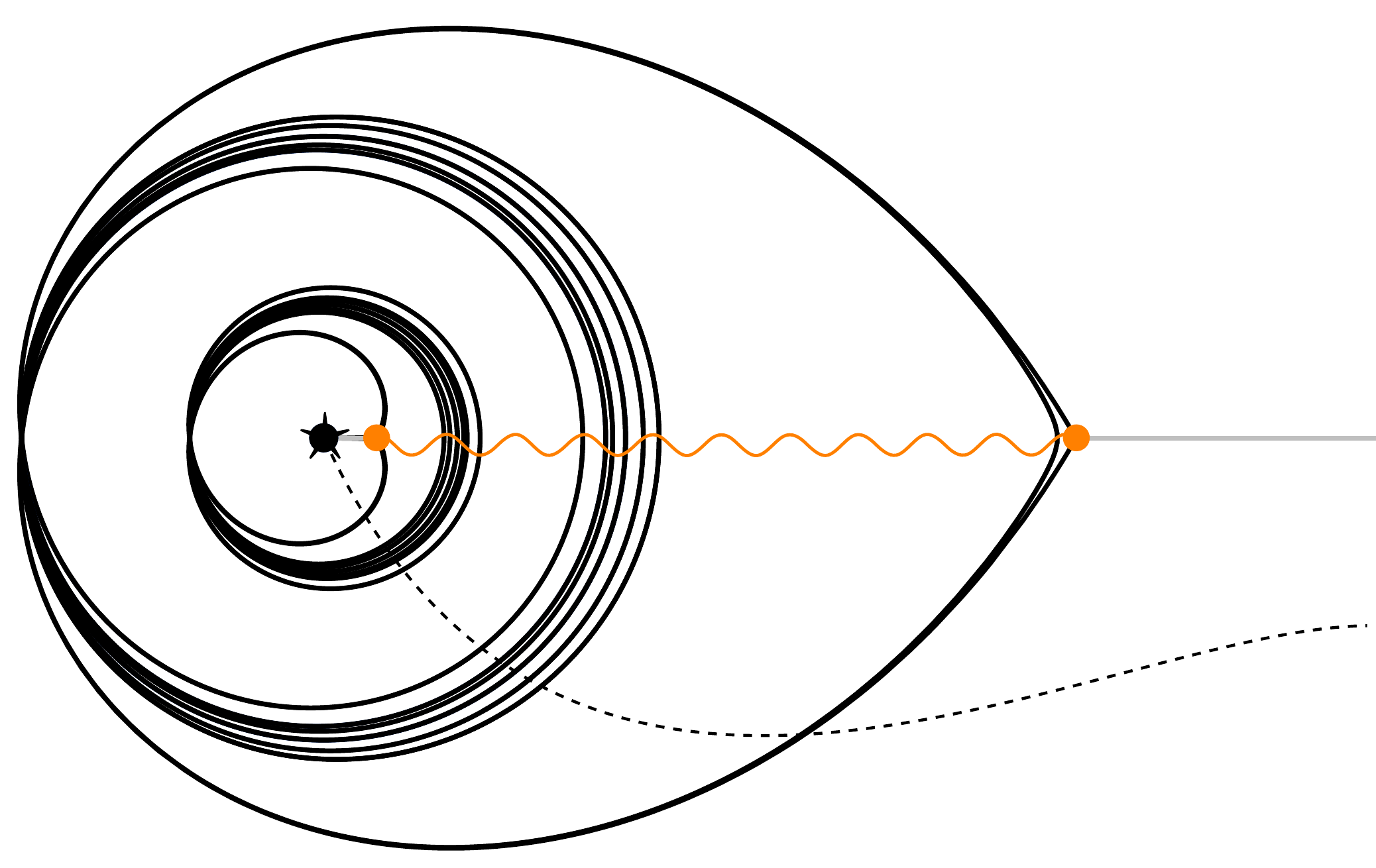}
\caption{D0 saddles of the conifold, for $Q=3/2$ at $\vartheta=0$, shown in black. Grey lines are one-way walls.
The yellow wavy line is the square-root cut of $\Sigma\to\IC^*_x$. The dashed line is the logarithmic cut on $\Sigma$, connecting one of the punctures at $x=0$ to one at $x=\infty$.
}
\label{fig:conifold-Q-3-2-theta-0}
\end{center}
\end{figure}

At $\vartheta = \arg Z_{D0} = 0$ we find the generalized saddle of Figure \ref{fig:conifold-Q-3-2-theta-0}. 
Thanks to the fact that the saddle consists of two disjoint components, we can analyze them separately.
In fact each of them resembles exactly the D0 brane saddle of the $\IC^3$ exponential networks studied in~\cite{Banerjee:2018syt}.
It follows immediately that these are two copies of the KK tower appearing in $\IC^3$.
Since saddles are disjoint the previous analysis of soliton data applies directly, leading to
\be
	\Omega(n D0) = -2 \qquad n\in \IZ_{>0}\,.
\ee
At $\vartheta=\pi$ we obtain the corresponding statement for $n\in\IZ_{<0}$.

Taking $Q\to\infty$ along the positive real axis, the curve degenerates as illustrated in subsection \ref{sec:conifold-factorization}.
In this limit, the two components of the saddle of Figure \ref{fig:conifold-Q-3-2-theta-0} simply survive and end up living either in the component (\ref{eq:mirr-conifold-factorized-1}) or in (\ref{eq:mirr-conifold-factorized-2}).
Following the decomposition of the mirror curve, one therefore ends up with two copies of $\IC^3$, each carrying its own tower of D0 branes, as described in \cite{Banerjee:2018syt}.

\subsubsection*{D2 states}

\begin{figure}[h!]
\begin{center}
\includegraphics[width=0.34\textwidth]{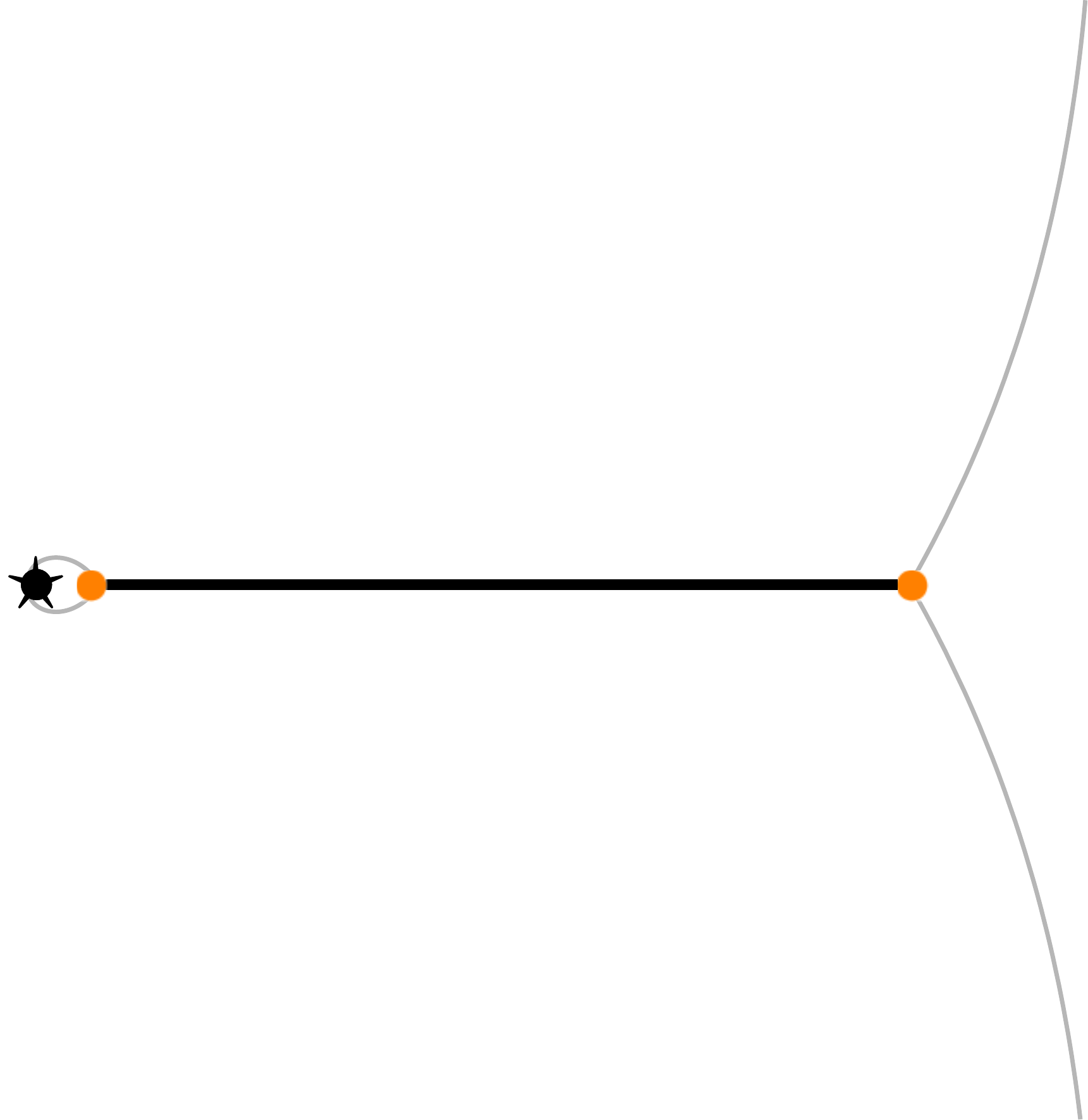}
\caption{D2 saddle of the conifold, for $Q=3/2$ at $\vartheta=\pi/2$. 
}
\label{fig:conifold-D2}
\end{center}
\end{figure}

At $\vartheta = \arg Z_{\text{D2}} = \frac{\pi}{2}$ we find the simple saddle of Figure \ref{fig:conifold-D2},
this is the D2 brane. 
The analysis of this saddle is very simple. There is a single two-way wall, made of one $\CE$-wall running left to right, and one in the opposite direction.
Let $x_0$ be a point on the double-wall. For definiteness, we shall work in American resolution by setting $\vartheta = \pi/2 - \epsilon$ for an infinitesimal $\epsilon > 0$.
In this case the $\CE$-wall emanating form the left branch point will run slightly below the one emanating from the right branch point, in opposite directions.
Let $a_N \in \Gamma_{ij,N,N}(x_0)$ be the relative homology classes of paths supported on the $\CE$-wall emanating from the left branch point, and $b_N\in \Gamma_{ji,N,N}(x_0)$ those supported on the wall emanating from the right one.\footnote{The types $ij,N,N$ have been assigned consistently with the choice of trivialization shown in Figure \ref{fig:conifold-Q-3-2-theta-0}.}
Then the generating function $Q$ for the double wall is
\be
	Q = \one +\sum_{N,M} X_{a_N} X_{b_M} 
	= \one  + \sum_{N} X_{a_N b_N}
	= \one \cdot (1+X_\gamma)
\ee
where $\gamma$ is the homology class on $\tSigma$ identified with $\gamma_N = a_N\circ b_N$ after the quotient by $\ker Z$ (see \cite{Banerjee:2018syt}). 
By (\ref{eq:Omega-formula-review}) this gives
\be
	\Omega(n \text{D2}) = \left\{
	\begin{array}{lr}
		1 \qquad & n=1 \\
		0 \qquad & n>1 
	\end{array}
	\right. \,.
\ee
At $\vartheta=-\frac{\pi}{2}$ we obtain the corresponding statement for $\overline{\text{D2}}$.

Taking $Q\to\infty$ along the positive real axis, the curve degenerates as illustrated in subsection \ref{sec:conifold-factorization}.
In this limit,  the saddle of Figure \ref{fig:conifold-D2} becomes infinitely long and gets lost, since one or the other branch point ends up absorbed into a collision of two punctures in both components (\ref{eq:mirr-conifold-factorized-1}) and (\ref{eq:mirr-conifold-factorized-2}).
Following the decomposition of the mirror curve the D2 brane disappears from the spectrum, as should be expected from the match with $\IC^3$.

\subsubsection*{D2-D0 states}

\begin{figure}[h!]
\begin{center}
\raisebox{15pt}{\includegraphics[width=0.34\textwidth]{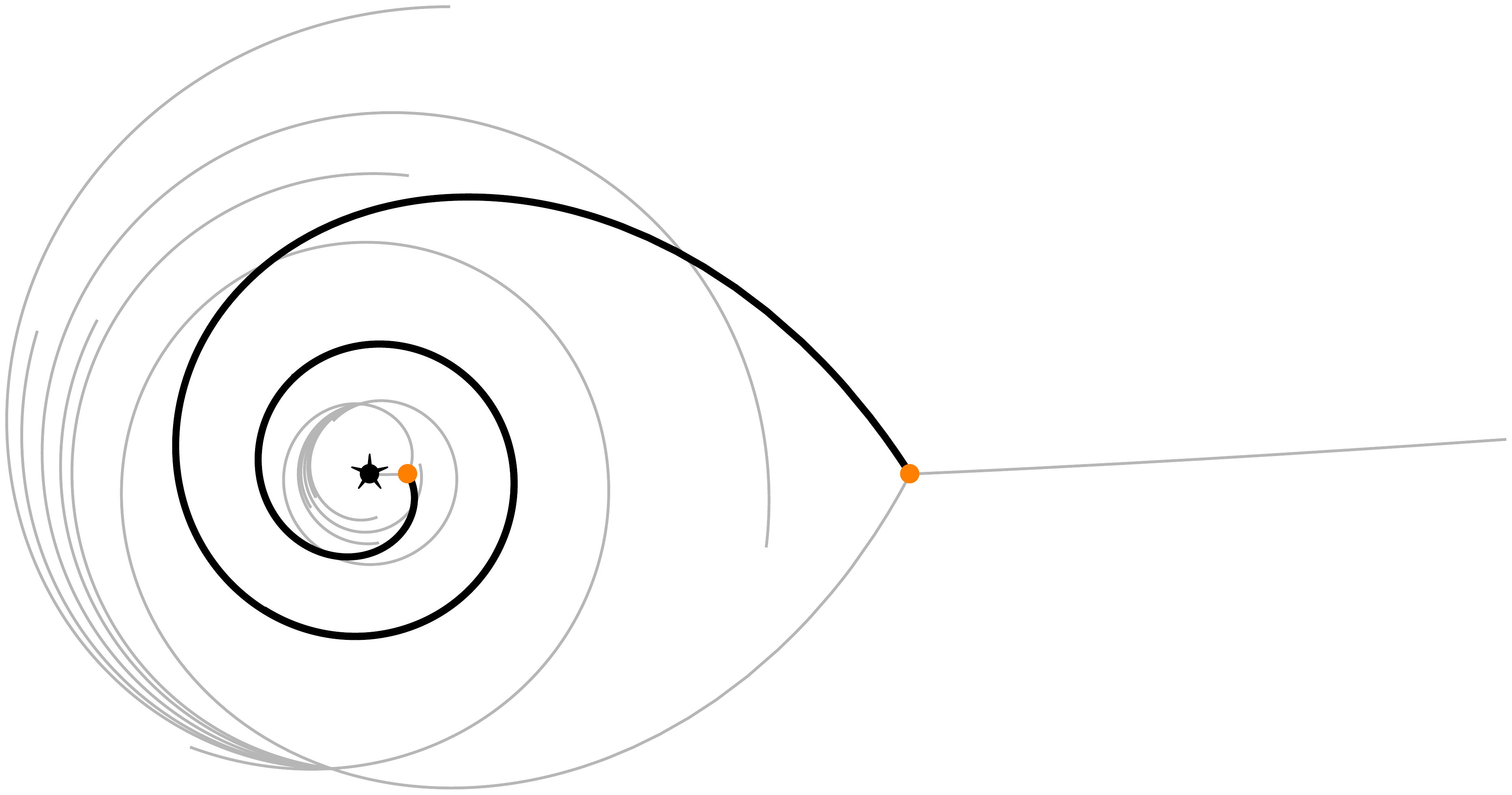}}
\hspace*{.1\textwidth}
\includegraphics[width=0.34\textwidth]{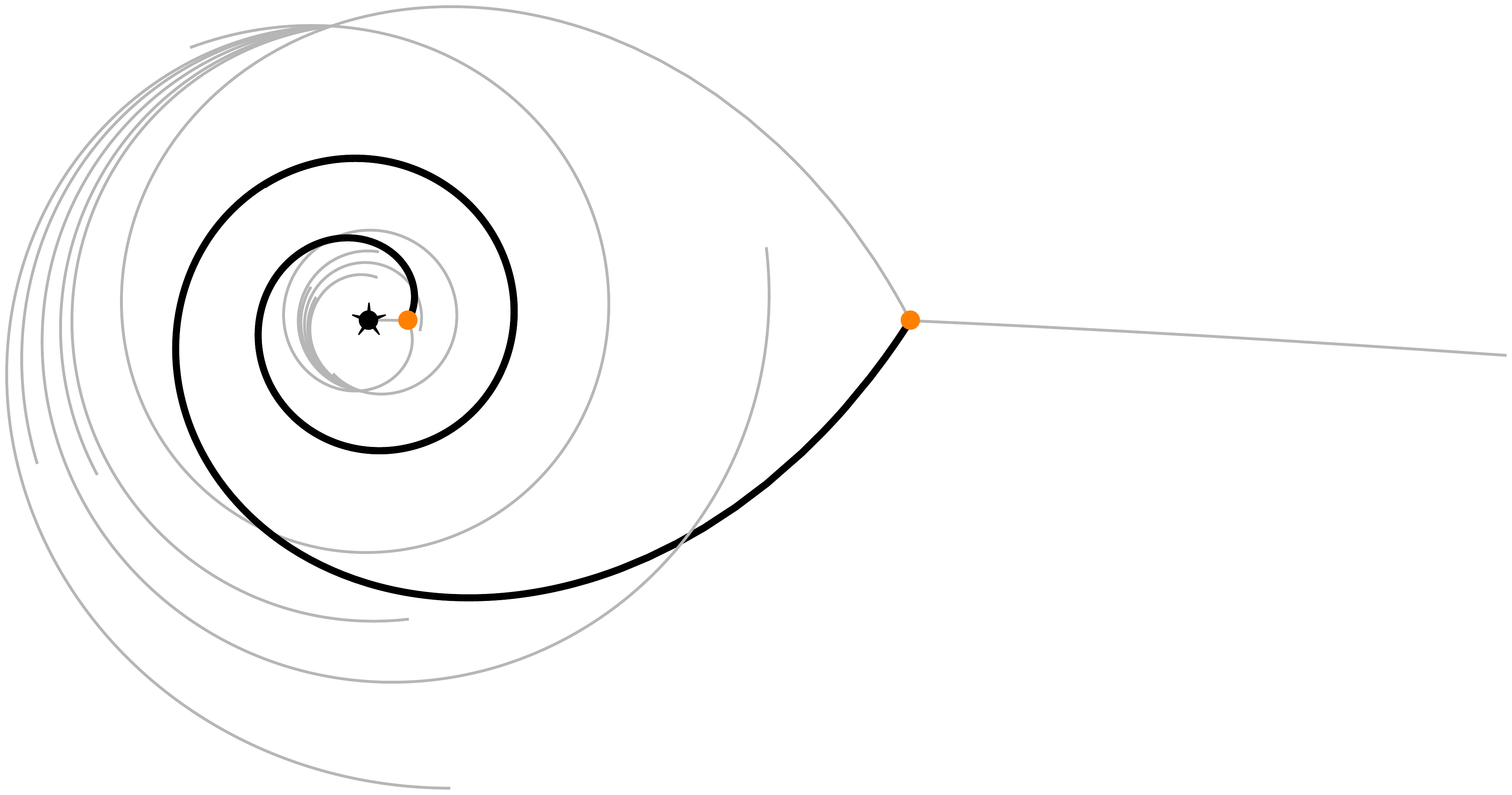}
\caption{Left: D2-D0 saddle of the conifold, for $Q=3/2$. Right: $\overline{\text{D2}}$-D0 saddle. 
}
\label{fig:conifold-D2-D0}
\end{center}
\end{figure}

At $\vartheta = \arg (Z_{\text{D2}} + Z_{D0})$ we find the simple saddle on the left in Figure \ref{fig:conifold-D2-D0}, this is the D2-D0 state. Likewise plotting at $\vartheta = \arg (-Z_{\text{D2}} + Z_{D0})$ we find the simple saddle on the right, this is the $\overline{\text{D2}}$-D0 state
The analysis of their BPS index follows through from the D2 case, since the saddle topology is the same:
\be
	\Omega(n (\text{D2-D0})) = \left\{
	\begin{array}{lr}
		1 \qquad & n=1 \\
		0 \qquad & n>1 
	\end{array}
	\right. \,.
\ee
and similarly for $\overline{\text{D2}}$-D0.

\begin{figure}[h!]
\begin{center}
\includegraphics[width=0.5\textwidth]{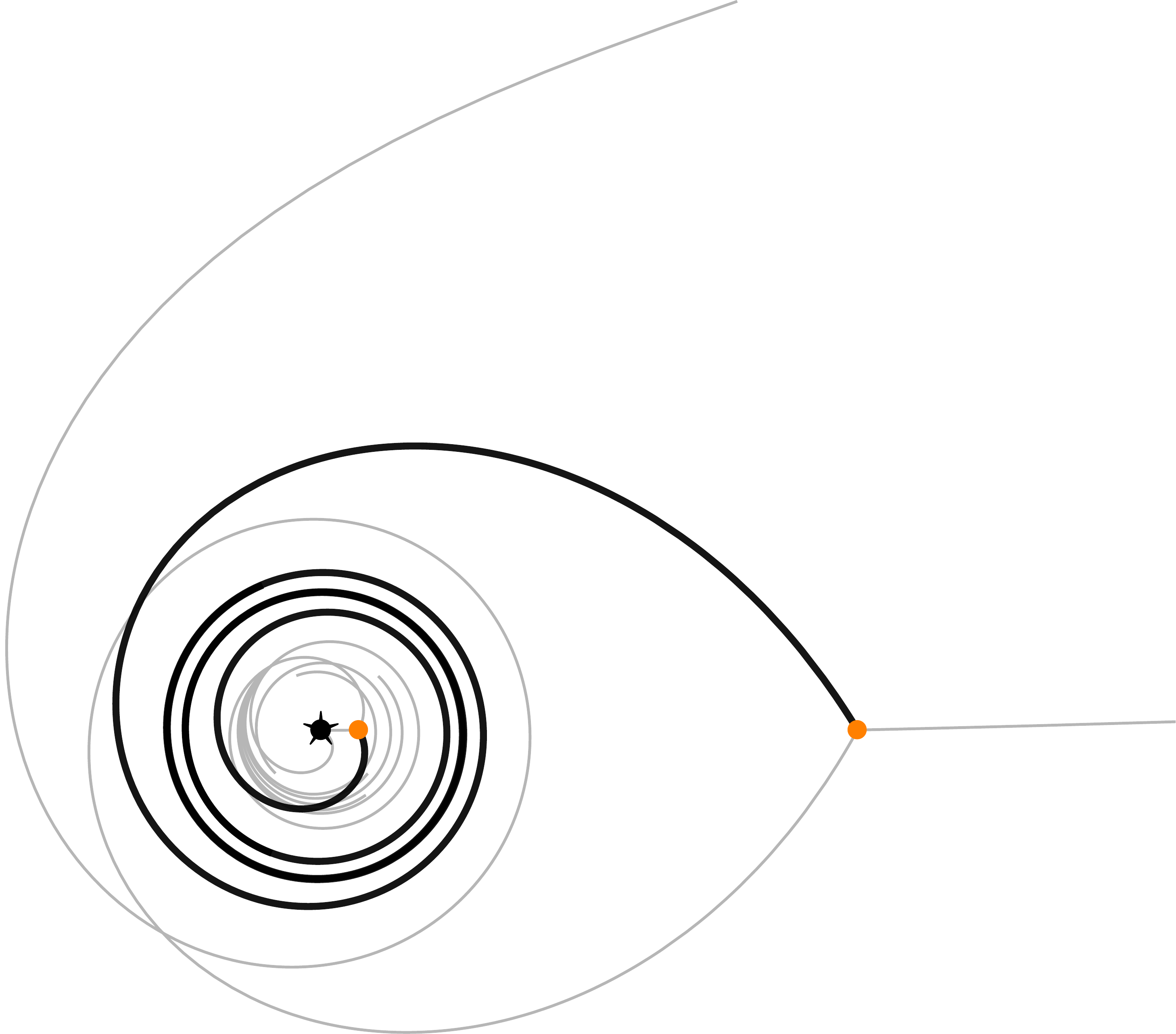}
\caption{D2-2D0 saddle of the conifold, for $Q=3/2$. 
}
\label{fig:conifold-D2-D0-D0}
\end{center}
\end{figure}

Proceeding to $\vartheta = \arg (Z_{\text{D2}} + k Z_{D0})$ we find again simple saddles for all $k>0$. For example, D2-$2$D0 is shown in Figure \ref{fig:conifold-D2-D0-D0}.
The BPS index is always 
\be
	\Omega(n (\text{D2-$k$D0})) = \left\{
	\begin{array}{lr}
		1 \qquad & n=1 \\
		0 \qquad & n>1 
	\end{array}
	\right. \,.
\ee

\subsubsection*{Full spectrum}

To summarize, we find the following BPS states (plus their CPT conjugates)
\be\label{eq:conifold-spectrum}
\begin{split}
	\Omega(n \text{D0}) & = -2 \\
	\Omega(\text{D2-$k$D0}) & = 1 \\
	\Omega(\overline{\text{D2}}\text{-$k$D0}) & = 1\\
\end{split}
\qquad
\begin{split}
	& n\geq 1\\
	& k\geq 0\\
	& k \geq 1
\end{split}
\ee

\subsection{Match with topological strings, framed wall-crossing and generalized DT invariants}\label{sec:conifold-match}

The most direct interpretation for the BPS spectrum we have computed above is that of instanton-particles and monopole-strings of a 5d $\CN=1$ theory engineered by the conifold, on $\IR^4\times S^1$. 
As reviewed above, and argued in detail in \cite{Banerjee:2018syt}, the framework of exponential networks was developed around the physical phenomenon of 2d-4d wall-crossing for a 3d-5d system on a circle, therefore the BPS spectrum really represents the KK tower of 5d BPS states.

\paragraph{Relation to Gopakumar-Vafa invariants}

We have indeed found that the BPS spectrum is arranged into KK towers: D0 charge corresponds to KK momentum (also recall that $Z_{\text{D0}}=2\pi / R$), therefore different towers are labeled by D2 charge, being either $0$ or $\pm1$. 
Each KK tower corresponds to a 5d BPS particle, and their spectrum should be captured by Gopakumar-Vafa invariants.

We can therefore compare our prediction (\ref{eq:conifold-spectrum}) with genus-zero GV invariants for the conifold. 
Our computation predicts a particle with zero D2 charge and BPS index $-2$, one particle with a unit of D2 charge, and its own antiparticle.
These BPS indices match respectively the expected contributions of D0 branes and the genus-zero GV invariants for the conifold, 
see for example \cite[eq. (18)]{Katz:2004js}
We therefore have a precise match with expectations based on other techniques for computing these invariants.

\paragraph{Relation to generalized Donaldson-Thomas invariants}

Let us also briefly comment on the relation to generalized Donaldson-Thomas invariants.
In physics these should count D6-D4-D2-D0 states, therefore to make contact with them one should consider each KK mode as a separate BPS state. Equivalently one may view the 5d theory on a circle as a 4d theory of its KK modes (keeping the radius finite).
Since the conifold is non-compact, only BPS states with zero D6 charge can have finite mass, these are sometimes known as \emph{degree-zero} generalized DT invariants. Likewise, due to the lack of compact four-cycles, we also expect only states with vanishing D4 charge.
The D2-D0 BPS spectrum we found in (\ref{eq:conifold-spectrum}) should therefore exhaust all generalized DT invariants for the conifold.

As a check, let us compare this with predictions from framed wall-crossing.
Standard DT invariants are a special case of generalized DT invariants, with a single unit of D6 charge.
In the case of conifold, these must be \emph{framed} BPS states, since the D6 would have infinite volume and therefore would give rise to an infinitely heavy, non-dynamical, defect in the 5d theory.
To begin with, let us make the following empirical observation. The generating series of DT invariants in the non-commutative chamber is as follows \cite{Szendroi:2007nu, 2007arXiv0709.3079Y} 
\be \label{con-nonc}
	\CZ = M(q)^{-2} \prod_{n\geq 0}(1-q^n Q )^n(1-q^n Q^{-1} )^n
\ee
where $M(q)$ is McMahon's function. 
Then we observe that the \emph{exponents} of this factorization coincide  precisely with the BPS indices we computed above in  (\ref{eq:conifold-spectrum}).

To clarify this observation, let us restrict to BPS states with non-negative D2 charge.
This makes contact with  Gromov-Witten invariants, and their well-known correspondence with Donaldson-Thomas \cite{2008arXiv0809.3976M}.\footnote{Incidentally, this restriction also coincides with the ``DT chamber'', although we stress that we are not really moving across any chambers in the moduli space of framed BPS states.}
Expanding this as a $q$-series gives
\be\label{eq:conifold-DT}
\begin{split}
	\CZ|_{\beta\geq 0} 
	& = M(q)^{-2} \prod_{n\geq 0}(1-q^n Q )^n\\
	& = \sum_{n,k\geq 0} DT_{D6,kD2,nD0} Q^k q^n 
	\\
	& = 
	1+2q+7q^2+18q^3+47q^4+110q^5+\dots
	\\
	& 
	+ Q\(- q - 4 q^2 - 14 q^3 - 42 q^4 - 117 q^5 +\dots\)
	+ \CO(Q^2)\,.
\end{split}
\ee
Now let us view these as \emph{boundstates} of a single, infinitely heavy, D6 brane with the BPS states carrying D2-D0 charges in (\ref{eq:conifold-spectrum}), in the spirit of \cite{Jafferis:2008uf}.
Given the spectrum of D2-D0 states, and the DSZ pairing between their charges and that of the D6 completely specifies the spectrum of boundstates by Kontsevich-Soibelman's wall-crossing formula.\footnote{In fact, this is an instance of the semi-primitive wall-crossing formula of Denef and Moore \cite{Denef:2007vg}.} Matching with (\ref{eq:conifold-DT}) would therefore provide a nontrivial check of our result (\ref{eq:conifold-spectrum}) for degree-zero generalized DT invariants.

To perform this check, one needs to note that the only non-trivial contribution to the DSZ pairing is between the D0 and the D6 charges
\be\label{eq:D0-D6-pairing}
	\langle \gamma_{\text{D6}}, \gamma_{\text{D0}}\rangle = 1\,.
\ee
Using this one can write the following wall-crossing identity\footnote{\label{ft:KSWFC}The operators $\CK_\gamma$ are Kontsevich-Soibelman symplectomorphisms acting on a complex 3-torus generated by charges $\{\gamma_{D6},\gamma_{D2},\gamma_{D0}\}$. For a well-suited review of conventions see \cite[Appendix A]{Longhi:2016wtv}.}
\be\label{eq:KSWCF-DT}
\begin{split}
	& \(  \prod_{n \geq 1} \CK_{n \gamma_{\text{D0}}}^{\Omega(n \text{D0})} \)
	\(  \prod^{\curvearrowleft}_{n \geq 0} \CK_{\gamma_{\text{D2}} + n \gamma_{\text{D0}}}^{\Omega(\text{D2-$n$D0})} \)
	\CK_{\gamma_{D6}}\\
	& \qquad\qquad
	=
	\CK_{\gamma_{D6}}
	\(  \prod_{\ell\geq 1}\prod_{k\geq 0}\prod_{n \geq 1} \CK_{\ell\gamma_{D6}+k\gamma_{D2}+n \gamma_{\text{D0}}}^{\fOmega(\text{$\ell$D6-$k$D2-$n$D}0)} \)
	\(  \prod^{\curvearrowright}_{n \geq 0} \CK_{\gamma_{\text{D2}} + n \gamma_{\text{D0}}}^{\Omega(\text{D2-$n$D0})} \)
	\(  \prod_{n \geq 1} \CK_{n \gamma_{\text{D0}}}^{\Omega(n \text{D0})} \)\,.
\end{split}
\ee
Here $\Omega(\text{D2-$n$D0})$ and $\Omega(\text{$n$D0})$ are the ones we found in (\ref{eq:conifold-spectrum}), while  $\fOmega$ denote framed BPS degeneracies of D2-D0 BPS states bound to $\ell$ D6 branes.
The latter are predicted by the former through the wall-crossing identity, the first few read:
\be\label{eq:conifold-gen-DT}
\begin{array}{llll}
	\fOmega_{1,0,1} = -2\,;
	\qquad
	&
	\fOmega_{1,1,1} = 1\,;
	\qquad
	&
	\fOmega_{1,0,2} = 7\,;
	\qquad
	&
	\fOmega_{2,0,2} = -2\,;
	\\
	\fOmega_{1,1,2} = -4\,;
	&
	\fOmega_{1,0,3} = -18\,;
	&
	\fOmega_{2,1,2} = 1\,;
	&
	\fOmega_{2,0,3} = -18\,;
	\\
	\fOmega_{1,1,3} = 14\,;
	&
	\fOmega_{1,0,4} = 47\,;
	&
	\fOmega_{3,0,3} = -2\,;
	&
	\fOmega_{2,1,3} = 14\,;
	\\
	\fOmega_{2,0,4} = -78\,;
	&
	\fOmega_{1,2,3} = -2\,;
	&
	\fOmega_{1,1,4} = -42\,;
	&
	\fOmega_{1,0,5} = -110\,;
	\\
	\dots
\end{array}
\ee
where $\fOmega_{\ell,k,n}\equiv \fOmega(\text{$\ell$D6-$k$D2-$n$D}0)$.
As expected, these match with (\ref{eq:conifold-DT}) 
\be
	DT_{D6,kD2,nD0} = (-1)^{n} \fOmega_{1,k,n}\,,
\ee
at least for $k=0,1$.
The wall-crossing formula predicts also higher-degree generalized Donaldson-Thomas invariants, those with $\ell>1$ in (\ref{eq:conifold-gen-DT}).
They are not captured by our framework, nor by the GW/DT correspondence. 
To compute them with our approach, one would first need to understand how to deal with framed BPS states.\footnote{In \cite{Eager:2016yxd}, a proposal for identifying framed BPS states was advanced.}

\subsection{Brane monodromy}\label{sec:PL-ST}

To continue our analysis of conifold networks, let us consider some global features of the mirror geometry over the $Q$-plane.

\subsubsection*{Monodromy around $Q=\infty$}

\begin{figure}[h!]
\begin{center}
\fbox{\includegraphics[width=0.30\textwidth]{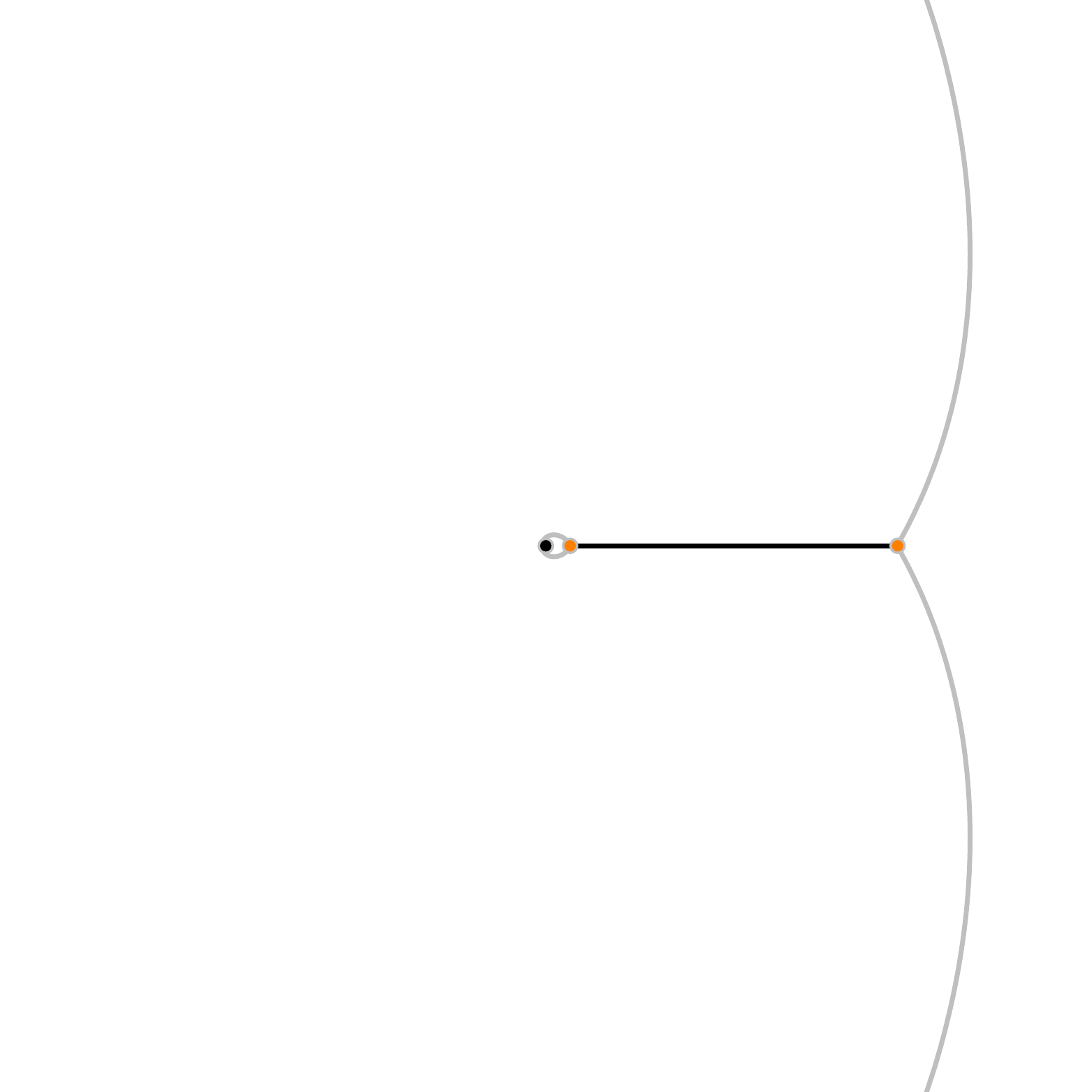}}
\hspace{.01\textwidth}
\fbox{\includegraphics[width=0.30\textwidth]{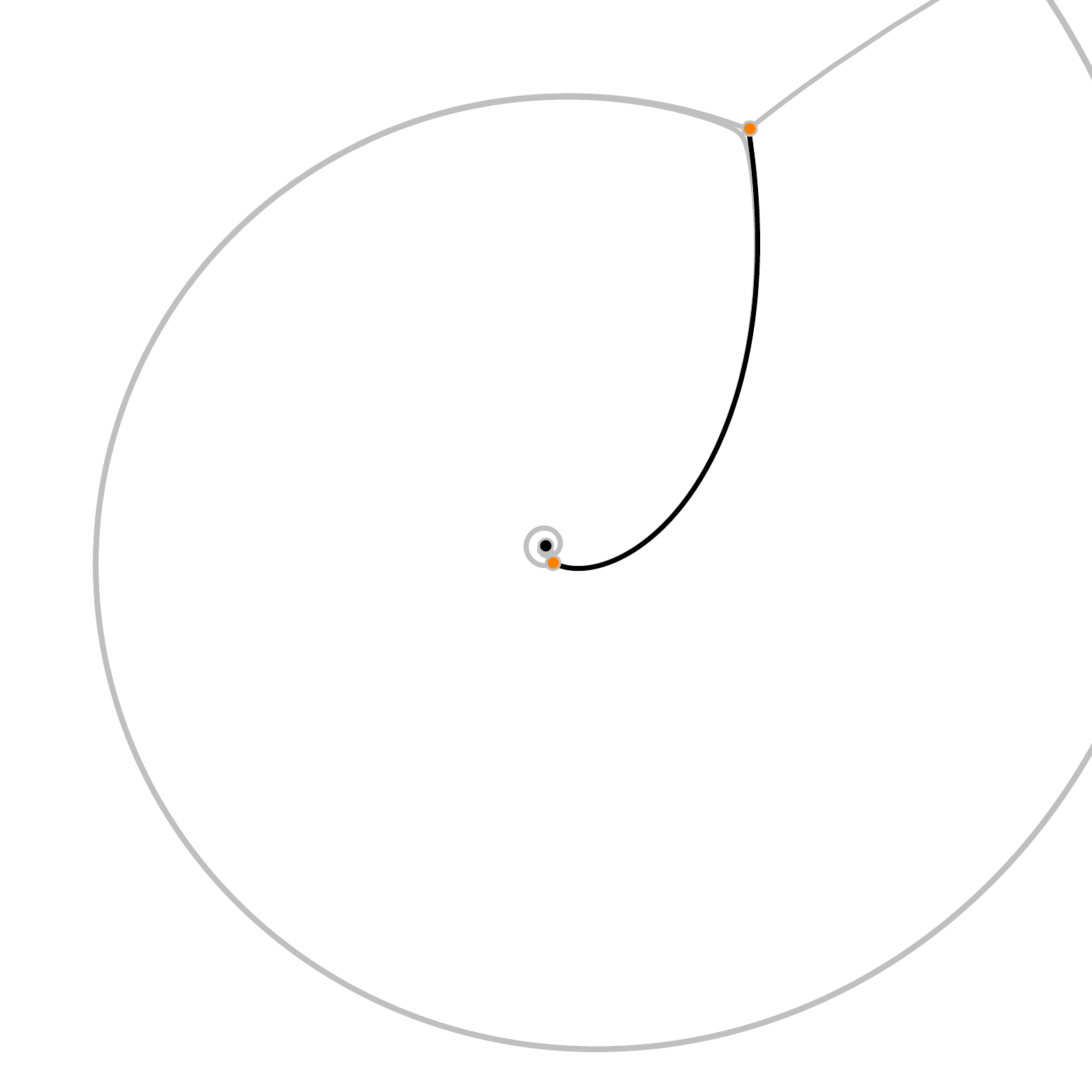}}
\hspace{.01\textwidth}
\fbox{\includegraphics[width=0.30\textwidth]{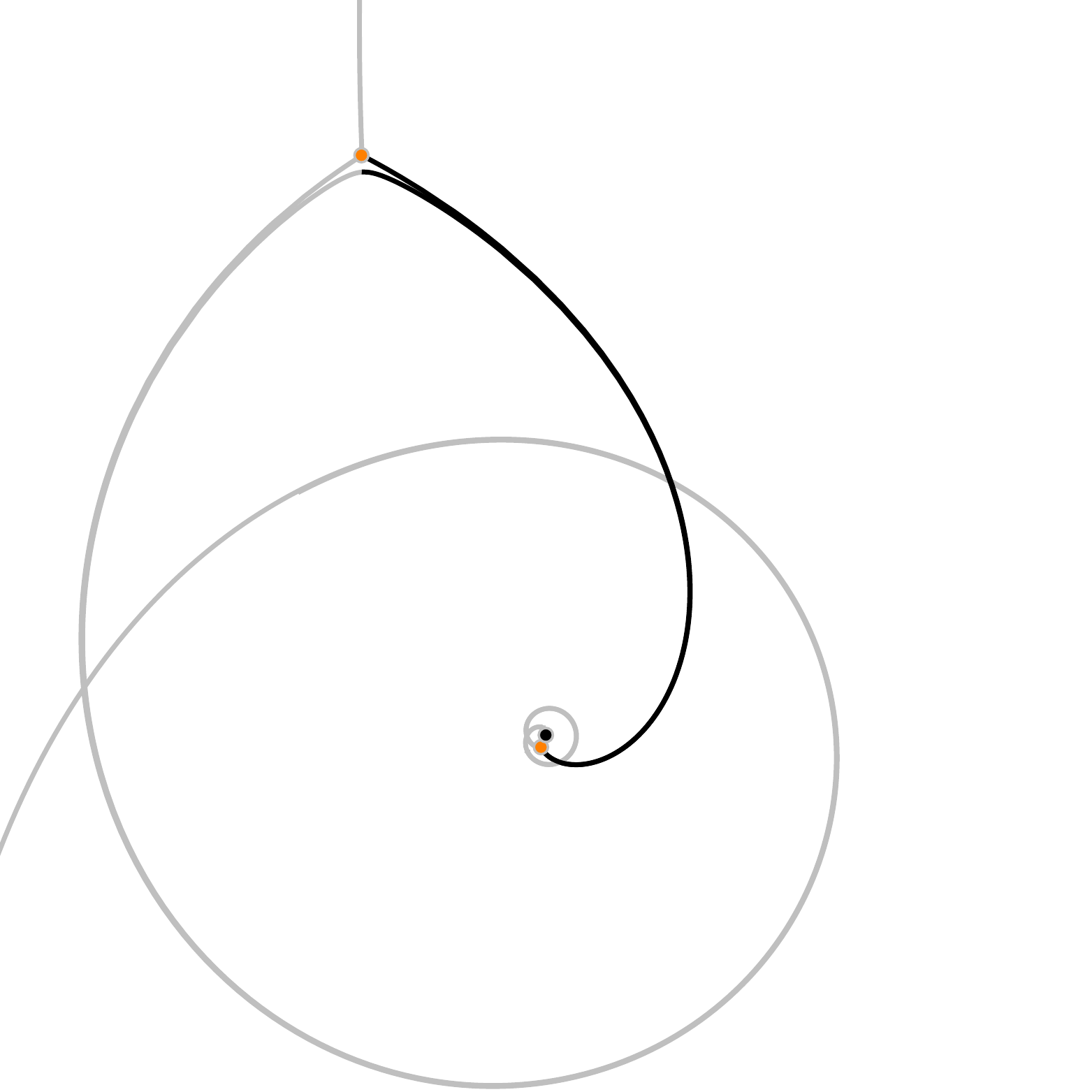}}
\\
\fbox{\includegraphics[width=0.30\textwidth]{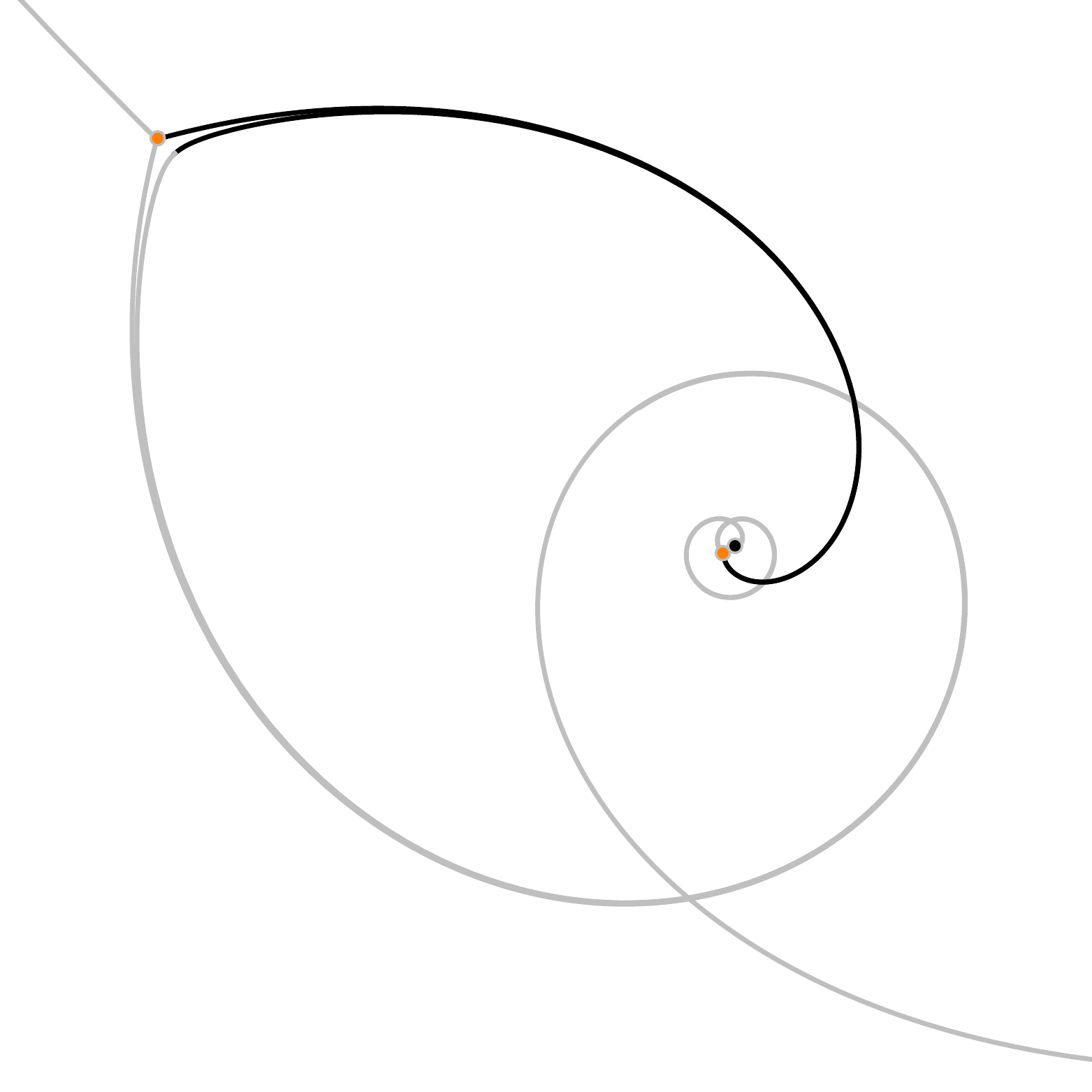}}
\hspace{.01\textwidth}
\fbox{\includegraphics[width=0.30\textwidth]{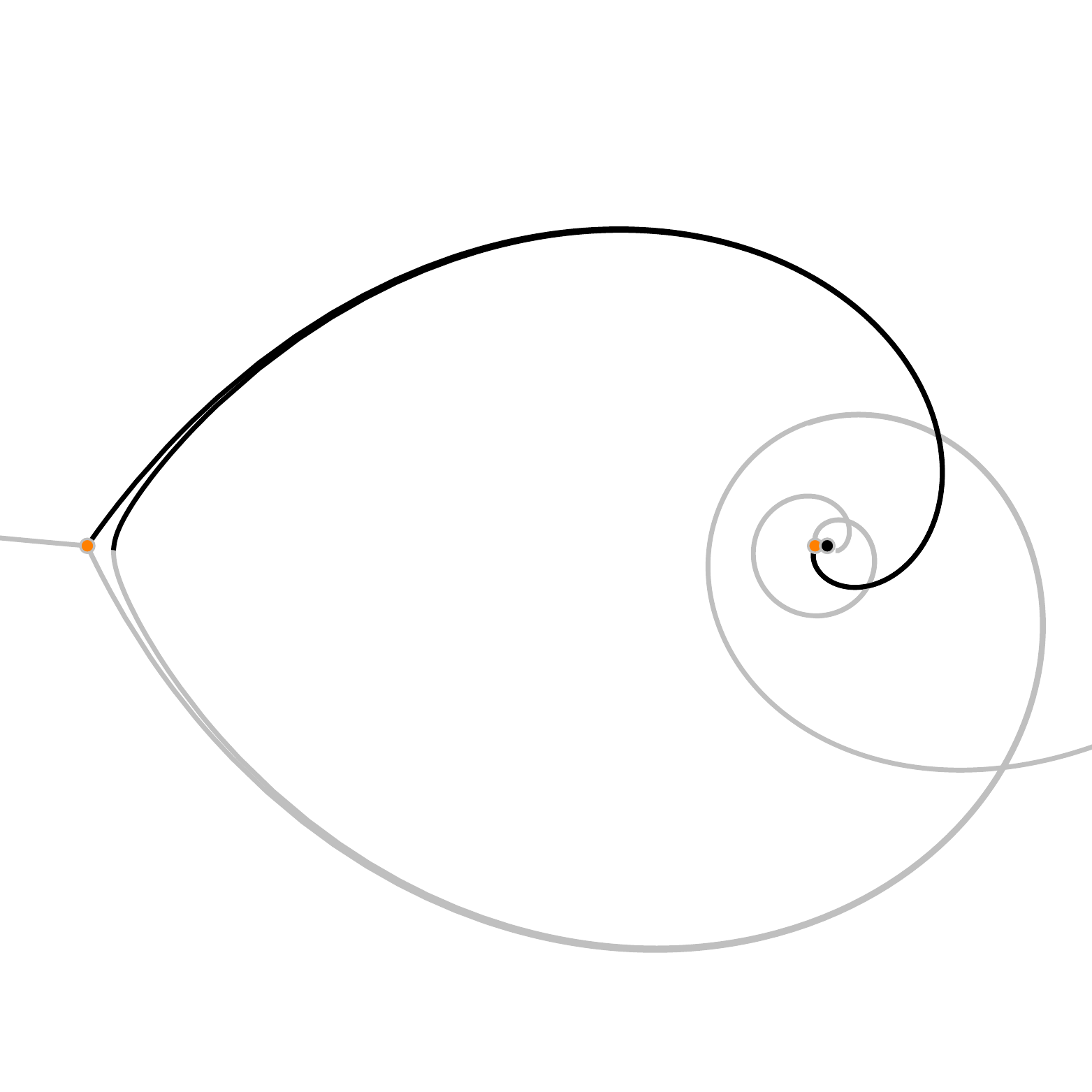}}
\hspace{.01\textwidth}
\fbox{\includegraphics[width=0.30\textwidth]{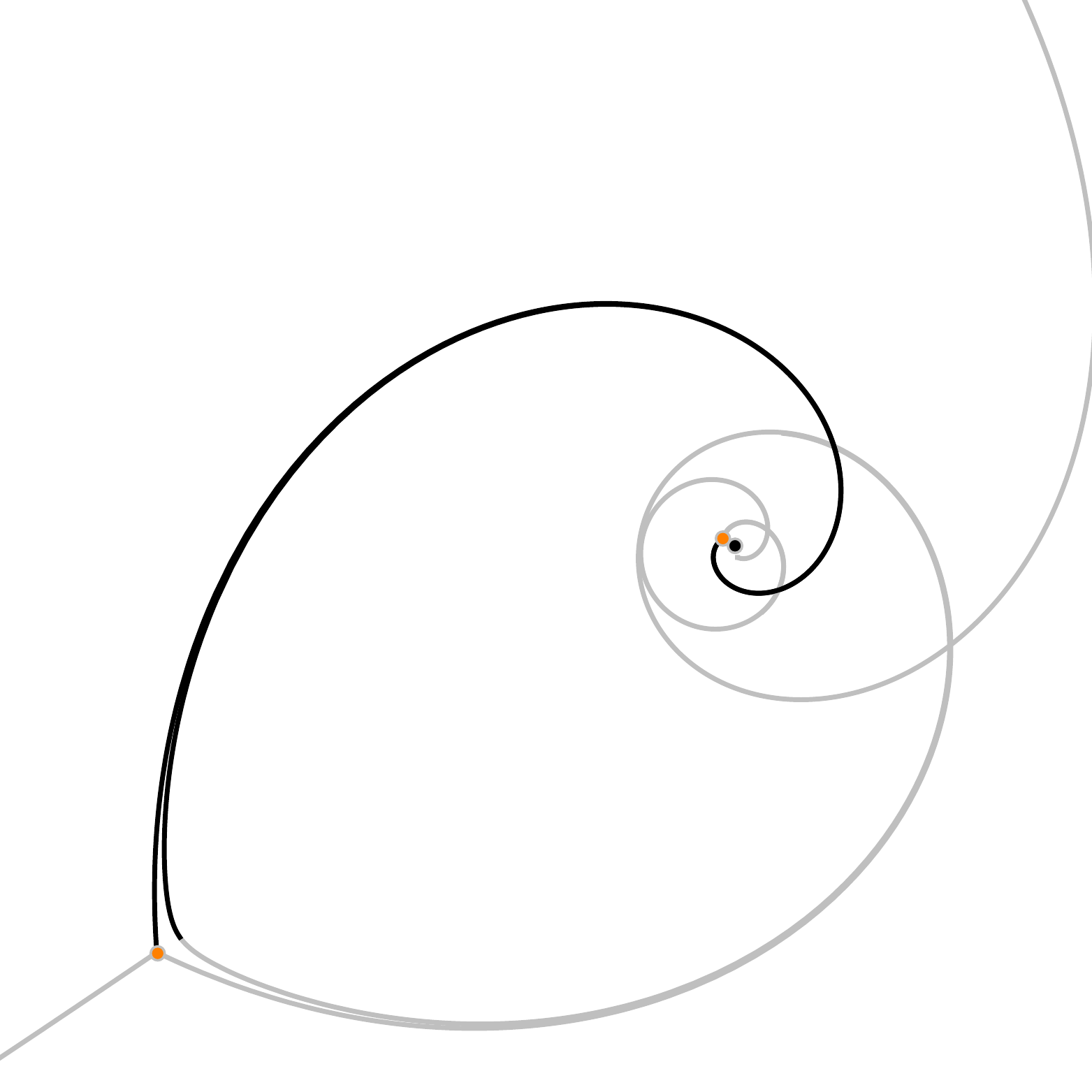}}
\\
\fbox{\includegraphics[width=0.30\textwidth]{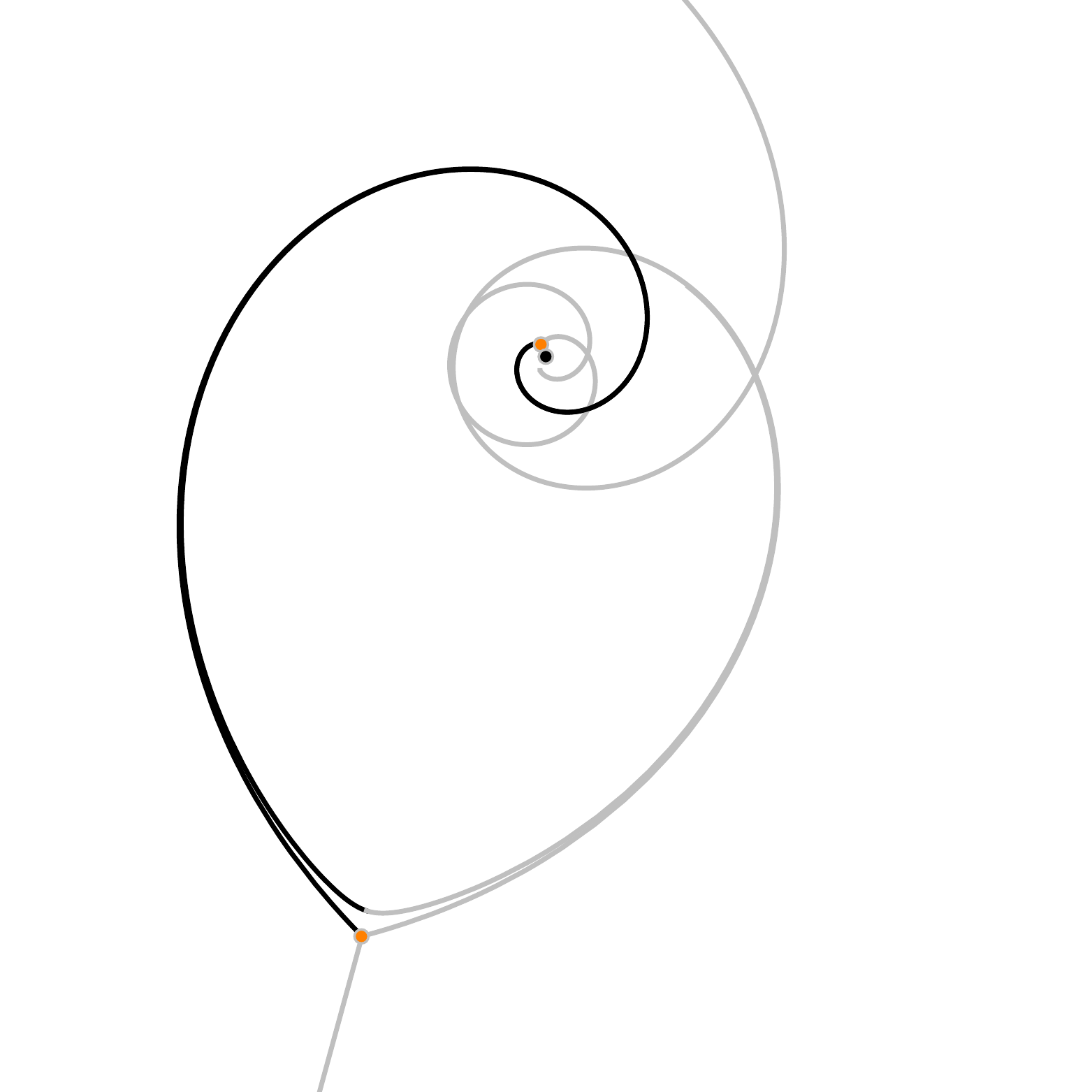}}
\hspace{.01\textwidth}
\fbox{\includegraphics[width=0.30\textwidth]{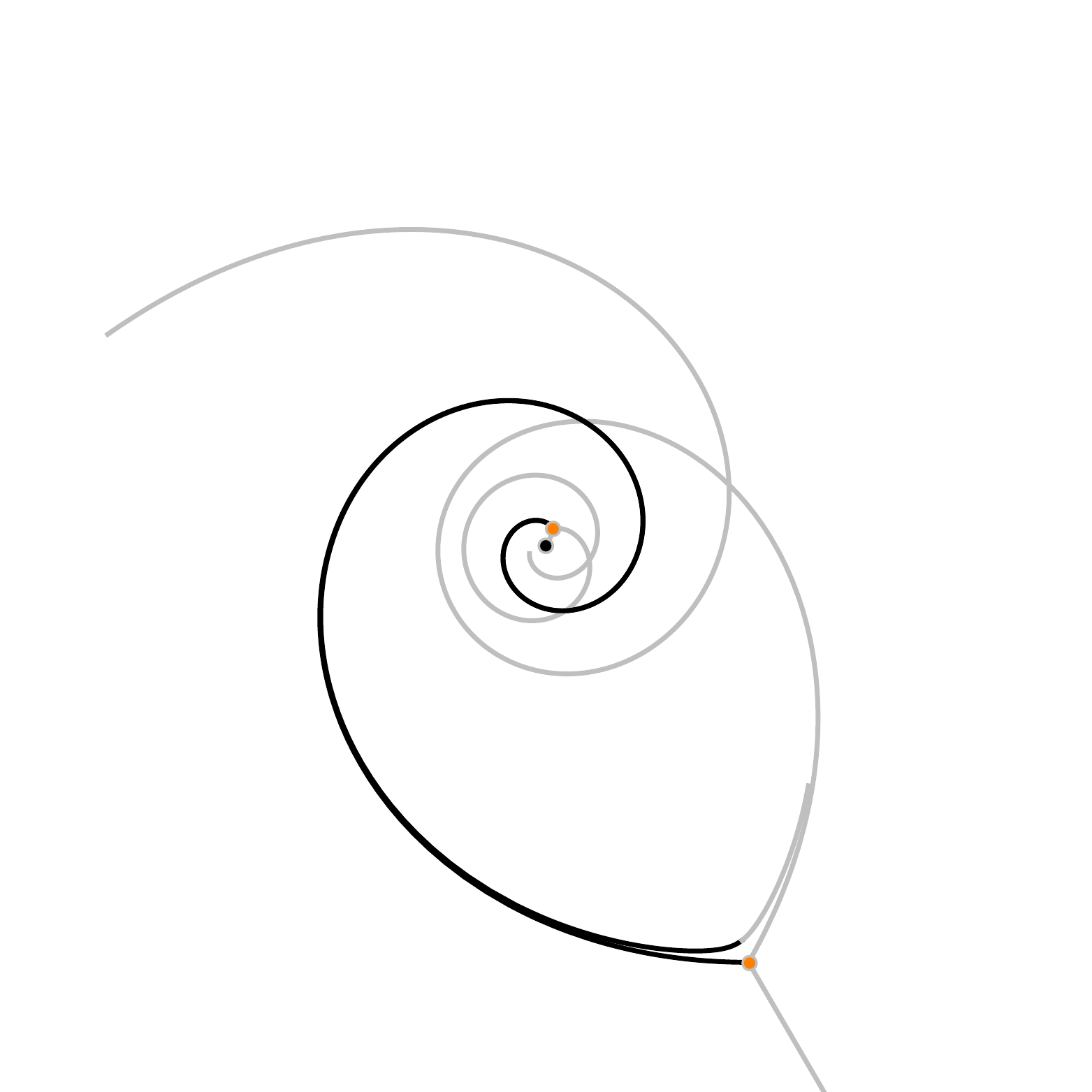}}
\hspace{.01\textwidth}
\fbox{\includegraphics[width=0.30\textwidth]{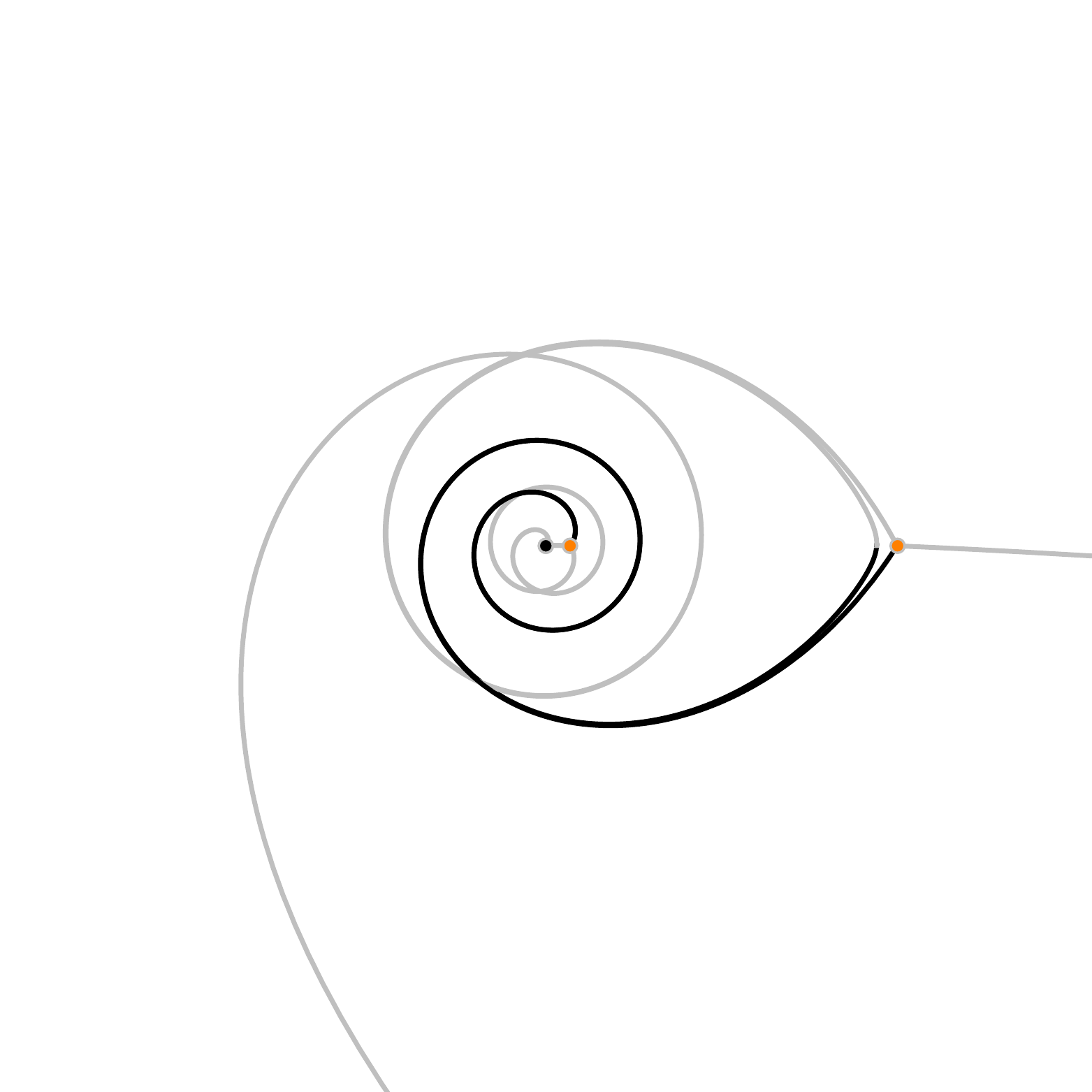}}
\caption{Monodromy around $Q=\infty$ for the D2 saddle  (shown in black, only primary walls are plotted) at $Q = \frac{3}{2}e^{\frac{2\pi i}{8}k}$ with $0\leq k \leq 8$. Overall, D2 becomes D2-D0.}
\label{fig:conifold-PL-monodromy}
\end{center}
\end{figure}

Near $Q\to\infty$ the D2 brane becomes infinitely heavy. By comparison one may think of the D0 as becoming parametrically light compared to D2, and expect some sort of Picard-Lefschetz monodromy, akin to phenomena related to singularities from massless particles.
Indeed, taking $Q\to  e^{2\pi i}\, Q$ the D2 central charge (\ref{eq:Z-A}) picks up a shift
\be
	Z_{\text{D2}} \to Z_{\text{D2}} + Z_{D0}\,.
\ee
This is a signal of monodromy for corresponding cycles $\gamma_{\text{D0}}$ and $\gamma_{\text{D2}}$. 
The occurrence of this monodromy can be made fully manifest by plotting the D2 saddle at several values of $Q$, see Figure \ref{fig:conifold-PL-monodromy}.

\subsubsection*{Monodromy around the conifold point}

As we have seen in subsection \ref{sec:conifold-saddles}, saddles corresponding to D0 branes naturally appear decomposed into two disjoint sets. 
Each set is ``supported'' by one branch point, see Figure \ref{fig:conifold-Q-3-2-theta-0}.
In the degeneration limit $Q\to\infty$ the two sets get separated and each one ends up furnishing the D0 spectrum on a copy of the $\IC^3$ mirror curve.
What happens to these D0 saddles near the conifold point, where the two branch points collide?

\begin{figure}[h!]
\begin{center}
\fbox{\includegraphics[width=0.30\textwidth]{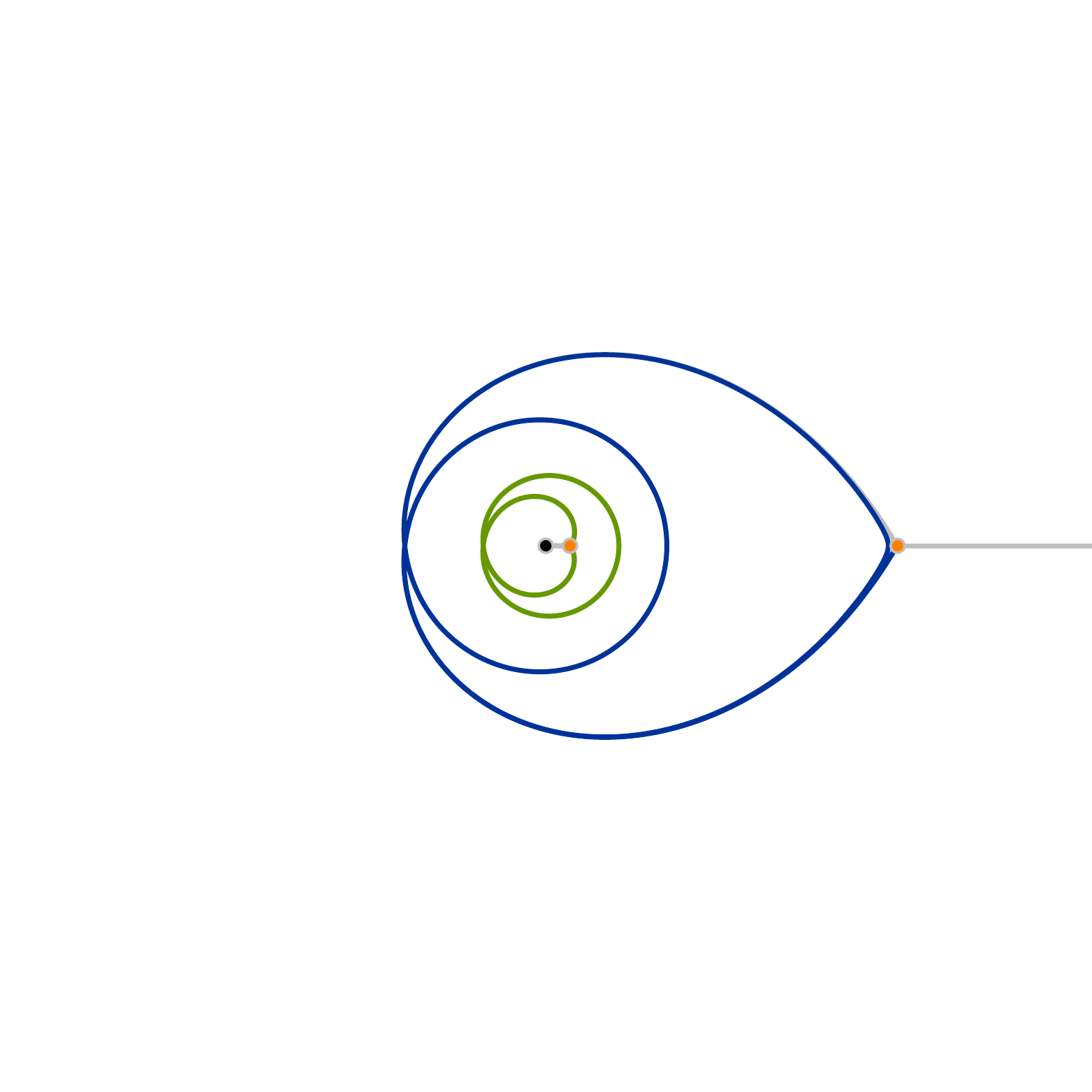}}
\hspace{.01\textwidth}
\fbox{\includegraphics[width=0.30\textwidth]{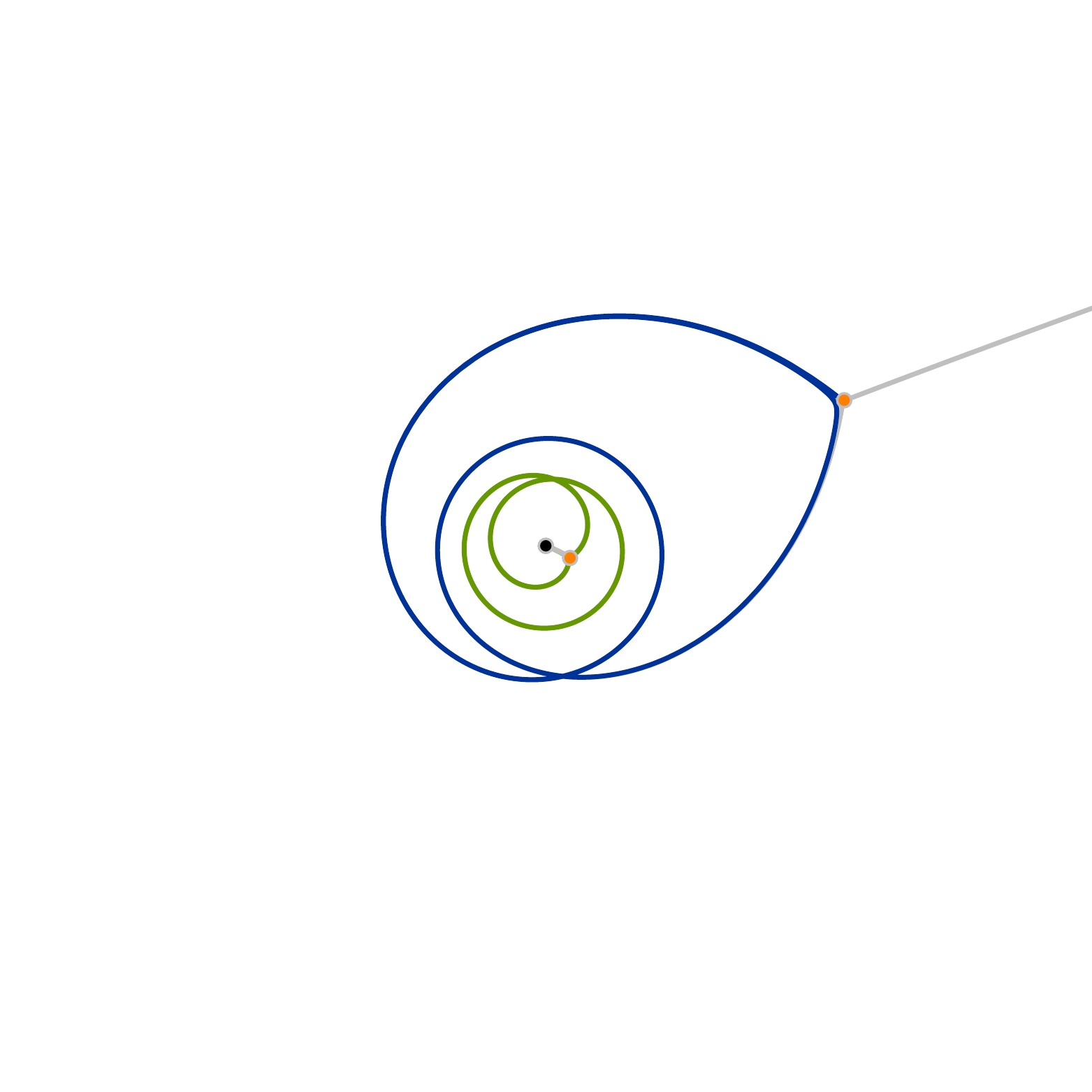}}
\hspace{.01\textwidth}
\fbox{\includegraphics[width=0.30\textwidth]{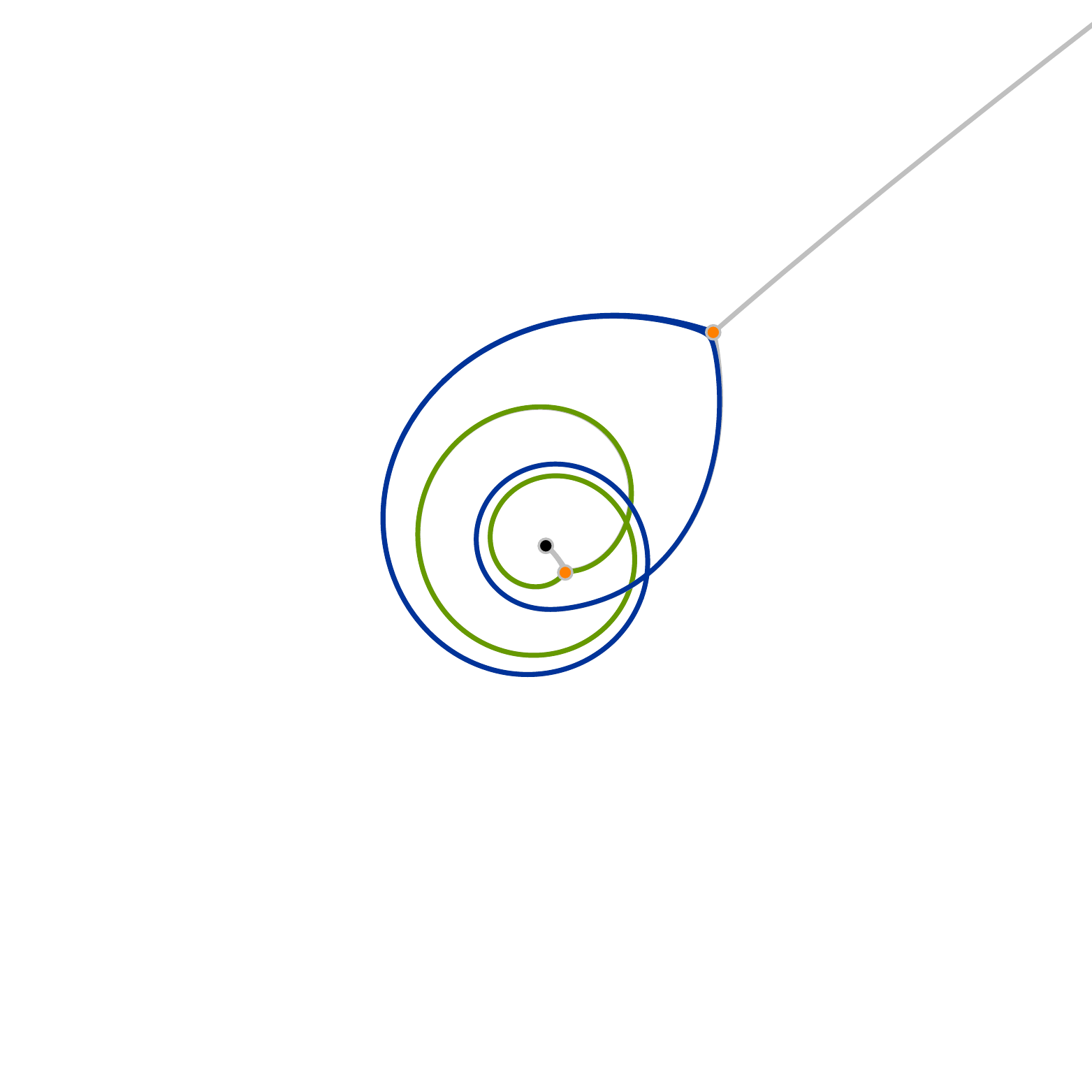}}
\\
\fbox{\includegraphics[width=0.30\textwidth]{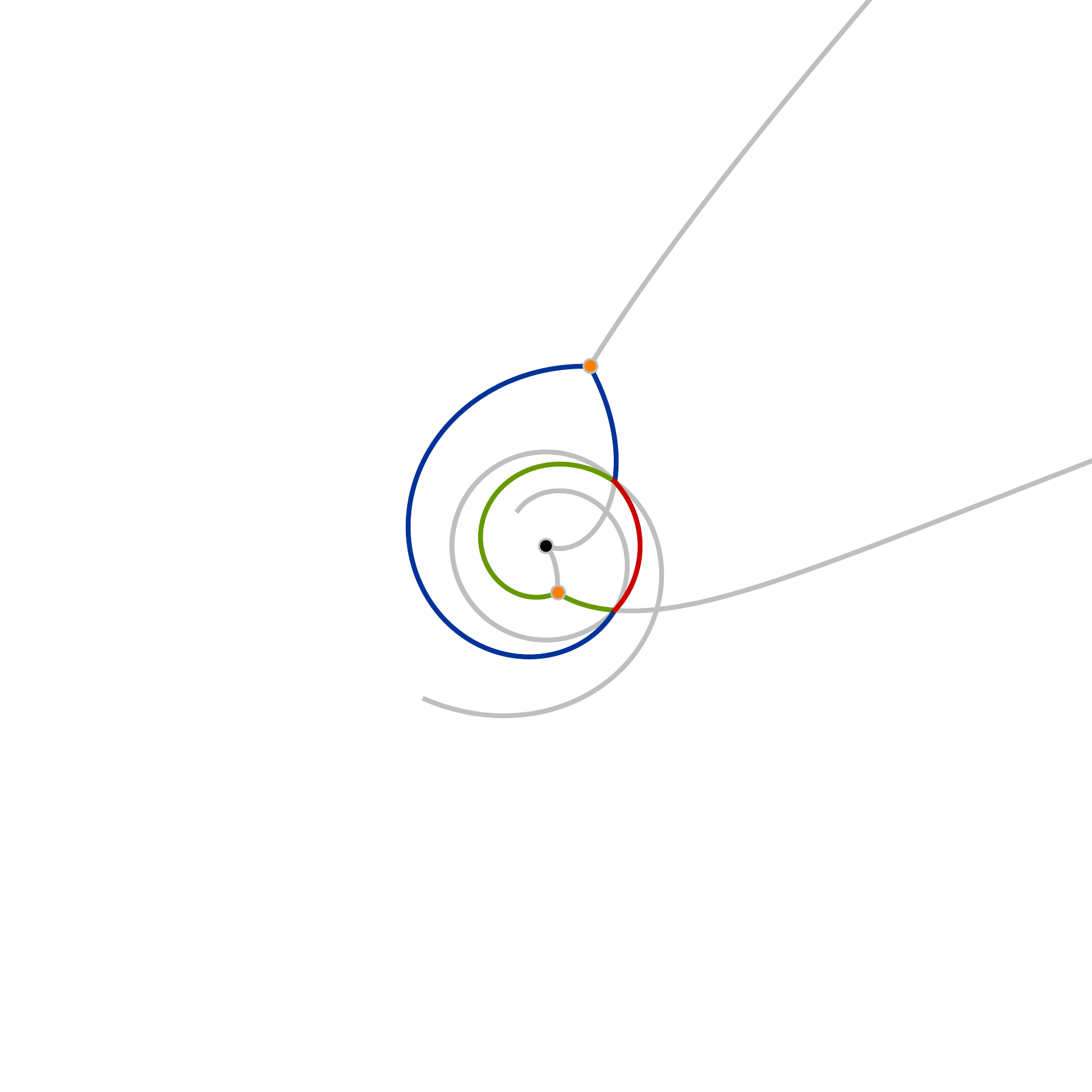}}
\hspace{.01\textwidth}
\fbox{\includegraphics[width=0.30\textwidth]{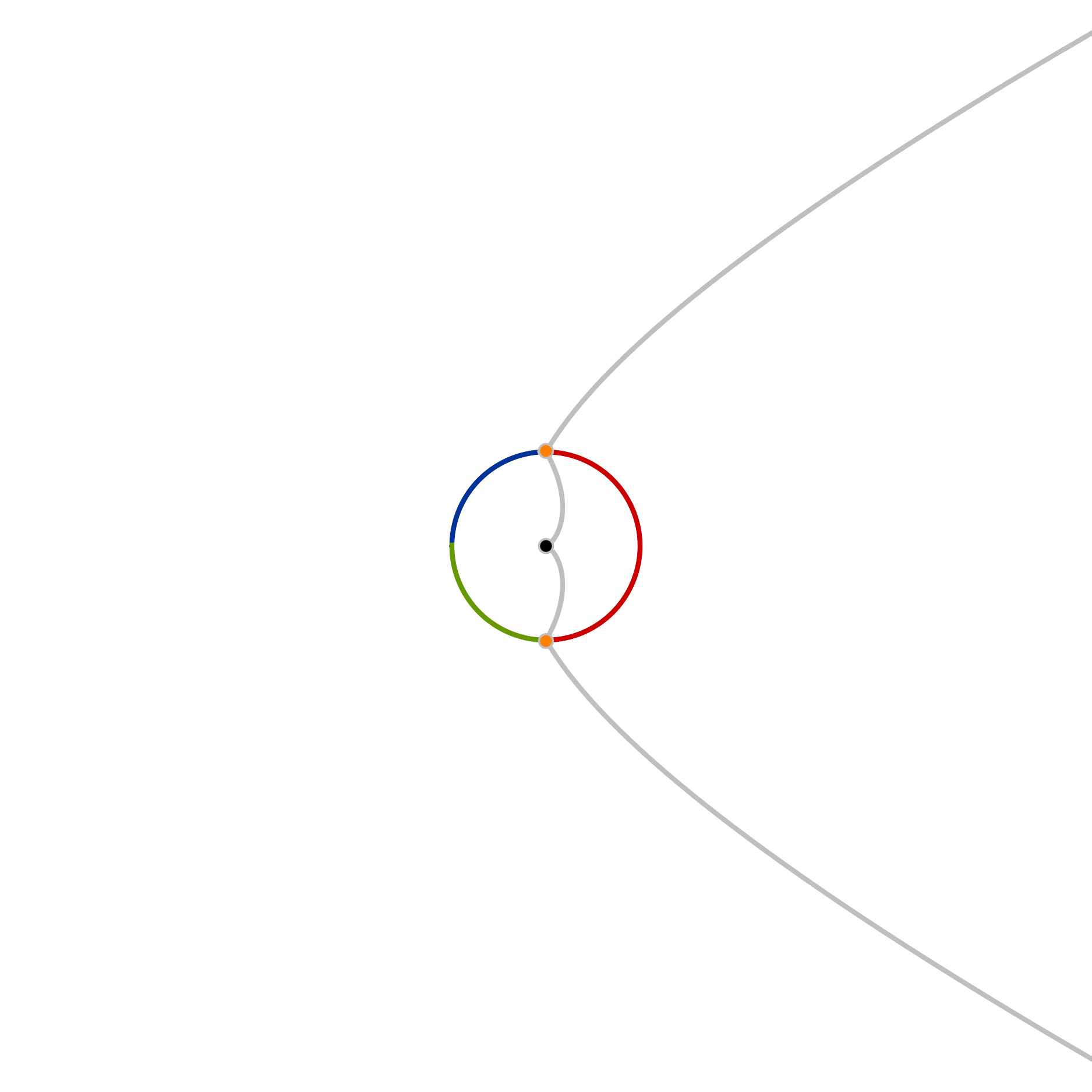}}
\hspace{.01\textwidth}
\fbox{\includegraphics[width=0.30\textwidth]{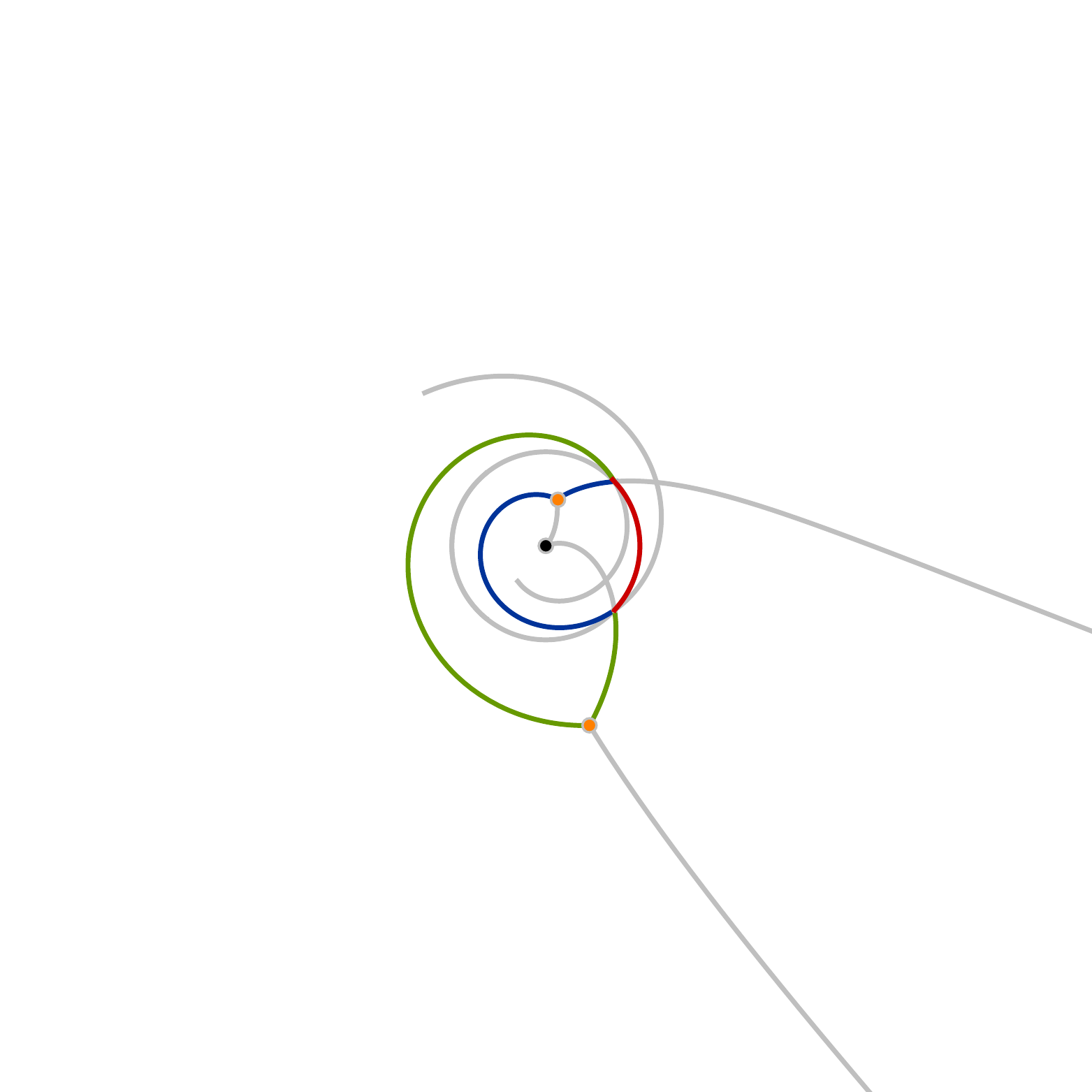}}
\\
\fbox{\includegraphics[width=0.30\textwidth]{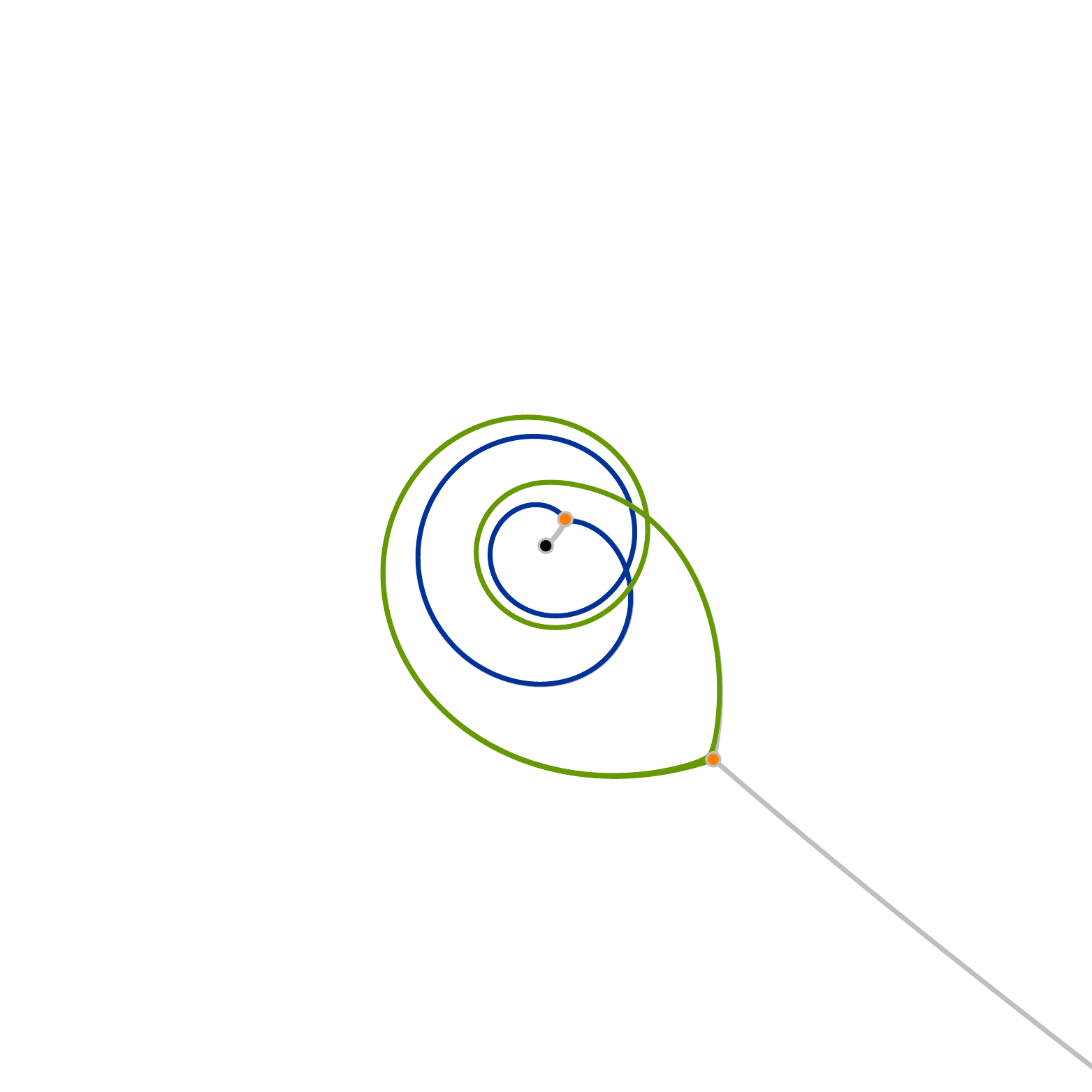}}
\hspace{.01\textwidth}
\fbox{\includegraphics[width=0.30\textwidth]{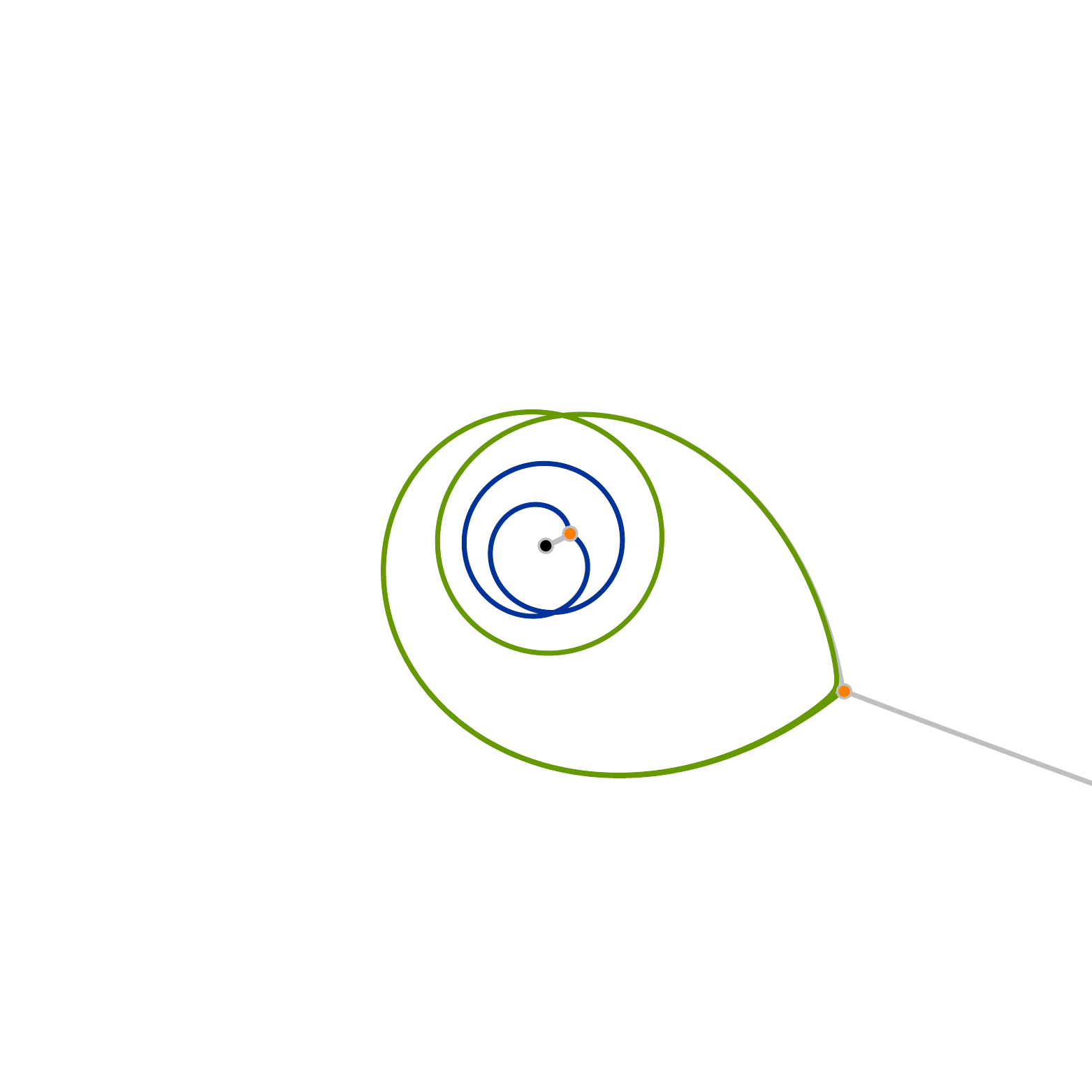}}
\hspace{.01\textwidth}
\fbox{\includegraphics[width=0.30\textwidth]{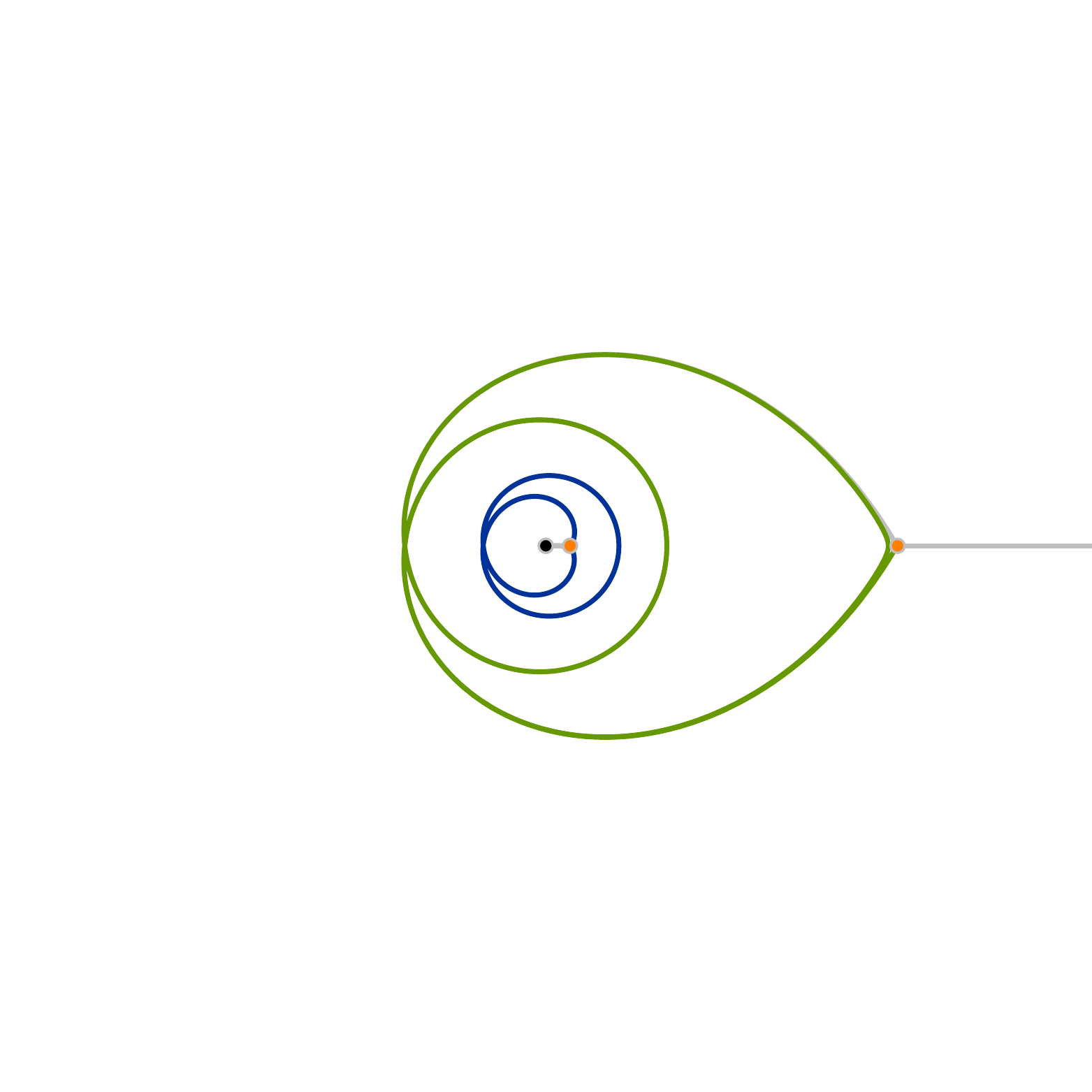}}
\caption{Monodromy of D0 cycles around the conifold point. Only some $\CE$-walls are shown (those in green, blue and red are two-way walls). The inner and outer D0 saddles get exchanged by the twist, through a topology-changing process involving the creation, and then destruction, of the red two-way wall connecting them.}
\label{fig:conifold-ST-twist}
\end{center}
\end{figure}

Let $Q = 1+\epsilon$ for some small $\epsilon>0$ and consider taking $\epsilon\cdot e^{2\pi i}$.
From (\ref{eq:conifold-branch-points}) it is clear that positions of branch points get switched $x_+\leftrightarrow x_-$. 
Naively, this implies that the inner D0 set of saddles gets exchanged with the outer D0 set of saddles.
But on the one hand, from Figure \ref{fig:conifold-Q-3-2-theta-0} it is clear that this cannot happen smoothly. On the other hand, both for $\epsilon = 1$ and $\epsilon = e^{2\pi i}$ we expect exactly the same plots.\footnote{$Z_{\text{D2}}$ remains unchanged, thanks to the fact that $y_\pm$ also get exchanged near the collison point $x=1$. Therefore we have two canceling effects on the periods: on the one hand $x_\pm$ get exchanged, on the other hand so do $y_\pm$. As a result, the integral $Z_{\text{D2}} = (2\pi R)^{-1}\int_{x_-}^{x_+}(\lambda_+ - \lambda_-)$ comes back to itself.} 
This hints to interesting transitions in the spectrum of BPS saddles, exchanging the inner D0 with the outer one through a sequence of topology-changing deformations.
This phenomenon can be fully appreciated by plotting the D0 saddles for different values of $\epsilon$, as shown in Figure \ref{fig:conifold-ST-twist}. 
In order for the exchange of the inner and outer saddles to occur, they first have to merge. In this process new two-way $\CE$-walls are created (those shown in red in Figure), and later destroyed, leaving behind the exchanged cycles.\footnote{In these plots we have taken $\epsilon=1/2$.}

The occurrence of these transitions is related to the Seidel-Thomas twist \cite{Seidel:2000ia}.
This twist happens at the conifold point where D2 becomes massless. However its physical effects differ significantly from those of the monodromy encountered at large $Q$.
In particular, since the central charge of the BPS states that remain massive (D0 branes) is independent of $Q$, it does not pick up any monodromy in this process.\footnote{At the level of vanilla BPS states, there doesn't seem to be any explicit manifestation of the Seidel-Thomas twist. We expect this to play an important role in the study of framed BPS states, such as D4-D2 boundstates \cite{Nishinaka:2010qk}.}

\subsection{Exponential BPS graph and BPS quiver}\label{sec:conifold-graph}

\begin{figure}[h!]
\begin{center}
\includegraphics[width=0.45\textwidth]{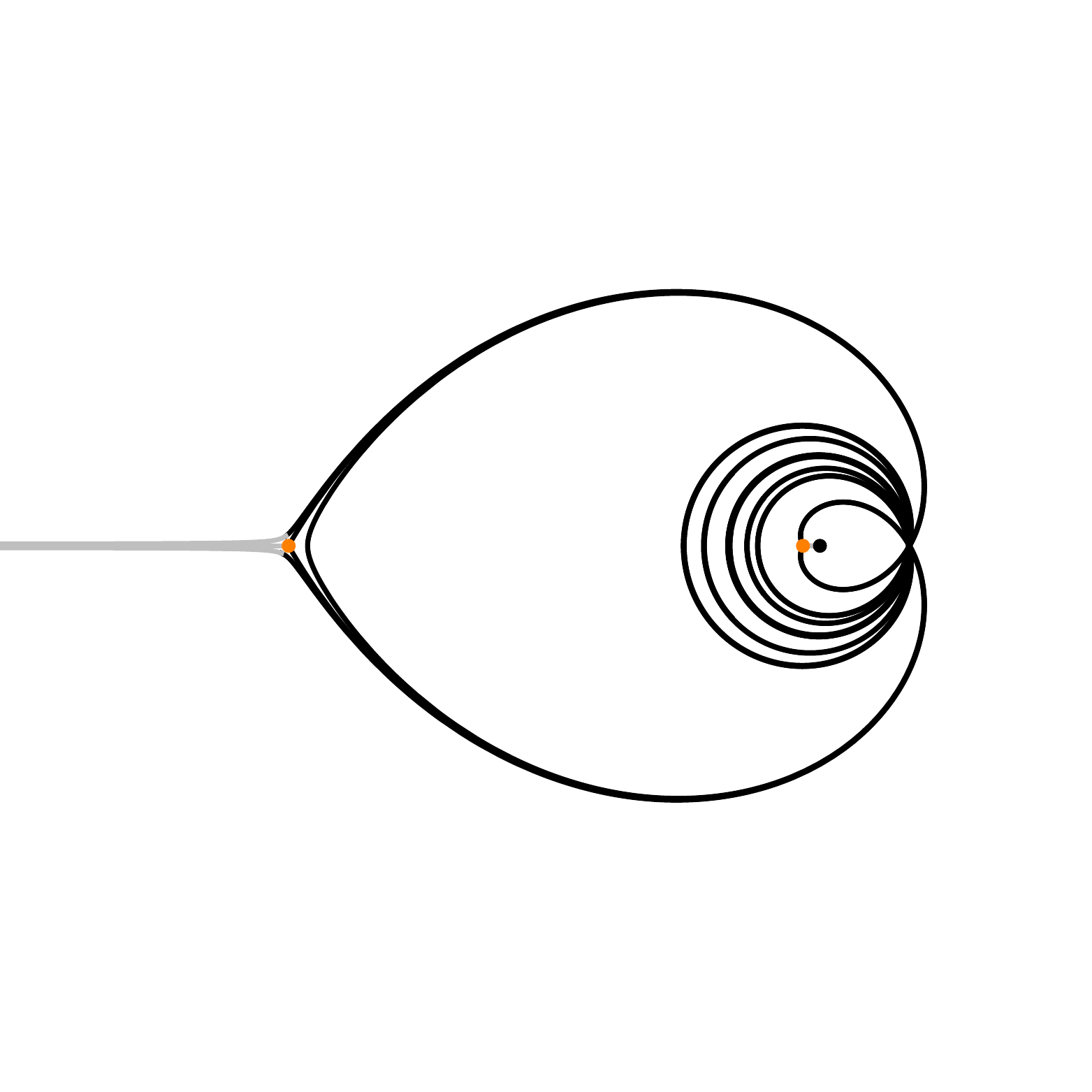}
\hspace*{.05\textwidth}
\includegraphics[width=0.45\textwidth]{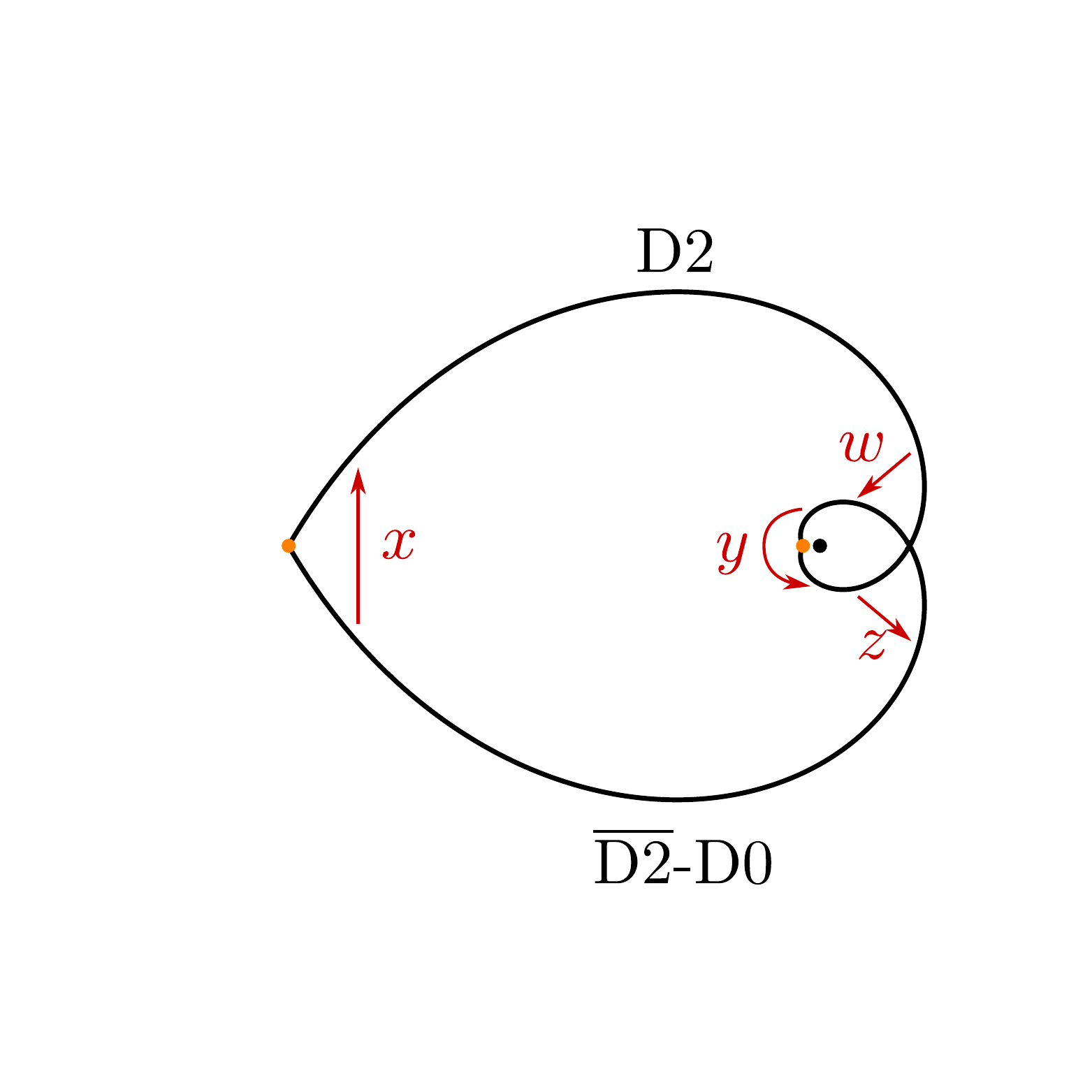}
\caption{Left: exponential network of the conifold at $Q=-1$. Right: exponential BPS graph.}
\label{fig:conifold-BPS-graph}
\end{center}
\end{figure}

We would now like to study the BPS graph of the conifold. Recall that BPS graphs, introduced in \cite{Gabella:2017hpz}, arise from maximally degenerate spectral networks. By ``maximally degenerate'' we mean that one chooses a point in the Coulomb branch of a class $\CS$ theory where central charges of all BPS states have the same phase (modulo $\pi$). 
By analogy, we define the \emph{exponential BPS graph} as the degenerate exponential network arising when 
\be
	\arg\, Z_{D2} = \arg Z_{D0}\,.
\ee
Given our explicit computation of central charges, this happens on the unit circle $|Q|=1$ in K\"ahler moduli space. 
To get the BPS graph from the degenerate exponential network, the first step is to decompose it into \emph{elementary webs}. 
The appropriate way of doing this involves studying the soliton data on each 2-way street, and identifying which ones are ``building blocks'' for others \cite{Longhi:2016wtv,Gabella:2017hpz}.
For the case of conifold these correspond to the two-way streets connecting the branch points, see Figure \ref{fig:conifold-BPS-graph}. In  a choice of trivialization for the charge lattice over the $Q$-plane, one of the two edges of the BPS graph lifts to the  D2 BPS cycle, the other to the $\overline{\text{D2}}$-D0 cycle.

Ordinary BPS graphs have several nice properties, in particular they encode the Kontsevich-Soibelman wall-crossing invariant of the theory \cite{Longhi:2016wtv} (a.k.a. motivic spectrum generator), as well as the BPS quiver with potential which provides a dual description of the BPS spectrum \cite{Alim:2011ae}. The first property carries over to exponential BPS graphs, since the whole reasoning in \cite{Longhi:2016wtv} is based on physical considerations of 2d-4d wall-crossing, which also apply to 3d-5d wall-crossing \emph{mutatis mutandis}. The analysis will however be more involved than the case of ordinary BPS graphs.
Here we focus on the second property, namely how to obtain quivers and their potential.

\begin{figure}[h!]
\begin{center}
\includegraphics[width=0.5\textwidth]{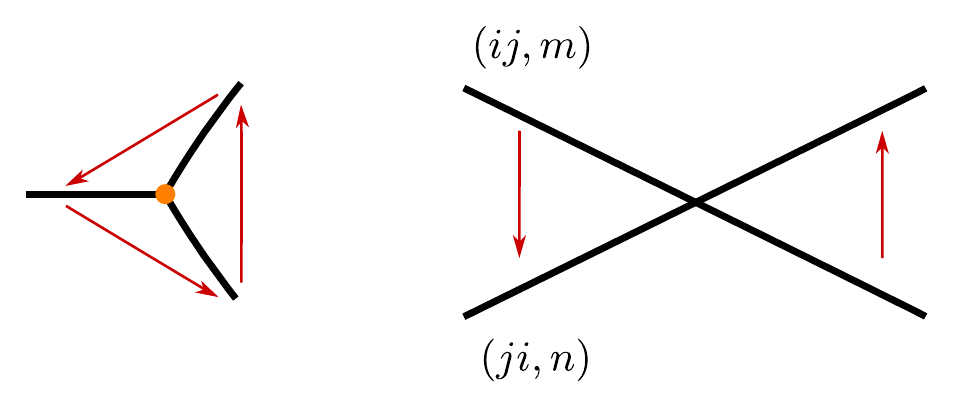}
\caption{Arrows for the BPS quiver arise from intersections of edges of the BPS graph. Left: a branch point, right: an $ij-ji$ junction (descendant walls are omitted).}
\label{fig:local-quiver-graph-rules}
\end{center}
\end{figure}

Nodes of the BPS quiver are dual to edges of the BPS graph, therefore we have two nodes in this case. Arrows are determined by intersections of the edges in the BPS graph, according to the local rules depicted in Figure \ref{fig:local-quiver-graph-rules}. These are obtained by studying intersections of the lifts of the two-way streets to the covering surface $\Sigma$.

\begin{figure}[h!]
\begin{center}
\includegraphics[width=0.35\textwidth]{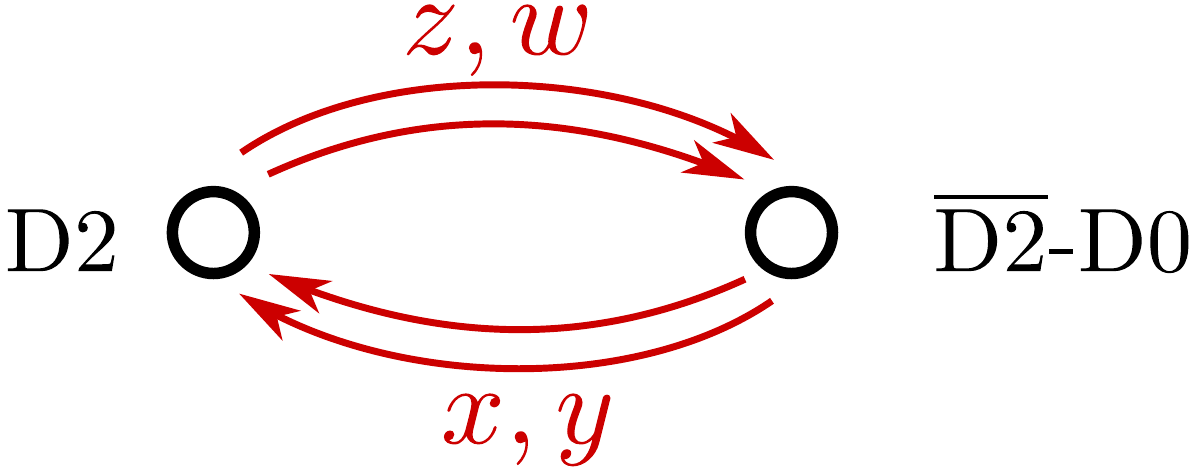}
\caption{BPS quiver of the conifold.}
\label{fig:conifold-quiver}
\end{center}
\end{figure}

The superpotential is generated by worldsheet disk instantons, which determine the $A_\infty$ structure of objects in the Fukaya category, for more details see \cite{Eager:2016yxd}.
An advantage of BPS graphs, is that the superpotential can be read off the graph directly. To do so, one picks a choice of trivialization for $\Sigma\to\IC^*_x$, this assigns labels $(ij,n)$ to each $\CE$-wall. Then for each connected component $U$ of $\IC^*_x\setminus \CG$ cut out by the exponential BPS graph $\CG$ one performs the following procedure. 
If the covering has $N$ sheets labeled by $i$, we fix an $i$. We then choose an arbitrary point inside $U$, and grow a ball from this point until we hit walls of type $i$. If such walls are met, this will give rise to a polygon bounded by edges of the exponential BPS graph, with corners corresponding to intersections of edges, therefore to arrows in the BPS quiver.
Moreover, the flow of the underlying exponential network assigns an orientation to the sheet $i$ on each $\CE$-wall: this is the same orientation that 3d-5d BPS solitons have when lifted to that sheet. 
Therefore we get a polygon with an oriented boundary, which induces an orientation of the polygon.\footnote{In fact, the requirement that different pieces of the boundary have compatible orientations under concatenation strongly constrains the admissible polygons. It turns out that at $ij$-branch points, a polygon of type $i$ or $j$ always has a corner. At $ij-jk$ junctions a polygon of  type $i,j$ or $k$ always has a corner. At junctions of type $ij-kl$ a polygon never has a corner. At junctions of type $ij-ji$ a polygon of type $i$ or $j$ may either have a corner, or its boundary may proceed straight (but often, in the second case one ends up with infinite-area polygons, so these are suppressed). These simple rules simplify the count holomorphic disk instantons, bypassing the  need for an explicit choice of trivialization.} 
Along the boundary we find a number of corners, let $a_1,\dots a_k$ their labels in the order fixed by the orientation of the polygon.
Each such polygon contributes to the superpotential by 
\be
	\Delta W =\pm\Tr  \( a_1\dots a_k\)
\ee 
where the sign depends on whether the orientation of the polygon agrees with that of $\IC^*_x$ or is opposite to it.
These rules agree with (and generalize) the ones in \cite{Gabella:2017hpz}, and are equivalent to the rules adopted in \cite{Eager:2016yxd} who worked directly on $\Sigma$. The point is that these rules produce polygons that lift to disks on $\Sigma$, by construction.
In fact, each of these contributions is weighted by $e^{-A}$ where $A$ is the area of the disk, for this reason we exclude regions containing punctures (they have infinite area).

\begin{figure}[h!]
\begin{center}
\includegraphics[width=0.85\textwidth]{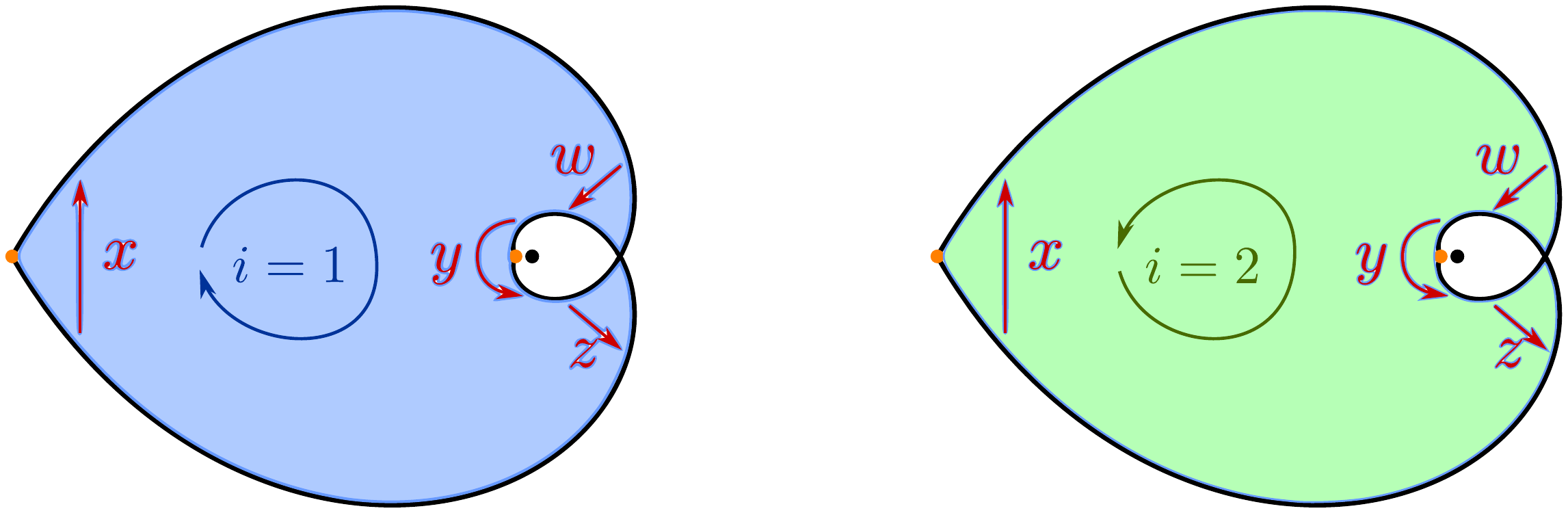}
\caption{Disk instantons generating the conifold superpotential.}
\label{fig:conifold-superpotential}
\end{center}
\end{figure}

For example, for the conifold's exponential BPS graph we have two sheets $i=1,2$ and only one bounded region in $\IC^*$. Their contributions are read off as illustrated in Figure \ref{fig:conifold-superpotential}, giving
\be
	W =\Tr\( \underbrace{-xwyz}_{i=1} + \underbrace{xzyw}_{i=2} \)\,.
\ee
This matches the known quiver with potential, see e.g. \cite{Aspinwall:2004bs} for derivation from D-branes.

\begin{figure}[h!]
\begin{center}
\includegraphics[width=0.99\textwidth]{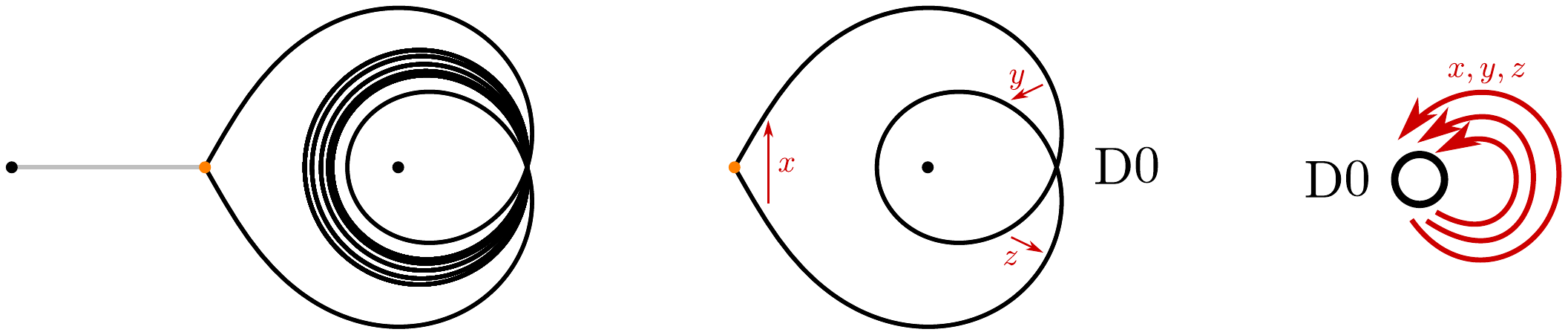}
\caption{Left: Exponential network of $\IC^3$ at $\vartheta=\arg Z_{\text{D0}}$. Center: BPS graph of $\IC^3$. Right: the dual BPS quiver.}
\label{fig:C3-graph-quiver}
\end{center}
\end{figure}

\paragraph{Excursus: exponential BPS graph and BPS quiver of $\IC^3$.}
Having set out the rules to obtain exponential BPS graphs and BPS quivers in full generality, let us check how they apply to the case of $\IC^3$. The charge lattice in this case is generated by the D0 brane, so the BPS graph is just the elementary web appearing at $\vartheta=0$. This is shown in Figure \ref{fig:C3-graph-quiver}, also see \cite[Figure 10]{Banerjee:2018syt}.
It is important to observe that, unlike in the case of spectral networks, exponential BPS graphs may give rise fo self-intersecting cycles. These would correspond to quivers with loops on their nodes. An example of this is indeed $\IC^3$ quiver, also shown in Figure \ref{fig:C3-graph-quiver}.
Again the covering map has degree 2, and one finds two polygons (triangles) giving a superpotential
\be
	W = \Tr\(xyz-xzy \)\,.
\ee
It is a nontrivial check, and an interesting exercise to study what happens in a different choice of framing. We have verified that our recipe for obtaining the BPS quiver and its superpotential still give the correct answer if we switch to cubic framing.\footnote{By cubic framing we mean one where the curve  is  $1+y+ xy^3=0$.}

\section{$\CO(0)\oplus\CO(-2)\to\IP^1$}\label{sec:C3modZ2}

This section is devoted to another local toric threefold, that may be regarded as a variant of the conifold studied above.
Our analysis will proceed in parallel with that of Section \ref{sec:conifold}, therefore we shall skip some details and focus on the salient differences.

\subsection{Geometry and its mirror}

The local toric threefold $\CO(0)\oplus\CO(-2)\to\IP^1$ starting from a $U(1)$ gauged linear sigma model with charges $(+1,+1,0,-2)$ for coordinates $(w_1,\dots ,w_4)\in \IC^4$. 
Physically one considers the zero-locus of the $D$-term equations
\be
\label{C2modZ2-sym-quot}
	|w_1|^2 + |w_2|^2  - 2|w_4|^2 =  r
\ee
and takes a $U(1)$ quotient for gauge-invariance.
Mathematically this corresponds to a symplectic quotient of $\IC^4$.
The mirror curve for this can be obtained by T-duality as in Hori-Vafa mirror symmetry
\be
x_1 x_2 x_4^{-2} = Q, \quad x_1 + x_2 + x_3 + x_ 4 = 0
\ee
where $|x_i| = e^{-|w_i|^2}$ and $|{Q}| = e^{r}$. 
Fixing the patch $x_1 = 1$, and setting $x_3 = x, x_4 = y$, one gets the following as the 
mirror curve 
\be
1 + x + y + Q y^{2} = 0. 
\ee
Changing framing $x \mapsto xy^{-1}$ one obtains the curve 
\be
\label{mirr-C3modZ2}
x + y + y^2 + Q \,y^3  = 0 \quad \subset\quad \IC^*_x \times \IC^*_y \,.
\ee
This curve satisfies the criteria explained in Appendix \ref{app:framing} regarding the choice of framing. 
We will study this form of the mirror curve with exponential networks.

There are three branches, locally labeled by three roots $y_i(x)$ $i=1,2,3$.
These meet in correspondence of two branch points, whose locations are
\be\label{eq:C2modZ2-branch-points}
	x_\pm = 
	\frac{9 Q-2 \pm 2 (1-3Q)^{3/2}}{27 Q^2}\,.
\ee
Branch points collide if $Q=1/3$, however no cycle pinches here, because the two branch points connect sheets $(12)$ and $(23)$ in a suitable choice of trivialization, and do not support a cycle.
The curve has a conifold point, this will be identified later when we compute periods.
Punctures on this curve are defined to be all those points where the curve intersects the lines $x=0,\infty$ or $y=0,\infty$.
There are four such points: 
\be \label{eq:C3modZ2-punctures}
\begin{split}
	&\mathfrak{p}_1:\ (x=0,y=-\frac{1}{2Q}(1+\sqrt{1-4Q}))
	\qquad
	\mathfrak{p}_3:\ (x=0,y=0)
	\\
	&
	\mathfrak{p}_2:\ (x=0,y=-\frac{1}{2Q}(1-\sqrt{1-4Q}))
	\qquad
	\mathfrak{p}_4: \ (x=\infty,y=\infty)
\end{split}
\ee

The curve is then a three-fold cover of the $x$-plane with two branch points of swuare-root type and a branch point with three-fold ramification, i.e. it is a sphere. 
Punctures clearly project to $x=0,\infty$. Above $x=0$ there is a puncture on each sheet.
Above $x=\infty$ one has $y\sim x^{1/3}$ and there is a single puncture in correspondence of a cubic ramification point where all three sheets meet. 
$\Sigma$ is thus a four-punctured sphere.

\begin{figure}[h!]
\begin{center}
\includegraphics[width=0.75\textwidth]{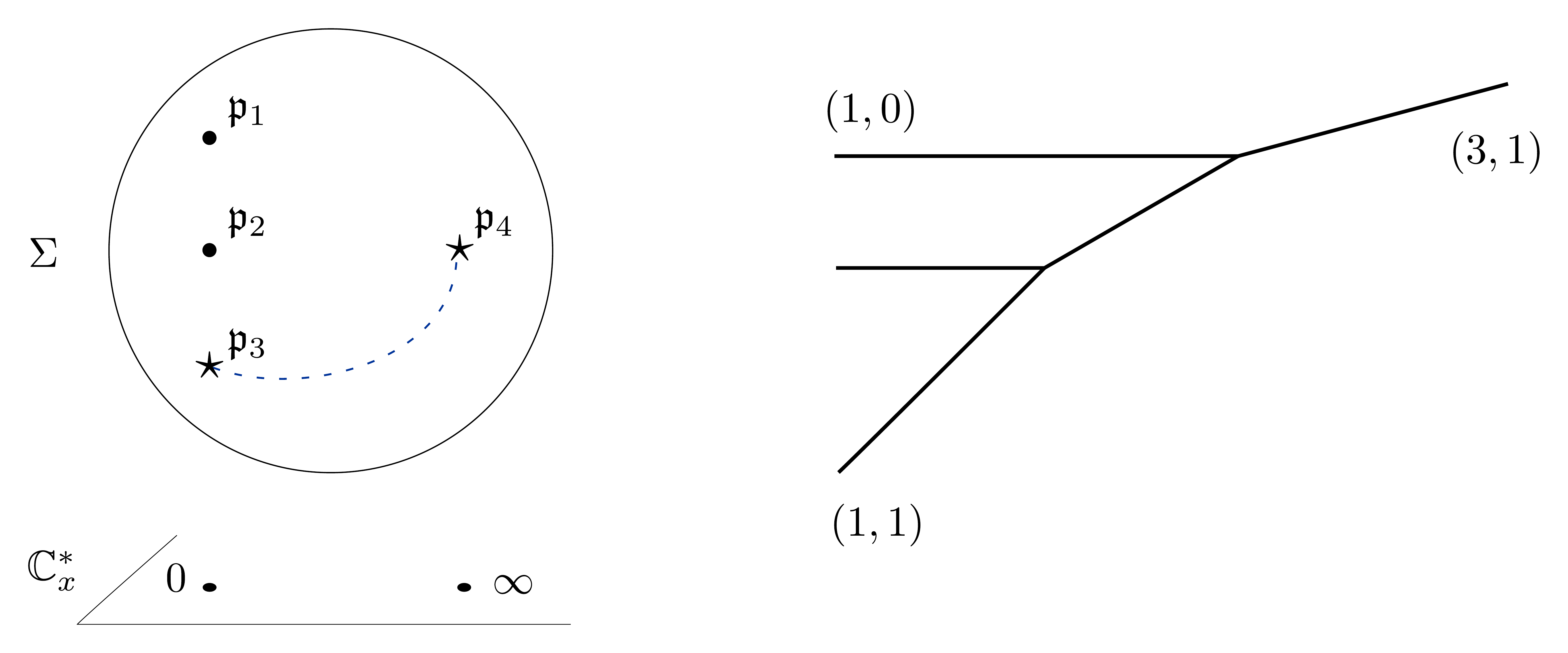}
\caption{Mirror curve and toric diagram for (\ref{mirr-C3modZ2}).}
\label{fig:C3modZ2-punctures}
\end{center}
\end{figure}

\subsubsection{Types of punctures}

According to the criterion explained in the previous section, we classify punctures according to the asymptotics of the logarithmic differential (\ref{eq:differential}) near each of them  
\be\label{eq:C3modZ2-differential-p1-p2-p3-p4}
\begin{split}
(\mathfrak{p}_1) \ \ &\lambda_1 = \log\( -\frac{1}{2Q}(1+\sqrt{1-4Q}))  \) \, d\log x + \dots
\\
(\mathfrak{p}_2) \ \ & \lambda_2 = \log\( -\frac{1}{2Q}(1-\sqrt{1-4Q}))  \) \, d\log x + \dots  
\end{split}
\qquad
\begin{split}
(\mathfrak{p}_3) \ \ &\lambda_3 = \log x \, d\log x + \dots 
\\ 
(\mathfrak{p}_4) \ \ &\lambda_{1,2,3} = \frac{1}{3} \log x \, d\log x + \dots 
\end{split}
\ee

Above $x=0$ on sheets 1 and 2 (punctures $\mathfrak{p}_1$ and $\mathfrak{p}_2$) the differential approaches $\frac{dx}{x}$ and is single valued, but on sheet 3 (puncture $\mathfrak{p}_3$) the differential approaches ${\log x}\frac{dx}{x}$ and is multivalued.
Likewise above $x=\infty$ we find that puncture $\mathfrak{p}_4$ is of logarithmic type.

\subsubsection{Periods}

Following the general discussion outlined for the conifold, once again we need to find suitable integration cycles whose periods do not depend on the basepoint. These cycles may then be promoted from homotopy classes to homology classes, and provide a suitable basis for charges of BPS states.

\subsubsection*{D2 cycle}

Let $p\in \Sigma$ be an arbitrary basepoint, such that it does not coincide with any of the punctures.
Let $C_i \in \pi_1(\Sigma,p)$ be a cycle based at $p$  and encircling  $\mathfrak{p}_i$ counterclockwise.
Periods of $C_1$ and $C_2$ do not depend on $p$, because these paths do not cross the logarithmic cut.
We define $A,\tilde A$ to be the homology classes
\be
\left\{
	\begin{split}
	A & : = [C_1\circ C_2^{-1}] = [C_1]-[C_2] \\
	\tilde A & : = [C_1\circ C_2] = [C_1]+[C_2] 
	\end{split}
\right.
\qquad
\in H_1(\Sigma,\IZ) \,.
\ee
Their periods are easily obtained as combinations of residues from (\ref{eq:C3modZ2-differential-p1-p2-p3-p4})
\be\label{eq:periods-C3modZ2}
\begin{split}
	Z_A 
	&= \frac{1}{2\pi R} \, \int_{C_1\circ C_2^{-1}} \lambda
	= \frac{i}{R} \log T
	\\
	Z_{\tilde A }
	&= \frac{1}{2\pi R} \, \int_{C_1\circ C_2} \lambda
	= -\frac{i}{R} \log Q 
\end{split}
\ee
where\footnote{The other root of $T(Q)$ would simply give $T^{-1}$, whose logarithm is the period of the cycle $-A$.} 
\be \label{eq:TandQrel}
	T = \frac{1-2 Q+\sqrt{1-4 Q}}{2 Q} 
	\,,
	\quad
	Q = \frac{T}{(1+T)^2}\,.
\ee
It will turn out by direct inspection that $Z_A$ is the D2 central charge, therefore we define
\be
	\gamma_{\text{D2}} = A\,.
\ee

\begin{figure}[h!]
\begin{center}
\includegraphics[width=0.55\textwidth]{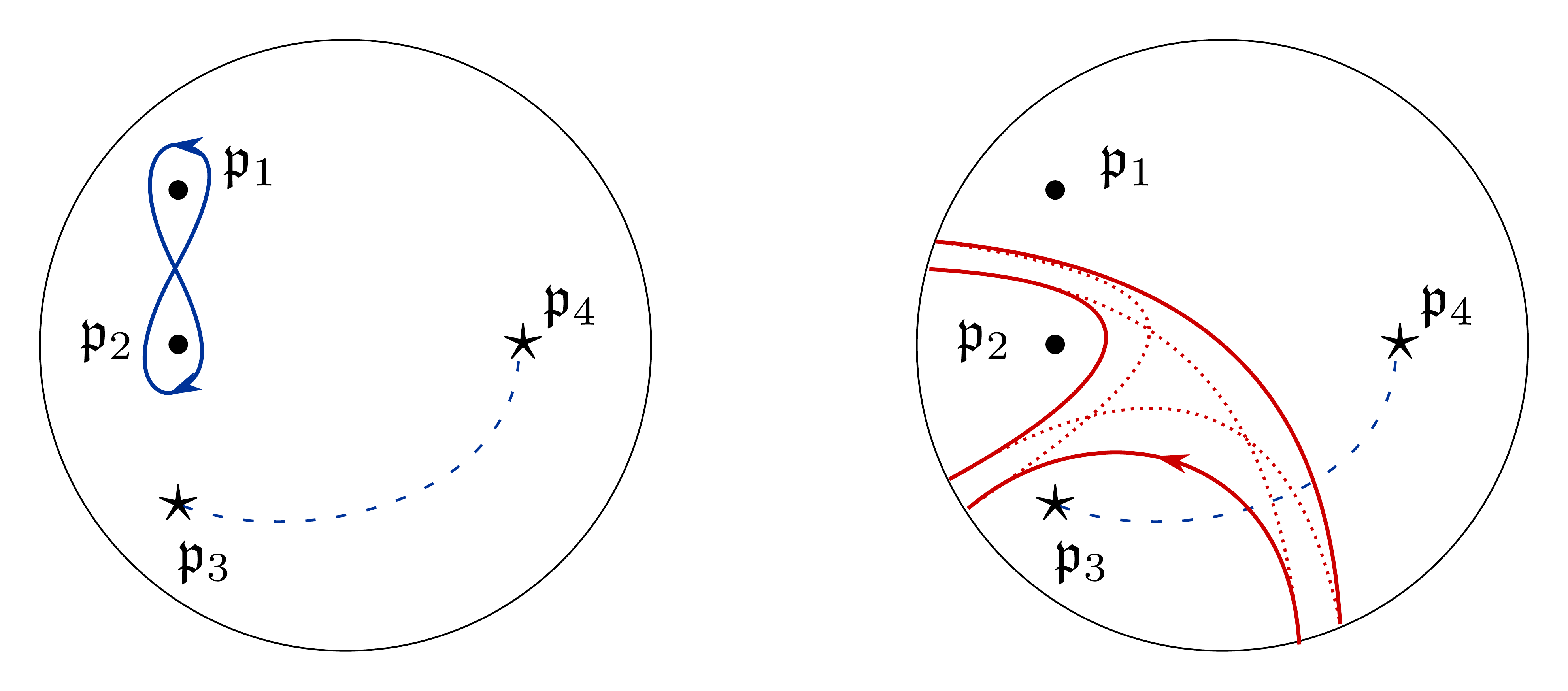}
\caption{Cycles corresponding D2 (blue) and D0 (red) branes on the mirror curve of $\CO(0)\oplus\CO(-2)\to\IP^1$. }
\label{fig:C3modZ2-cycles}
\end{center}
\end{figure}

\subsubsection*{D0 cycle}

The D0 cycle is more subtle than D2, but its description is very similar to the case of the conifold, as shown in Figure \ref{fig:conifold-cycles}.
By the same arguments as those illustrated in the previous section, one can show that its period is independent of the basepoint, and that it descends to a functional of the homology class of the cycle.
\be\label{eq:C3modZ2-D0-period}
	Z_{D0} = \frac{1}{2\pi R}\oint_{\gamma_{D0}}\lambda = \frac{2\pi}{R}\,.
\ee

\subsubsection{Distinguished values of $Q$ and curve degeneration}\label{sec:C3modZ2-factorization}

There are three distinguished regions in the moduli space of the curve (\ref{mirr-con}): the region near $T=1$ where $A$ pinches, and those near $Q=0,\infty$.

\subsubsection*{Conifold point}

$T=1$ corresponds to the conifold point. 
One branch point moves to $x_- = 0$ the other to $x_+ = 8/27$.
Above $x=0$, the three sheets are positioned at $y_1 = y_2 = -2$ and $y_0=0$. Punctures $\mathfrak{p}_1$ and $\mathfrak{p}_2$ on $\Sigma$ now collide, this is why $A = [C_1] -[ C_2]$ pinches. On the toric diagram, this corresponds to shrinking the internal leg, then the two $(1,0)$ legs become coincident.

\subsubsection*{Degeneration at large $T$}

Near $T= \infty$ one has $Q\to 0$, therefore branch points (\ref{eq:C2modZ2-branch-points}) become infinitely separated with $x_-\to \infty$ and $x_+\to 1/4$.
The curve (\ref{mirr-C3modZ2}) becomes
\be\label{eq:mirr-C3modZ2-factorized-1}
	x+ y + y^2 = 0
\ee
which has the same form as (\ref{eq:mirr-conifold-factorized-2}), hence giving a copy of the $\IC^3$ mirror curve. 
Above the branch point $x_+$, two sheets remain at finite distance $y = -1/2$, while the third one goes to infinity $y\sim 1/Q$. 
On the toric diagram this can be understood as sending the length of the internal leg to infinity, which results in one of the $(1,0)$ legs moving up and infinitely far away. On the curve, we have a degeneration similar to the one shown in Figure \ref{fig:conifold-degeneration-Qeq0}. 
In this case the curve develops a long thin tube separating $\mathfrak{p}_{1}$ and $\mathfrak{p}_{4}$ from $\mathfrak{p}_{2}$ and $\mathfrak{p}_{3}$.
As $Q\to 0$ the remnants correspond to $\mathfrak{p}_2$, $\mathfrak{p}_3$ and the puncture created by the pinching cycle.

So far we have observed the local behavior near the bottom-left vertex of the $(2,1)$ internal leg.
To zoom into the other corner we should rescale $x\to Q^{-2} x$ and $y\to Q^{-1} y$, which leads to 
\be\label{eq:mirr-C3modZ2-factorized-2}
	x + y^2 + y^3 = 0
\ee
which is the mirror curve of $\IC^3$ in quadratic framing.\footnote{This can be obtained by substitution $x\to x y^{-2}$ in $1+x+y=0$.}
In this scaling limit we retain punctures $\mathfrak{p}_1$ and $\mathfrak{p}_4$, while $\mathfrak{p}_2$ and $\mathfrak{p}_3$ have coalesced.

\subsubsection*{Degeneration at small $T$}

For $T\to 0$ one again finds $Q\to 0$, therefore the curve degenerates exactly in the same way as above. This is related to the fact that there is no flop transition for this Calabi-Yau.

\subsection{BPS states} 

Let us now analyze the BPS spectrum of this curve by plotting exponential networks at various phases, and detecting BPS states of the theory as generalized saddles. 
The BPS spectrum is not expected to undergo any jumps by wall-crossing\footnote{Due to the absence of compact 4-cycles and 6-cycles, the spectrum of D2-D0 states cannot undergo wall-crossing, which would require the presence of mutually non-local BPS states.}
therefore we will fix a convenient value of $T$ and analyze the spectrum at a single point in moduli space. We will later comment on the global features of exponential networks in other regions.

\subsubsection{Networks at various phases}
\begin{figure}[h!]
\begin{center}
\fbox{\includegraphics[width=0.228\textwidth]{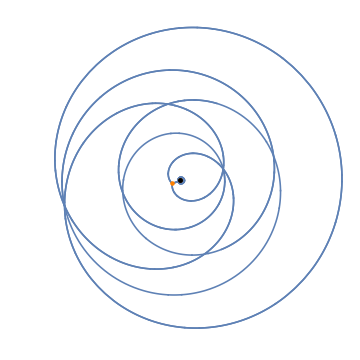}}
\hfill
\fbox{\includegraphics[width=0.228\textwidth]{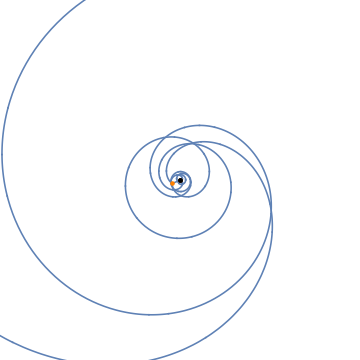}}
\hfill
\fbox{\includegraphics[width=0.228\textwidth]{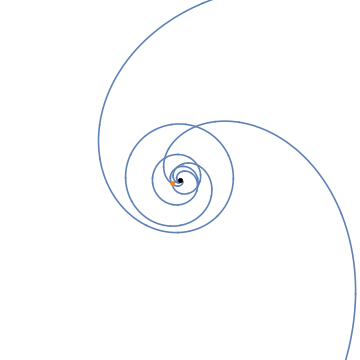}}
\hfill
\fbox{\includegraphics[width=0.228\textwidth]{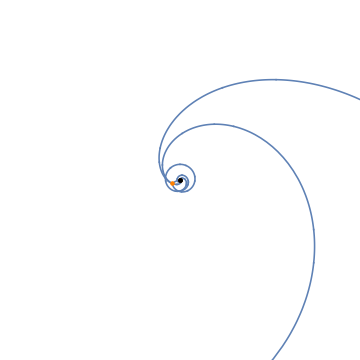}}
\\
\fbox{\includegraphics[width=0.228\textwidth]{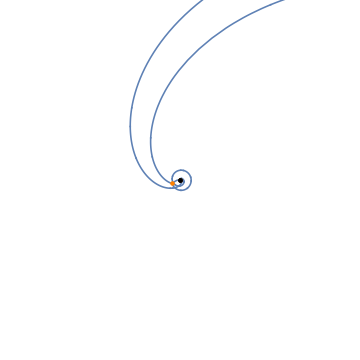}}
\hfill
\fbox{\includegraphics[width=0.228\textwidth]{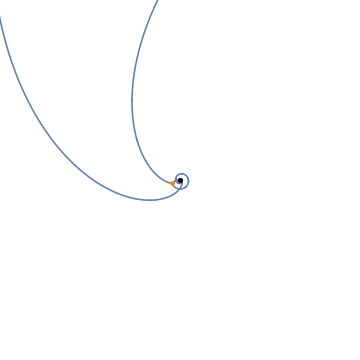}}
\hfill
\fbox{\includegraphics[width=0.228\textwidth]{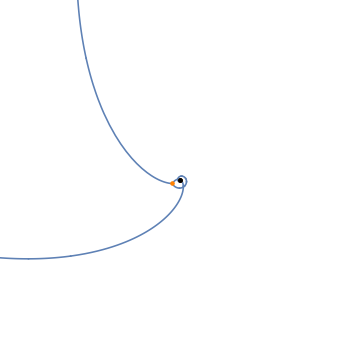}}
\hfill
\fbox{\includegraphics[width=0.228\textwidth]{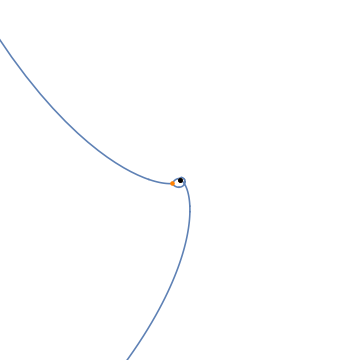}}
\\
\fbox{\includegraphics[width=0.228\textwidth]{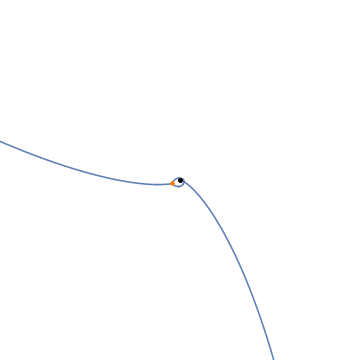}}
\hfill
\fbox{\includegraphics[width=0.228\textwidth]{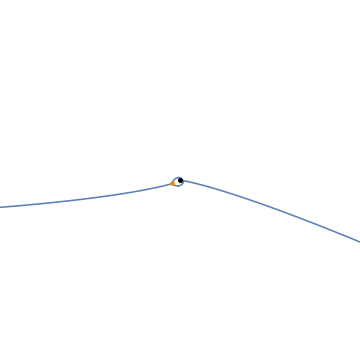}}
\hfill
\fbox{\includegraphics[width=0.228\textwidth]{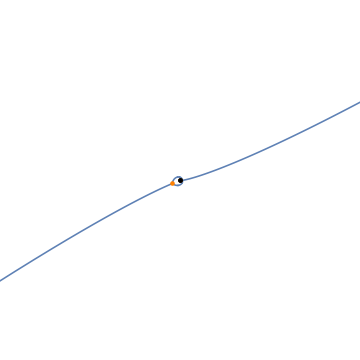}}
\hfill
\fbox{\includegraphics[width=0.228\textwidth]{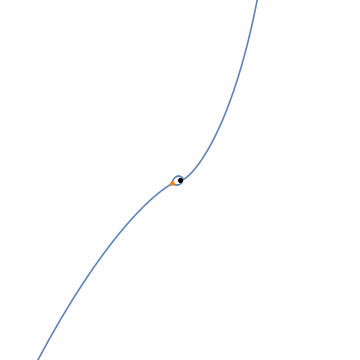}}
\\
\fbox{\includegraphics[width=0.228\textwidth]{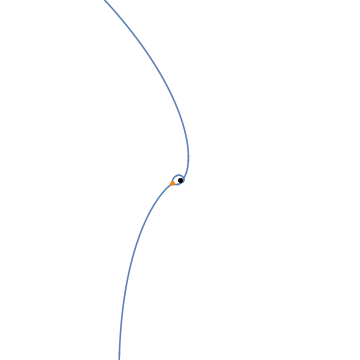}}
\hfill
\fbox{\includegraphics[width=0.228\textwidth]{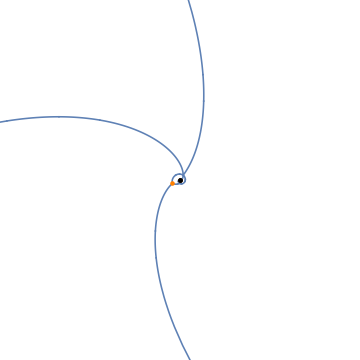}}
\hfill
\fbox{\includegraphics[width=0.228\textwidth]{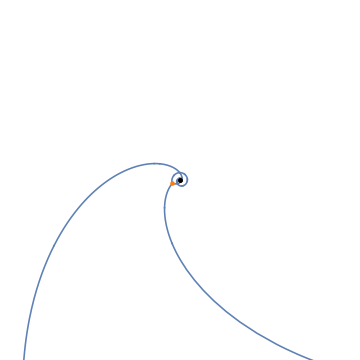}}
\hfill
\fbox{\includegraphics[width=0.228\textwidth]{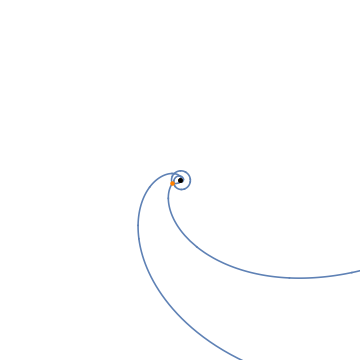}}
\\
\fbox{\includegraphics[width=0.228\textwidth]{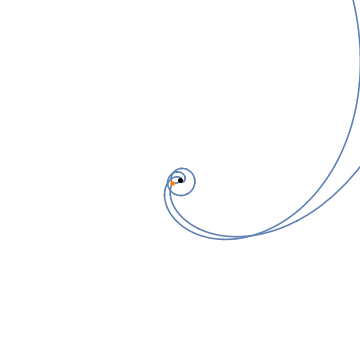}}
\hfill
\fbox{\includegraphics[width=0.228\textwidth]{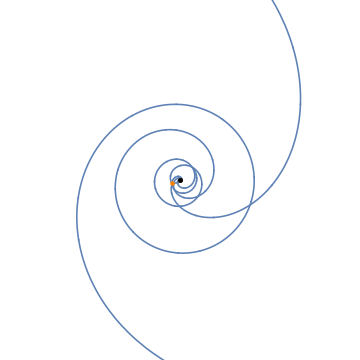}}
\hfill
\fbox{\includegraphics[width=0.228\textwidth]{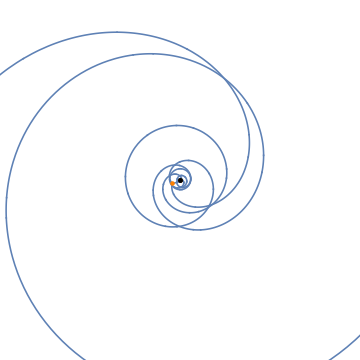}}
\hfill
\fbox{\includegraphics[width=0.228\textwidth]{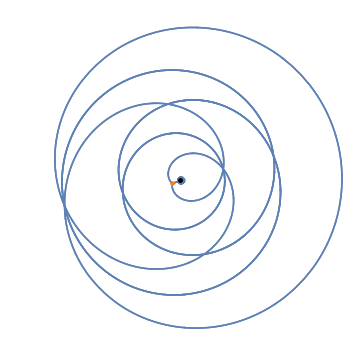}}
\caption{Exponential networks of $\CO(0)\oplus\CO(-2)\to\IP^1$ for $T=6+i$ with $\vartheta\in (0,\pi)$}
\label{fig:C3modZ2_T_6_i}
\end{center}
\end{figure}

Let us fix $T=6+i$, a sample of exponential networks is shown in Figure \ref{fig:C3modZ2_T_6_i}.
To look for BPS states we must, as usual, tune $\vartheta$ to the phases of their central charges and verify the presence of a saddle.

\subsubsection{Generalized saddles}\label{sec:C3modZ2-saddles}

At $T=6+i$ we find the following saddles through the evolution (partially) depicted in Figure \ref{fig:C3modZ2_T_6_i}.

\subsubsection*{D0 states}

\begin{figure}[h!]
\begin{center}
\includegraphics[width=0.85\textwidth]{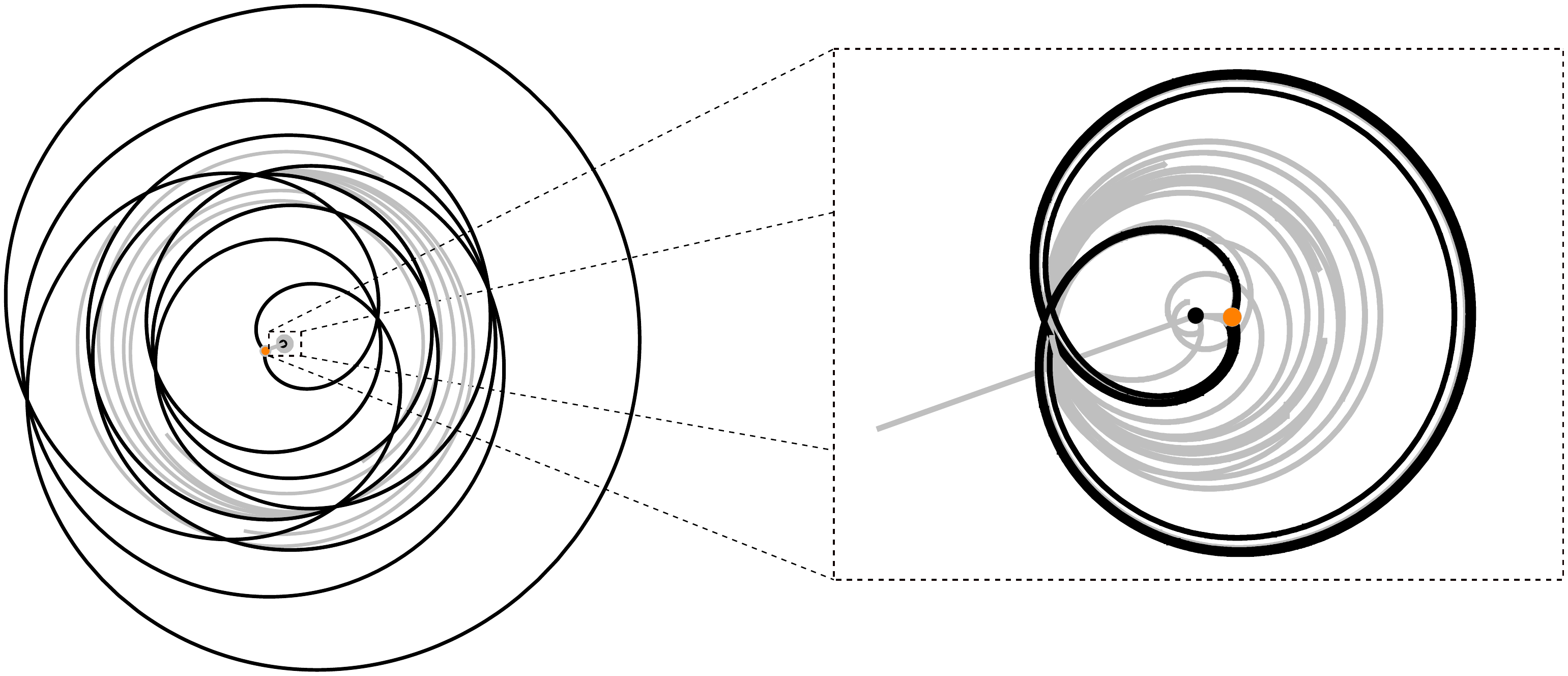}
\caption{
D0 saddles of $\CO(0)\oplus\CO(-2)\to\IP^1$, for $T=6+i$ at $\vartheta=0$. In this picture only some two-way walls are shown in black.
}
\label{fig:C3modZ2-T-6-i-theta-0}
\end{center}
\end{figure}

At $\vartheta = \arg Z_{D0} = 0$ we find the generalized saddles of Figure \ref{fig:C3modZ2-T-6-i-theta-0}. 
Thanks to the fact that the saddle consists of two disjoint components, we can analyze them separately.
In fact one of them resembles exactly the D0 brane saddle of the $\IC^3$ exponential networks studied in \cite{Banerjee:2018syt}. This is the saddle supported by the branch point $x_+$, which is the one closest to $x=0$.\footnote{Indeed we have seen above that in the degeneration limit, the D0 near $x=0$ remains of finite mass and wraps around two punctures. For this reason it resembles the cycle studied previously.}
The other saddle is supported by the branch point $x_-$ and looks like a D0 brane for the $\IC^3$ curve in cubic framing (compare with \cite[Figure 51]{Eager:2016yxd}).\footnote{Wa have also checked that the D0 cycle appearing here resembles exactly that of $\IC^3$ with curve $1+y+x y^3=0$.}

It follows immediately that these are two copies of the KK tower appearing in $\IC^3$.
Since saddles are disjoint, the previous analysis of soliton data applies directly, leading to
\be
	\Omega(n D0) = -2 \qquad n\in \IZ_{>0}\,.
\ee
At $\vartheta=\pi$ we obtain the corresponding statement for $n\in\IZ_{<0}$.

Taking $T\to 0$ or $T\to\infty$ the curve degenerates as illustrated in subsection \ref{sec:C3modZ2-factorization}.
In this limit, the two components of the saddle of Figure \ref{fig:C3modZ2-T-6-i-theta-0} simply survive and end up living either in the component (\ref{eq:mirr-C3modZ2-factorized-1}) or in (\ref{eq:mirr-C3modZ2-factorized-2}).
Following the decomposition of the mirror curve, one therefore ends up with two copies of $\IC^3$, each carrying its own tower of D0 branes, as described in \cite{Banerjee:2018syt}.
The fact that the curve degenerates into copies of the $\IC^3$ curve in quadratic and cubic framing is reflected even at finite $T$ by the topology of the saddles of D0 branes, as shown in Figure \ref{fig:C3modZ2-T-6-i-theta-0}.

\subsubsection*{D2 states}

\begin{figure}[h!]
\begin{center}
\includegraphics[width=0.5\textwidth]{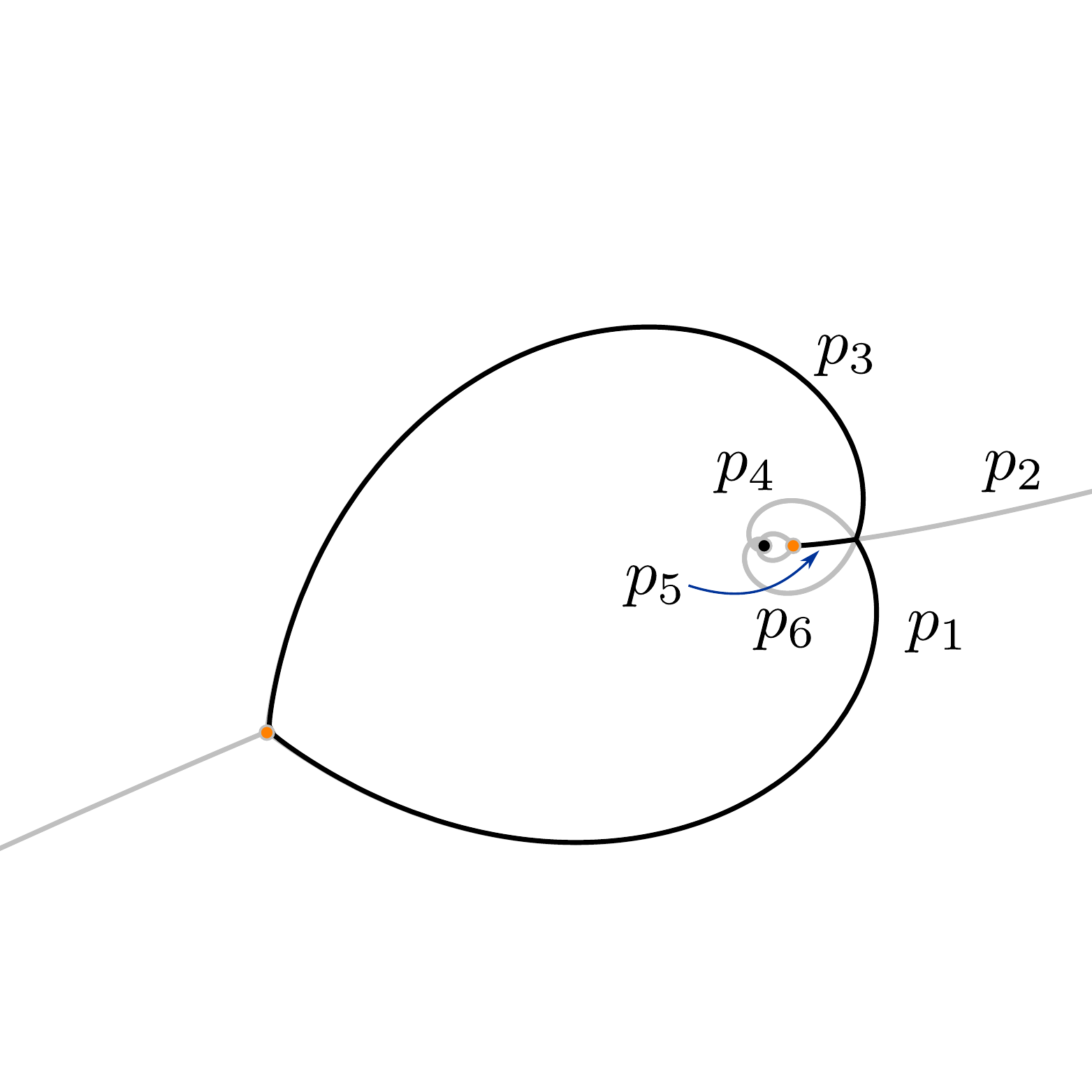}
\caption{
D2 saddle of $\CO(0)\oplus\CO(-2)\to\IP^1$, for $T=6+i$ at $\vartheta=\arg Z_{\text{D2}}$. 
}
\label{fig:C3modZ2-D2}
\end{center}
\end{figure}

At $\vartheta = \arg Z_{\text{D2}}$ we find the saddle of Figure \ref{fig:C3modZ2-D2},
this is the D2 brane. 
To compute the BPS index one has to go through a brief analysis for this saddle, since its topological type is novel by the present literature on spectral networks. 

There are three two-way walls, meeting at a junction with three additional one-way walls. 
Choosing the American resolution by taking $\vartheta = \arg Z_{\text{D2}} - \epsilon$ for small $\epsilon>0$, each two-way wall is decomposed into two one-way walls running in opposite directions.
Recall that each one-way wall carries certain combinatorial ``soliton'' data.
To uniformize the treatment, one can also pretend that all six walls meeting at the junction are are of two-way type, and assign trivial soliton content to three of the incoming one-way walls.

To set notation, we shall label these six 2-way walls by $p_i$ with $i=1\dots 6$. 
The generating functions encoding soliton data of the in-going one-way walls will be denoted by $\nu_i$, while those of out-going one-way walls by $\tau_i$.
In the following, it will be implicit that solitons of a wall of type $(ij,n)$ are relative homology classes starting at sheet $(i,N)$ and ending at sheet $(j,N+n)$ in keeping with conventions from \cite{Banerjee:2018syt}. 
In this particular case, the basepoints are taken to be in the fibre of $\tSigma\to \IC^*_x$ above the junction itself.\footnote{In this way, all paths can be naturally concatenated.}

The six-way junction rules in American resolution are
\be\label{eq:2-way-junction-AM}
\begin{split}
& \tau_1 = \nu_4 + \tau_6 \nu_5\,,
\quad 
\tau_2 = \nu_5 + \nu_6\tau_1\,, 
\quad 
\tau_3 = \nu_6 + \tau_2\nu_1\,,
\\ 
& \tau_4 = \nu_1 + \nu_2\tau_3\,, 
\quad 
\tau_5 = \nu_2 + \tau_4\nu_3\,, 
\quad 
\tau_6 = \nu_3 + \nu_4\tau_5
\end{split}
\ee
The fact that $p_2,p_4,p_6$ have no incoming soliton data means that we impose
\be
	\nu_2 = \nu_4 = \nu_6 = 0\,.
\ee
The other incoming data is sourced by branch points. 
For $p_5$ there is simply the sum of `simpletons' supported on the branch point closer to $x=0$ (see \cite[Section 3.3]{Banerjee:2018syt}) which we denote by $b_N$.\footnote{By `simpleton' we mean a simple path obtained by directly lifting the wall in question, running from sheet $(i,N)$ to $(j,N+\delta)$ for a suitable fixed shift $\delta$, determined by the choice of trivialization.} 
Likewise for $p_1$ one has only a sum of simpletons supported on the branch point farther from $x=0$, which we denote by $a_{1,N}$. 
For incoming soliton data on $p_3$ one has again simpletons sourced by the outer branch point, which we denote by $a_{2,N}$, but there are additional contributions due to the global topology of the saddle.
The point is that both $p_3$ and $p_1$ are attached to the bottom-left branch point, and $p_1$ is attached to the slot clock-wise from $p_3$.
The contribution of $p_1$ to soliton data of $p_3$ from the shared branch point have a universal form (see \cite[Appendix A]{Gaiotto:2012rg}).
Essentially what happens is the following: since we have chosen the american resolution, the out-going solitons in $\tau_1$ will come out of the junction, extend along $p_1$, and once they get to the branch point they ``veer-off'' onto $p_3$, ending up in a contribution to $\nu_3$. Overall we package this information as follows 
\be 
\begin{split}
	& \nu_1 = \sum_N X_{a_{1,N}}\,,
	\quad 
	\nu_3 = \sum_N X_{a_{2,N}} +  {\hat \tau}_1\,,
	\quad 
	\nu_5 = \sum_N X_{b_N}\,.
\end{split}
\ee
here ${\hat \tau}_1$ denotes the generating function of solitons counted by $\tau_1$, after transport along $p_1$ and then along $p_3$, as described above. 
Since this transport eventually takes a soliton back to the junction, and since one of the endpoints of solitons in $\tau_1$ must lie on the same sheet $(i,N)$ as one of the endpoints of solitons in $\nu_3$, we may represent the effect of transport by left- (or right-) concatenation of solitons in $\tau_1$ with a path at the \emph{other} endpoint.
A direct analysis quickly leads to identifying this with concatenation by $\alpha_N \equiv a_{1,N}\circ a_{2,N}$, therefore we set
\be
	{\hat \tau}_1 = \(\sum_{N}X_{\alpha_N}\) \, {\tau}_1\,.
\ee
The path algebra sets to zero products of $X$-variables whenever their endpoints don't match, so exactly one $\alpha_N$ from the sum acts as transport on each soliton in $\tau_1$.

At this point we have all the necessary equations, and we just need to solve them to determine $\tau_{i}$ and eventually $Q_i = 1+\nu_i\tau_i$. 
It will be useful to note that $a_{1,N}\circ a_{2,N} \circ b_N$ is a closed cycle, that we may call $\gamma_N$ upon forgetting the basepoint.
Let us begin from $p_1$:
\be
\begin{split}
Q_1 & = 1+ \nu_1\tau_1 
\\
& = 1+ \nu_1  \(  \sum_N X_{a_{2,N}} + \(\sum_{N'} X_{\alpha_{N'}}\)\tau_1\)   \sum_{N''} X_{b_{N''}}
\\ & 
= 1+ \sum_N X_{\gamma_N}+ (Q_1-1) \sum_N X_{\gamma_N}  
\\ & 
= \(1-\sum_N X_{\gamma_N}\)^{-1}
= \one\cdot  \(1- X_{\gamma}\)^{-1}\,,
\end{split}
\ee
where $\gamma$ is the homology class on $\tSigma$ identified with $\gamma_N$ after the quotient by $\ker Z$ (see \cite{Banerjee:2018syt}). 
Similarly for $p_3$
\be
\begin{split}
& Q_3 = 1+ \tau_3\nu_3 
\\ & 
= 1+ (\tau_2\nu_1) \(  \sum_N X_{a_{2,N}} + \(\sum_{N'} X_{\alpha_{N'}}\)\tau_1\)  
\\ & 
= 1+ \sum_{N} X_{\gamma_{N}} + \(\sum_{N} X_{\gamma_{N}} \)  (Q_1-1) 
\\ & 
= \one\cdot  \(1- X_{\gamma}\)^{-1}\,,
\end{split}
\ee
and for $p_5$
\be
\begin{split}
& Q_5  = 1+ \nu_5\tau_5 
\\ & 
= 1 + \(\sum_N X_{b_N}\) (\tau_4\nu_3) 
\\ & 
= 1+ \(\sum_N X_{b_N}\) \nu_1 \(\sum_{N'} X_{a_{2,N'}} + \( \sum_{N''}X_{\alpha_{N''}} \){\tau}_1\) 
\\ & 
= 1+ \(\sum_N X_{\gamma_N}\)  + \(\sum_{N'} X_{\gamma_{N'}}\) \(Q_1 - 1\)
\\ & 
= \one\cdot  \(1- X_{\gamma}\)^{-1}\,.
\end{split}
\ee
Then the lift of $\gamma_N$ is given by $\pi_N^{-1} (-p_1 \cup -p_3\cup -p_5)$ leading to 
\be
	\Omega(n \text{D2}) = \left\{
	\begin{array}{lr}
		-1 \qquad & n=1 \\
		0 \qquad & n>1 
	\end{array}
	\right. \,.
\ee
At $\vartheta=\arg\(- Z_{\text{D2}}\)$ the network has exactly the same topology, and we obtain the corresponding statement for $\overline{\text{D2}}$.

Taking $Q\to0$ the curve degenerates as illustrated above.
In this limit  the saddle of Figure \ref{fig:C3modZ2-D2} becomes infinitely long and gets lost, since the D2 cycle encircles both $\mathfrak{p}_1$ and $\mathfrak{p}_1$ (recall Figure \ref{fig:C3modZ2-cycles}), and these punctures end up in separate connected components  (\ref{eq:mirr-C3modZ2-factorized-1}) and (\ref{eq:mirr-C3modZ2-factorized-2}).
Following the decomposition of the mirror curve the D2 brane disappears from the spectrum, as should be expected from the match with $\IC^3$.

\subsubsection*{D2-D0 states}

\begin{figure}[h!]
\begin{center}
\includegraphics[width=0.75\textwidth]{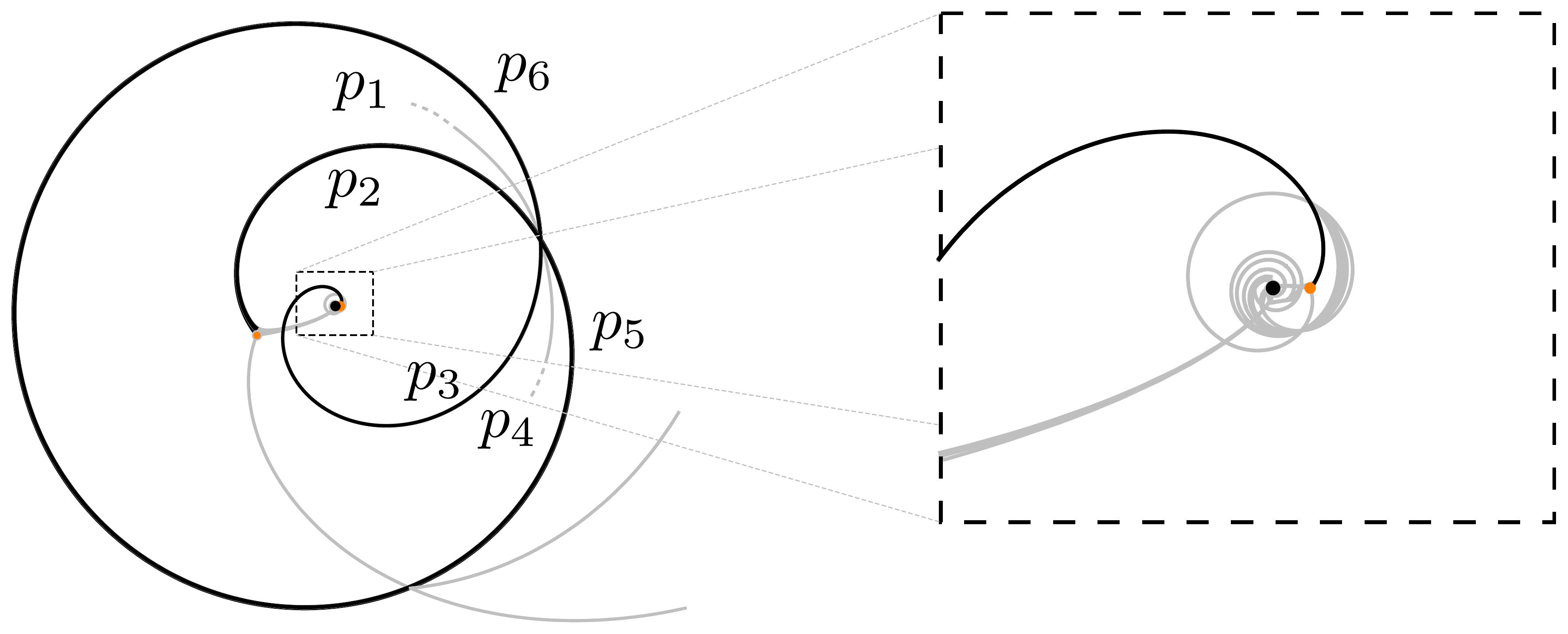}
\caption{D2-D0 saddle of $\CO(0)\oplus\CO(-2)\to\IP^1$ at $T=6+i$. }
\label{fig:C3modZ2-D2-D0}
\end{center}
\end{figure}

At $\vartheta=\arg\(Z_{\text{D0}}+Z_{\text{D2}}\)$ we find the saddle shown in Figure \ref{fig:C3modZ2-D2-D0}. The topology of this saddle is also new, so we will develop a fairly thorough analysis.

The saddle is attached to both branch points, and features a six-way junction with double-walls denoted $p_i$ for $i=1,\dots, 6$.
Walls $p_1$ and $p_4$ are actually one-way walls, hence their incoming data must be trivial
\be
	\nu_1 = 0, \qquad \nu_4 = 0\,.
\ee
For the other incoming soliton generating functions, we have
\be
\nu _2 = \sum_{N} X_{a_N} \,,
\quad
\nu_3 = \sum_N X_{b_N}\,, 
\quad 
\nu_5 = \(\sum_N X_{\delta_N}\) \tau_6\,, 
\quad 
\nu_6 = \(\sum_N X_{\delta_N}\) \tau_5\,.
\ee
Here $a_N$ are simpletons sourced by the outer branch point along $p_2$ and $b_N$ are simpletons sourced by the inner branch point along $p_3$.
For $\nu_5, \nu_6$ we simply took into account the global periodicity of the saddle, and introduced placeholders $X_{\delta_N}$ to keep track of transport of solitons along the lift of $p_5\simeq p_6$. (Similarly to the role of $X_{\alpha_N}$ from the analysis of D2).
The concatenation $a_N\circ \delta_N\circ b_N$ gives a closed cycle $\gamma_N \in  H_1(\tSigma,\IZ)$ whose equivalence class upon taking the quotient by $\ker Z$ we denote by $\gamma$.

\begin{figure}[h!]
\begin{center}
\includegraphics[width=0.45\textwidth]{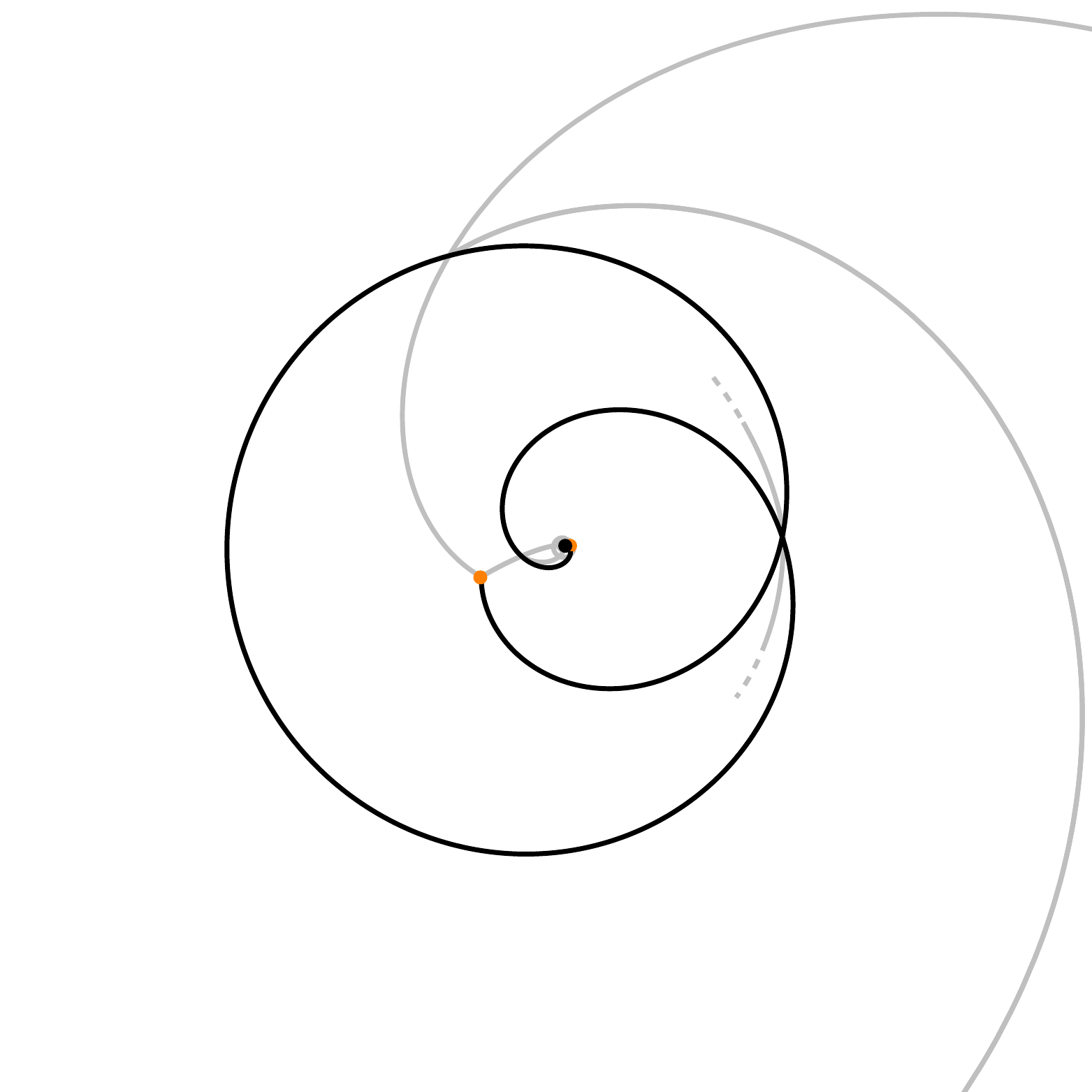}
\caption{$\overline{\text{D2}}$-D0 saddle of $\CO(0)\oplus\CO(-2)\to\IP^1$ at $T=6+i$.}
\label{fig:C3modZ2-D2bar-D0}
\end{center}
\end{figure}

In American resolution, the equations for the six-way junction are again (\ref{eq:2-way-junction-AM}). 
Their solutions encode the out-going soliton data. 
For example, for $p_2$ one obtains easily
\be
	\tau_2 = \nu_5+\nu_6\tau_1 = \(\sum_NX_{\delta_N}\) \,\tau_6 \,Q_5
	= \(\sum_NX_{\delta_N}\) \,\(\sum_{N'} X_{b_{N'}}\)  \,Q_5
\ee
where $Q_5 = 1+\tau_5\nu_5$.
In a similar way, one can show that
\be
\begin{split}
	\tau_3 & =  \(\sum_NX_{\delta_N}\) \,  \(\sum_{N'} X_{a_{N'}}\)  \, Q_3 \,,
	\qquad
	\tau_5 = \( \sum_{N} X_{a_N} \) \, Q_3 \,, 
	\qquad
	\tau_6 =   \sum_NX_{b_N} \,.
\end{split}
\ee
One immediately gets 
\be
\begin{split}
	Q_3 
	& = 1+\nu_3\tau_3 
	= 1 +\( \sum_N X_{b_N\circ \delta_N\circ a_N} \)Q_3
	\\
	&= 1+ \( \sum_N X_{\gamma_N}\) Q_3
	= \one\cdot \(1- X_{\gamma}\)^{-1}\,,
\end{split}
\ee
and likewise
\be
\begin{split}
	Q_5 
	& = 1+\tau_5\nu_5
	= 1 +\( \sum_N X_{a_N\circ \delta_N\circ b_N} \)Q_3
	= \one\cdot \(1- X_{\gamma}\)^{-1}\,,
\end{split}
\ee
\be
	Q_2 
	= 1+\nu_2\tau_2 
	= 1 +\( \sum_N X_{a_N\circ \delta_N\circ b_N} \)Q_5
	= \one\cdot \(1- X_{\gamma}\)^{-1}\,,
\ee
\be
	Q_6 
	= 1+\tau_6\nu_6 
	= 1 +\( \sum_N X_{b_N\circ \delta_N\circ a_N} \)Q_3
	= \one\cdot \(1- X_{\gamma}\)^{-1}\,.
\ee
Since $Q_2=Q_3=Q_4=Q_5 = (1-\sum_N X_{\gamma_N})^{-1}$ the saddle gets lifted to  $\pi_N^{-1} (-p_2 \, \cup\,- p_3\, \cup \, -p_5 \, \cup \,-  p_6 )$. Its BPS index is therefore
\be
	\Omega(n (\text{D2-D0})) = \left\{
	\begin{array}{lr}
		-1 \qquad & n=1 \\
		0 \qquad & n>1 
	\end{array}
	\right. \,.
\ee
A similar analysis applies to $\overline{\text{D2}}$-D0, whose saddle has a very similar topology shown in Figure \ref{fig:C3modZ2-D2bar-D0}.

Proceeding to $\vartheta = \arg (Z_{\text{D2}} + k Z_{D0})$ we find again saddles (of increasing complexity) for all $k>0$. 
The BPS index is always 
\be
	\Omega(n (\text{D2-$k$D0})) = \left\{
	\begin{array}{lr}
		-1 \qquad & n=1 \\
		0 \qquad & n>1 
	\end{array}
	\right. \,.
\ee
%

\subsubsection*{Full spectrum}

Our analysis stops here, although nothing prevents one from going further in the study of higher boundstates. It is natural to guess the following BPS states (plus their CPT conjugates)
\be\label{eq:C3modZ2-spectrum}
\begin{split}
	\Omega(n \text{D0}) & = -2 \\
	\Omega(\text{D2-$k$D0}) & = -1 \\
	\Omega(\overline{\text{D2}}\text{-$k$D0}) & = -1\\
\end{split}
\qquad
\begin{split}
	& n\geq 1\\
	& k\geq 0\\
	& k \geq 1
\end{split}
\ee

\subsection{Match with topological strings, framed wall-crossing and generalized DT invariants}

As for the conifold, we can compare our results to related computations of Gopakumar-Vafa and Donaldson-Thomas invariants.

\paragraph{Relation to Gopakumar-Vafa invariants}
The BPS spectrum is arranged into KK towers: D0 charge corresponds to KK momentum and different towers are labeled by D2 charge, being either $0$ or $\pm1$. 
Each KK tower corresponds to a 5d BPS particle, whose spectrum should be encoded by Gopakumar-Vafa invariants.

Our computation predicts a particle with zero D2 charge and BPS index $-2$, one particle with a unit of D2 charge and BPS index $-1$, and the antiparticle of the latter. These predictions match respectively with the expected number of D0 branes and with known results on genus-zero GV invariants (see e.g. \cite{Iqbal:2007ii}).
We therefore have a precise match with expectations based on other techniques for computing these invariants.

\paragraph{Relation to generalized Donaldson-Thomas invariants}

Let us also briefly comment on the relation between our BPS indices (\ref{eq:C3modZ2-spectrum}) and degree-zero generalized Donaldson-Thomas invariants.
As for the conifold, we will adopt (framed) wall-crossing arguments to obtain a prediction for the latter from the standard Donaldson-Thomas invariants.
Standard DT invariants are a special case of generalized DT invariants, with a single unit of D6 charge.
In the case of local Calabi-Yau threefolds, these must be \emph{framed} BPS states, since the D6 would have infinite volume and therefore would give rise to an infinitely heavy, non-dynamical, defect in the 5d theory engineered via M theory.
To begin with, let us make the following empirical observation. The generating series of DT invariants in the non-commutative chamber 
\be \label{C3-Z2-nonc}
	\CZ = M(q)^{-2} \prod_{n\geq 0}(1-q^n T )^{-n}(1-q^n T^{-1} )^{-n}
\ee
where $M(q)$ is McMahon's function. 
Then we observe that the \emph{exponents} of this factorization coincide  precisely with the BPS indices we computed above in  (\ref{eq:C3modZ2-spectrum}).

To clarify this observation, let us restrict to BPS states with non-negative D2 charge.
This makes contact with  Gromov-Witten invariants, their well-known correspondence with Donaldson-Thomas \cite{2008arXiv0809.3976M}.\footnote{Incidentally, this restriction also coincides with the ``DT chamber'', although we stress that we are not really moving across any chambers in the moduli space of framed BPS states.}
Expanding this as a $q$-series gives
\be\label{eq:C3modZ2-DT}
\begin{split}
	\CZ |_{\beta\geq 0}
	& = M(q)^{-2} \prod_{n\geq 0}(1-q^n T )^{-n} \\
	& = \sum_{n,k\geq 0} DT_{D6,kD2,nD0} T^k q^n 
	\\
	& = 
	1+2q+7q^2+18q^3+47q^4+110q^5+\dots
	\\
	& 
	+ T\(q + 4 q^2 +14 q^3 + 42 q^4 + 117 q^5 +\dots\)
	+ \CO(T^2)\,.
\end{split}
\ee
Now let us view these as \emph{boundstates} of a single, infinitely heavy, D6 brane with the BPS states carrying D2-D0 charges in (\ref{eq:C3modZ2-spectrum}).
Given the spectrum of D2-D0 states, and the DSZ pairing between their charges and that of the D6 completely specifies the spectrum of boundstates by Kontsevich-Soibelman's wall-crossing formula.\footnote{In fact, this is an instance of the semi-primitive wall-crossing formula of Denef and Moore \cite{Denef:2007vg}.} Matching with (\ref{eq:C3modZ2-DT}) would therefore provide a nontrivial check of our result (\ref{eq:C3modZ2-spectrum}) for degree-zero generalized DT invariants.

To perform this check, one needs to note that the only non-trivial contribution to the DSZ pairing is between the D0 and the D6 charges, as in (\ref{eq:D0-D6-pairing}).
Using this we derive again a wall-crossing identity like (\ref{eq:KSWCF-DT})
where  $\fOmega$ denote framed BPS degeneracies of D2-D0 BPS states bound to $\ell$ D6 branes.
The latter are predicted by the former through the wall-crossing identity, the first few read:
\be\label{eq:C3modZ2-gen-DT}
\begin{array}{llll}
	\fOmega_{1,0,1} = -2\,;
	\qquad
	&
	\fOmega_{1,1,1} = -1\,;
	\qquad
	&
	\fOmega_{1,0,2} = 7\,;
	\qquad
	&
	\fOmega_{2,0,2} = -2\,;
	\\
	\fOmega_{1,1,2} = 4\,;
	&
	\fOmega_{1,0,3} = -18\,;
	&
	\fOmega_{2,1,2} = -1\,;
	&
	\fOmega_{2,0,3} = -18\,;
	\\
	\fOmega_{1,2,2} = 1\,;
	&
	\fOmega_{1,1,3} = -14\,;
	&
	\fOmega_{1,0,4} = 47\,;
	&
	\fOmega_{3,0,3} = -2\,;
	\\
	\fOmega_{2,1,3} = -14\,;
	&
	\fOmega_{2,0,4} = -78\,;
	&
	\fOmega_{1,2,3} = -4\,;
	&
	\fOmega_{1,1,4} = 42\,;
	\\
	\fOmega_{1,0,5} = -110\,;
	&
	\dots
\end{array}
\ee
where $\fOmega_{\ell,k,n}\equiv \fOmega(\text{$\ell$D6-$k$D2-$n$D}0)$.
As expected, these match with (\ref{eq:C3modZ2-DT}) 
\be
	DT_{D6,kD2,nD0} = (-1)^{n} \fOmega_{1,k,n}\,,
\ee
at least for $k=0,1$.
Once again, the wall-crossing formula predicts also higher-degree generalized Donaldson-Thomas invariants, those with $\ell>1$ in (\ref{eq:C3modZ2-gen-DT}).
They are not captured by our framework, nor by the GW/DT correspondence and to compute them with our approach, one would first need to understand how to deal with framed BPS states.

\subsection{Exponential BPS graph and BPS quiver}\label{sec:C3modZ2-graph}

\begin{figure}[h!]
\begin{center}
\includegraphics[width=0.45\textwidth]{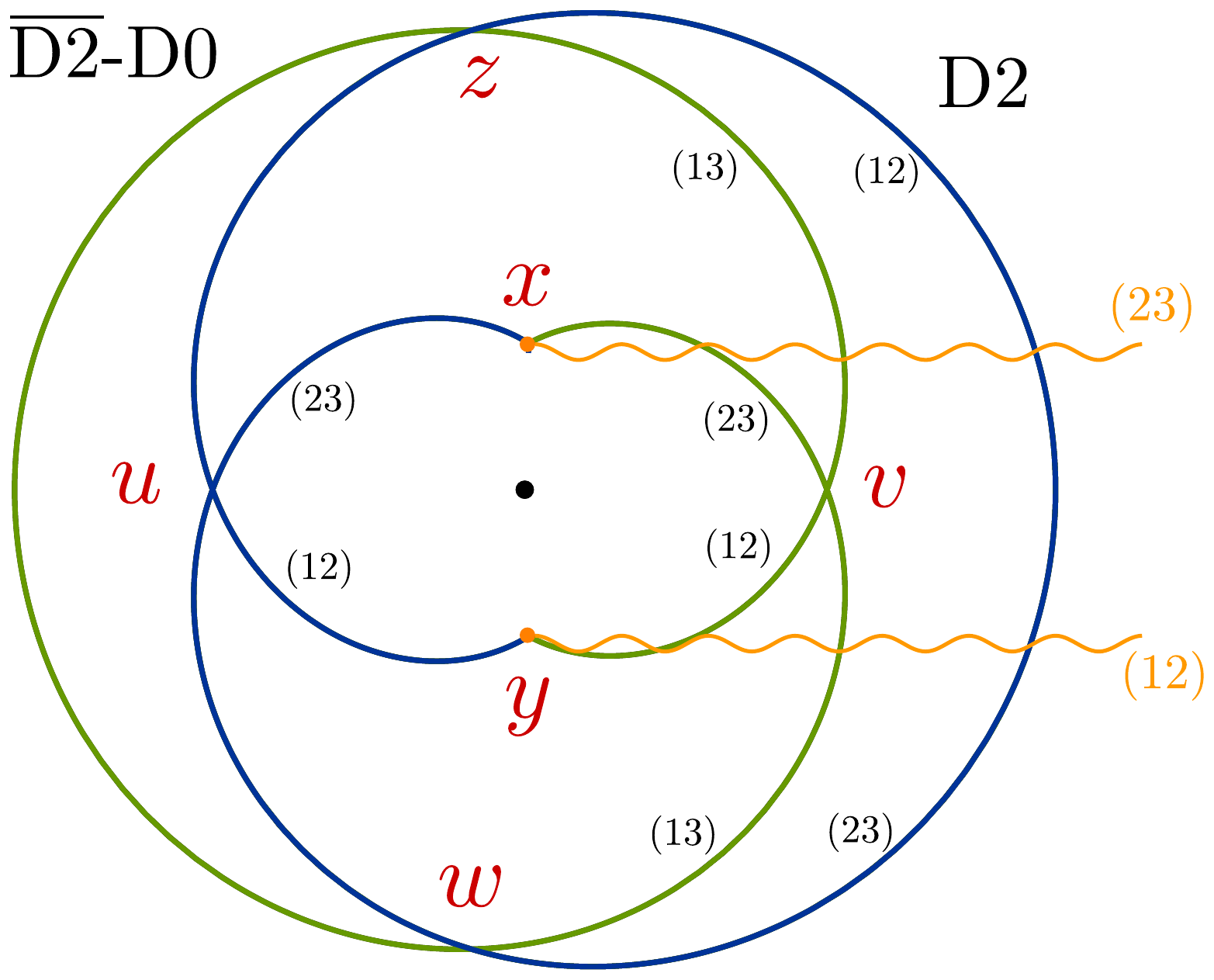}
\caption{Exponential BPS graph of $\CO(0)\oplus\CO(-2)\to\IP^1$ for $T\to-1$, with a choice of trivialization made explicit.}
\label{fig:C3modZ2-BPS-graph}
\end{center}
\end{figure}

We would now like to study the BPS graph of $\CO(0)\oplus\CO(-2)\to\IP^1$. We follow closely the procedure explained in Section \ref{sec:conifold-graph}, and here provide only the key points.
The exponential BPS graph arises when we fix moduli in such a way that 
\be
	\arg\, Z_{D2} = \arg Z_{D0}\,.
\ee
Given our explicit computation of central charges, this happens on the unit circle $|T|=1$ in K\"ahler moduli space. 
We fix $T$ very close to $-1$, then the elementary webs that make up the BPS graph are just the two-way streets connecting the branch points, see Figure \ref{fig:C3modZ2-BPS-graph}. In  a choice of trivialization for the charge lattice over the $Q$-plane, one of the two edges of the BPS graph lifts to the  D2 BPS cycle, the other to the $\overline{\text{D2}}$-D0 cycle.

\begin{figure}[h!]
\begin{center}
\includegraphics[width=0.35\textwidth]{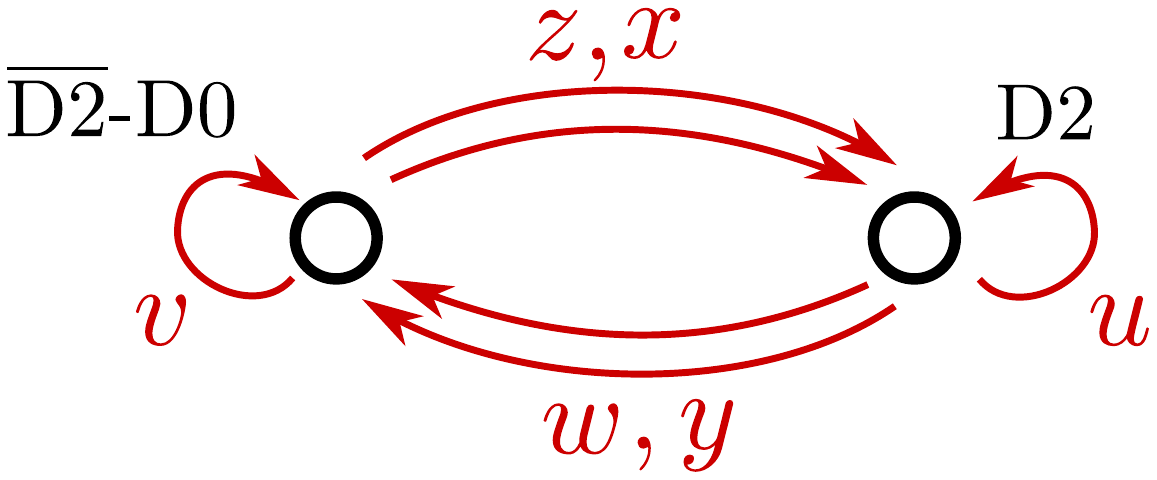}
\caption{BPS quiver of $\CO(0)\oplus\CO(-2)\to\IP^1$.}
\label{fig:C3modZ2-quiver}
\end{center}
\end{figure}

\begin{figure}[h!]
\begin{center}
\includegraphics[width=0.85\textwidth]{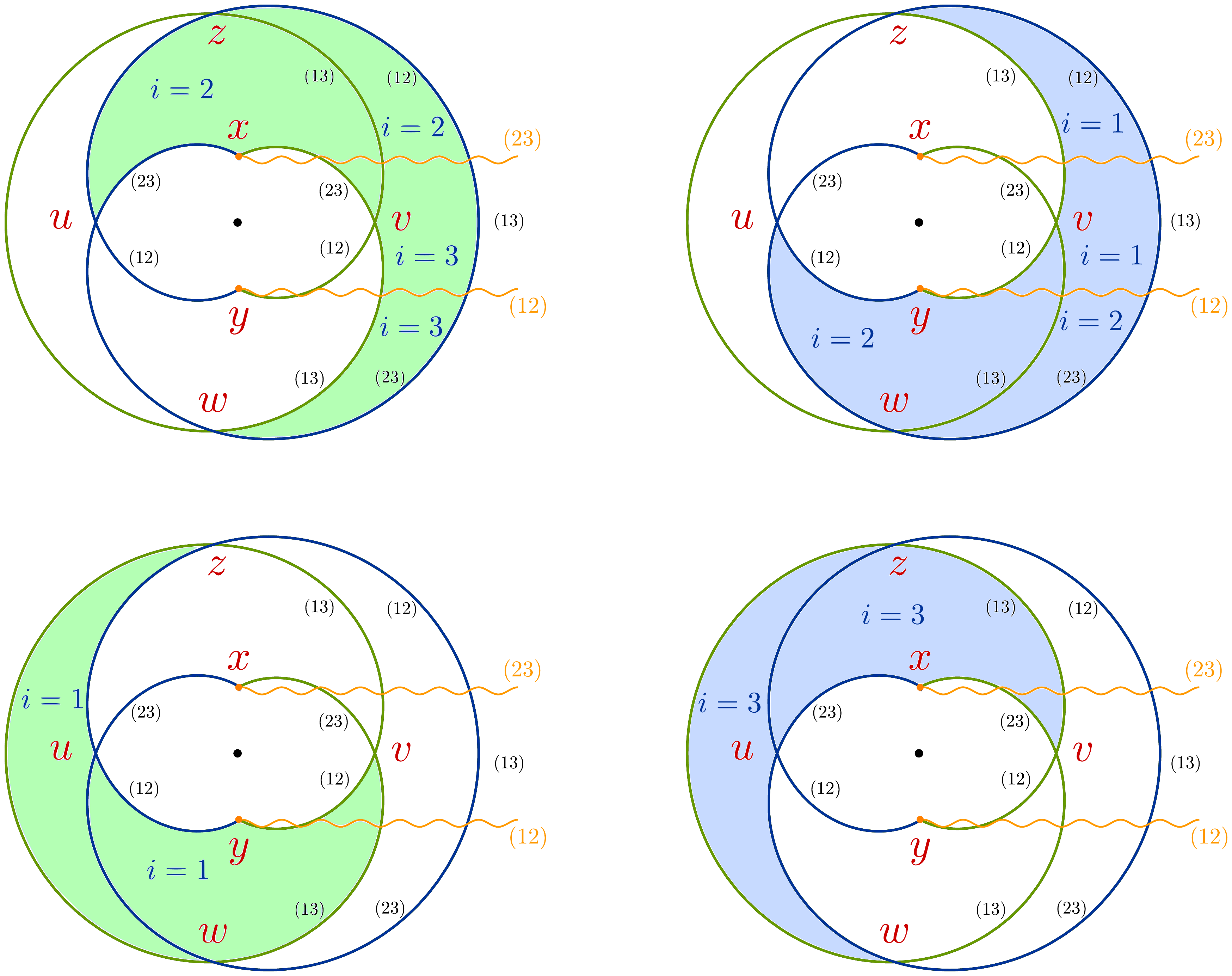}
\caption{Disk instantons generating the quiver superpotential for  $\CO(0)\oplus\CO(-2)\to\IP^1$.}
\label{fig:C3modZ2-superpotential}
\end{center}
\end{figure}

For the superpotential we find four contributions, shown in Figure \ref{fig:C3modZ2-superpotential}, giving
\be
	W =\Tr\( {xwu} - {yzu} +  z v y - wvx \)\,,
\ee
with the first two contributions coming from the diagrams in the top row, and the last two from the bottom row.
This matches the known quiver with potential \cite{Aspinwall:2004bs}. 

\cleardoublepage

\appendix

\section{The partition function for the DT invariants}
In this appendix, we derive the DT partition starting from the wall-crossing formula \eqref{eq:KSWCF-DT}.  The Kontsevich-Soibelman operators $\CK_\gamma$. are constructed from the generators of the 
Lie algebra of infinitesimal symplectomorphisms of the complex torus \cite{Kontsevich:2008fj}
\be
\CK_\gamma = \exp\(\sum_{n\ge 1} \frac{e_{n\gamma}}{n^2}\)
\ee
where 
\be
[e_\gamma, e_{\gamma'}] = (-1)^{\langle \gamma,\gamma'\rangle} \langle \gamma,\gamma'\rangle \, e_{\gamma+\gamma'}. 
\ee
We derive the partition function in the case when $\ell=1$ in the formula \eqref{eq:KSWCF-DT}. It then reads 
\be
\begin{split}
	& \(  \prod_{n \geq 1} \CK_{n \gamma_{\text{D0}}}^{\Omega(n \text{D0})} \)
	\(  \prod^{\curvearrowleft}_{n \geq 0} \CK_{\gamma_{\text{D2}} + n \gamma_{\text{D0}}}^{\Omega(\text{D2-$n$D0})} \)
	\(  \prod^{\curvearrowleft}_{n \geq 0} \CK_{\gamma_{\overline{\text{D2}}} + n \gamma_{\text{D0}}}^{\Omega({\overline{\text{D2}}}- n \text{D0})} \)
	\CK_{\gamma_{D6}}\\
	& \qquad\qquad
	=
	\CK_{\gamma_{D6}}
	\( \prod_{k\geq 0}\prod_{n \geq 1} \CK_{\gamma_{D6}+k\gamma_{D2}+n \gamma_{\text{D0}}}^{\fOmega(\text{D6-$k$D2-$n$D}0)} \)
	\(  \prod_{k\geq 0}\prod_{n \geq 1} \CK_{\gamma_{D6}+k\gamma_{\overline{D2}}+n \gamma_{\text{D0}}}^{\fOmega(\text{D6}-k\overline{\text{D2}}- n\text{D0})} \)  \\ 
	& \qquad \qquad
	\(  \prod^{\curvearrowright}_{n \geq 0} \CK_{\gamma_{\text{D2}} + n \gamma_{\text{D0}}}^{\Omega(\text{D2-$n$D0})} \)
	\(  \prod^{\curvearrowright}_{n \geq 0} \CK_{\gamma_{\overline{\text{D2}}} + n \gamma_{\text{D0}}}^{\Omega({\overline{\text{D2}}}- n \text{D0})} \)
	\(  \prod_{n \geq 1} \CK_{n \gamma_{\text{D0}}}^{\Omega(n \text{D0})} \)\,.
\end{split}
\ee
Let us define a commutative product $\circ$ such that $e_{\gamma} \circ e_{\gamma'} = e_{\gamma+\gamma'}$.
In its terms, one finds ($\hat{e}$ is the transformed basis) 
\be
\begin{split}
& \sum_{n\ge0, k\ge 1} \fOmega(\text{D6-$k$D2-$n$D}0) \hat{e}_{\gamma_{\mathrm{D6}}+ k\gamma_{\mathrm{D2}}+ n\gamma_{\mathrm{D0}}}
\\ & \qquad 
= \circ \prod_{n\ge 1} \bigg(\bigg\{1+(-1)^{\langle \gamma_{\mathrm{D6}}, \gamma_{\mathrm{D2}}+n\gamma_{\mathrm{D0}}\rangle} e_{\gamma_{\mathrm{D2}}+n\gamma_{\mathrm{D0}}}\bigg\}^{
\circ - \Omega(\mathrm{D2}-n\mathrm{D0}) \langle \gamma_{\mathrm{D6}}, \gamma_{\mathrm{D2}}+n\gamma_{\mathrm{D0}}\rangle}
\\ & \qquad \qquad\qquad 
\circ \prod_{m\ge 1} \bigg\{1+(-1)^{\langle \gamma_{\mathrm{D6}}, m\gamma_{\mathrm{D0}}\rangle} e_{m\gamma_{\mathrm{D0}}}\bigg\}^{\circ - \Omega(m\mathrm{D0})\langle \gamma_{\mathrm{D6}}, m\gamma_{\mathrm{D0}}\rangle }
\end{split}
\ee
and similarly 
\be
\begin{split}
& \sum_{n\ge0, k\ge 1} \fOmega(\mathrm{D6}-k\overline{\mathrm{D2}}-n\mathrm{D0}) \hat{e}_{\gamma_{\mathrm{D6}}+ k\gamma_{\overline{\mathrm{D2}}}+ n\gamma_{\mathrm{D0}}}
\\ & \qquad
= \circ \prod_{n\ge 1} \bigg(\bigg\{1+(-1)^{\langle \gamma_{\mathrm{D6}}, \gamma_{\mathrm{D2}}+n\gamma_{\mathrm{D0}}\rangle} e_{\gamma_{\mathrm{D2}}+n\gamma_{\mathrm{D0}}}\bigg\}^{
\circ - \Omega(\mathrm{D2}-n\mathrm{D0}) \langle \gamma_{\mathrm{D6}}, \gamma_{\mathrm{D2}}+n\gamma_{\mathrm{D0}}\rangle}
\\ & \qquad\qquad \qquad 
\circ \prod_{m\ge 1} \bigg\{1+(-1)^{\langle \gamma_{\mathrm{D6}}, m\gamma_{\mathrm{D0}}\rangle} e_{m\gamma_{\mathrm{D0}}}\bigg\}^{\circ - \Omega(m\mathrm{D0})\langle \gamma_{\mathrm{D6}}, m\gamma_{\mathrm{D0}}\rangle }.
\end{split}
\ee
The BPS partition function is defined by 
\be
\sum_{k\ge0,n\ge 1}\fOmega(\text{D6-$k$D2-$n$D}0) q^n Q^k
\ee
where $-q$ and $-Q$ are fugacities corresponding to the D0 and D2 charges respectively. 

Plugging in the values for the BPS indices for conifold in the DT chamber, one gets 
\be
\sum_{k\ge0,n\ge 1}\fOmega(\text{D6-$k$D2-$n$D}0) q^n Q^k = M(q)^{-2} \prod_{n\ge 1} (1-q^n Q)^n . 
\ee
Similarly for $\CO(-2) \oplus \CO(0) \rightarrow \IP^1$ one has 
\be
\sum_{k\ge0,n\ge 1}\fOmega(\text{D6-$k$D2-$n$D}0) q^n T^k = M(q)^{-2} \prod_{n\ge 1} (1-q^n T)^{-n} . 
\ee
Changing the $B$-field, one can go to the non-commutative chamber \cite{Szendroi:2007nu} and one gets for the conifold and $\CO(-2) \oplus \CO(0) \rightarrow \IP^1$ respectively 
\be
 M(q)^{-2} \prod_{n\ge 1} (1-q^n Q)^n (1-q^n Q^{-1})^n . 
\ee
and  
\be
 M(q)^{-2} \prod_{n\ge 1} (1-q^n T)^{-n} (1-q^n T^{-1})^{-n} . 
\ee
as the non-commutative partition functions.

\section{Admissible framings for exponential networks}\label{app:framing}
Exponential networks are defined relative to a presentation of the  algebraic curve  $\Sigma\subset\IC^*_x \times \IC^*_y $ as a $K:1$ ramified covering over $\IC^*_x$. 
Different choices of framing give different networks, however results about 5d BPS states are expected to be independent of framing choice.
Nevertheless, not all choices of framing are on the same footing, and some of them should be excluded. In this appendix we clarify why this is the case, and what the criteria are.

\subsection{Euler characteristics}

The first criterion is based on Euler characteristics, reviewing an argument by \cite{Bouchard:2011ya}. 
By genericity, we assume all ramification points of $\pi:\Sigma\to \IC^*_x$ are simple (punctures however can be sources of higher-degree ramification). 
If $X$ is the toric threefold whose mirror curve is $\Sigma$, then the number of ramification 
points is given by $n= \chi(X)$, which is also the number of torus fixed points of $X$. 
In particular note that this doesn't depend on the choice of framing, i.e. on the degree $d$ of the $\pi$-map.

This fact follows from the Riemann-Hurwitz formula  
$\chi(\Sigma) = \chi(\IC^*) d - n$, where $n$ is the number of ramification points.
Since  $\chi(\IC^*) = 0$, and since $\chi(\Sigma) = -\chi(X)$ by Mirror Symmetry,\footnote{This can be also observed by matching the number of trinions in a pants decomposition of $\Sigma$ to the number of vertices in the toric diagram of $X$.} it follows that $n$ is independent of $d$.
Now comes the point: certain choices of framing produce mirror curves that appear to violate this simple requirement, therefore such choices of framing should be avoided for 5d BPS counting. We now outline how this works out in the three examples considered in this and our previous paper \cite{Banerjee:2018syt}.

\begin{itemize}

\item 
For $\IC^3$ with framing $f=2$ the mirror curve is $xy^2 + y + 1 = 0$. The discriminant is $\Delta = 1-4x $, therefore there is one branch point at $x=1/4$. 
For cubic framing the mirror curve becomes $xy^3  + y + 1 = 0$. The discriminant is now $\Delta = - x^2 (4x + 27)$, which again has only one solution 
 $x=-27/4$. 
We observe the same property for higher framings too. 
However, in the linear framings $f=0,1$ the curve becomes $x+y+1 =0$ or $xy+y+1=0$, and there is no branch point at all. Linear framings are therefore not allowed according to the criteria set out above (besides, there is no exponential network without branch points).

\item
For the resolved conifold the curve is  $Qxy^2 + xy + y + 1 =0$, with discriminant $\Delta = (1+x)^2 -4Qx$, which has two solutions. 
This is clearly in line with our expectations. The mirror curve of the conifold is a four punctured sphere which has Euler characteristic $-2$, and the resolved conifold itself has Euler characteristic $2$. 
By dint of a very similar explanation as above, the linear framings such as $Qxy + y + x + 1 =0$ are not allowed. 
By contrast, the discriminant for the cubic framing for which the mirror curve is $Qxy^3 + xy^ 2 + y + 1=0$ is given by 
$\Delta = x(x-4Q -4x^2 -27Q^2 x+ 18 Qx)$ which again has two solutions in $\IC^*_x$. For higher framings it is also the case, and all such framings are allowed. 

\item
For $\CO(-2) \oplus \CO(0) \rightarrow \IP^1$ the mirror curve is $1+x+y + Q y^2 = 0$. This has discriminant $\Delta = 1-4Q(1+x)$ which has only one solution, hence  this covering has  only one branch point. 
However, this is also a four punctured sphere. Hence this choice of framing is no good (note that the framing is not linear!). 
In our example \eqref{mirr-C3modZ2}, we perform the change of framing $x \rightarrow xy^{-1}$ to the above curve
to obtain a cubic equation in $y$. As is computed, it indeed has two branch points. 

Let us also note that the change of framing $x\rightarrow xy$ also works, as it gives rise to two branch points again. To understand why this not such a suitable choice, we have to understand the punctures of the mirror curve, as we explain next.

\end{itemize}

\subsection{Positions of punctures}

We now consider  an alternative argument, based on positioning of punctures under the projection map $\pi$.

Consider the first example again of $\IC^3$. If we are to choose the linear framing to
have the curve as $x+y+1 = 0$, the positions of the punctures will be at $(x=0, y =-1), (x=-1, y =0)$ and $(x=\infty, y = \infty)$. 
Let us analyze the differential $\lambda = \log y d\log x$ near the second puncture $(x=-1, y =0)$. 
One can parametrize the curve as $x = -1-t, y = t$. Then for small $t$, the differential becomes $\lambda = \log t d\log(-1-t) \, \sim \, \log t dt$. Near this puncture, the integral curve is is given by$\int^w \log t dt = w\log w - w = e^{i\vartheta}$. 
As $w$ approaches $0$, one can approximate this as $w\log w \sim e^{i\vartheta}$. This is a transcendental equation which is very complicated to work with from the point of view of the exponential networks rendering such punctures
which appear for finite $x$ are not amenable to our analysis. 
But there is an easy systematic way to circumvent it. 
For $\IC^3$, we have seen that the change of framing does the job. Then all the punctures lie either above $x=0$ or $x=\infty$. 
The same is the situation with either of conifold and $\CO(-2) \oplus \CO(0) \rightarrow \IP^1$. For other curves (even higher genus ones that support compact D4-branes), we arrived at the same conclusion as well. Empirically, we have found that it fits in line very well 
with the statements we made before, concerning allowed choices of framings. 

Let us therefore propose an algorithm on how to circumvent such situation, where the punctures appear for finite $x$. If there is a constant term in the curve, we can perform the next step directly. However, imagine 
a mirror curve where all terms are either of $x$ or $y$-dependent or depend on both. There must be a term which is a monomial in $y$ in such a curve. Otherwise, one could divide by the lowest power of $x\in \IC^*$, 
to generate such a term. Then change the framing first $x\rightarrow xy^f$ choosing $f$ as the  degree of the aforementioned monomial. Then divide throughout by this monomial since $y\in \IC^*$ to generate a 
constant term in the curve. 

Then in this form of the curve, there can arise the ``pathological punctures''. One can remove them too, through framing transformation again. All that one needs to ensure is that the highest degree term ($\ge 2$) in $y$ is multiplied with 
some powers of $x$ (probably also some constant coefficient). This guarantees that the punctures can not be pathological. 
\footnote{Note that it is only a sufficient condition, but not necessary. Consider $\IC^3$ in framing $-1$. The mirror curve is of the form $y^2 + y + x = 0$ which has one branch point and three punctures 
$(x=0, y =0), (x= 0, y = -1)$ and $(x=\infty, y =\infty)$, none of which is pathological. However, our algorithm above does not create any additional complication. We change the framing $x\rightarrow xy$ and obtain 
$x+y+1 =0$. We notice as before that it has pathological puncture. Hence, we remove it through framing transformation again, say $x\rightarrow xy^2$. One could perform $x\rightarrow x/y$ which would have led 
to the curve before $y^2+y+x = 0$.  }

Lastly, we provide on more empirical criterion and illustrate with an example. The punctures which correspond to $y=0, \infty$ are of logarithmic nature.
\footnote{The differential behaves as $\lambda \sim  \pm d (\log x)^2$ in this case.}
 It is hard to analyze the decoupling limit if the presentation curve 
makes them both on top of either of $x=0,\infty$. In fact, the networks that represent D0-branes would be too complicated to make an accurate counting. Even though, there is nothing wrong with such a presentation, however, 
we can change the framing such that the two logarithmic punctures lie above $x=0$ and $x=\infty$ respectively. For example, whereas the mirror curve of $\CO(-2) \oplus \CO(0) \rightarrow \IP^1$ \eqref{mirr-C3modZ2} has one puncture each above 
$x=0$ and $x=\infty$, changing the framing by $x\rightarrow xy^2$, one would have obtained $Qy^2 + xy + y + 1 = 0$ which has two branch points at $x_{\pm} = -1 + 2\sqrt{Q}$ and has four punctures, two above $x=0$ and two above $x=\infty$. 
The two above $x=\infty$ correspond to $y\to 0,\infty$ and hence they are logarithmic. Due to the reason mentioned above, that is very hard to track the copies of D0's. Therefore, we were led to work with a curve cubic in $y$ 
for this case.

\bibliography{biblio}{}

\providecommand{\href}[2]{#2}\begingroup\raggedright\begin{thebibliography}{10}

\bibitem{2003math.....12059M}
D.~{Maulik}, N.~{Nekrasov}, A.~{Okounkov}, and R.~{Pandharipande}, {\it
  {Gromov-Witten theory and Donaldson-Thomas theory, I}},  {\em ArXiv
  Mathematics e-prints} (Dec., 2003)
  [\href{http://arxiv.org/abs/math/0312059}{{\tt math/0312059}}].

\bibitem{2004math......6092M}
D.~{Maulik}, N.~{Nekrasov}, A.~{Okounkov}, and R.~{Pandharipande}, {\it
  {Gromov-Witten theory and Donaldson-Thomas theory, II}},  {\em ArXiv
  Mathematics e-prints} (June, 2004)
  [\href{http://arxiv.org/abs/math/0406092}{{\tt math/0406092}}].

\bibitem{2008arXiv0809.3976M}
D.~{Maulik}, A.~{Oblomkov}, A.~{Okounkov}, and R.~{Pandharipande}, {\it
  {Gromov-Witten/Donaldson-Thomas correspondence for toric 3-folds}},  {\em
  ArXiv e-prints} (Sept., 2008) [\href{http://arxiv.org/abs/0809.3976}{{\tt
  arXiv:0809.3976}}].

\bibitem{Dijkgraaf:2006um}
R.~Dijkgraaf, C.~Vafa, and E.~Verlinde, {\it {M-theory and a topological string
  duality}},  \href{http://arxiv.org/abs/hep-th/0602087}{{\tt hep-th/0602087}}.

\bibitem{Kontsevich:2008fj}
M.~Kontsevich and Y.~Soibelman, {\it {Stability structures, motivic
  Donaldson-Thomas invariants and cluster transformations}},
  \href{http://arxiv.org/abs/0811.2435}{{\tt arXiv:0811.2435}}.

\bibitem{Joyce:2008pc}
D.~Joyce and Y.~Song, {\it {A Theory of generalized Donaldson-Thomas
  invariants}},  \href{http://arxiv.org/abs/0810.5645}{{\tt arXiv:0810.5645}}.

\bibitem{Gaiotto:2008cd}
D.~Gaiotto, G.~W. Moore, and A.~Neitzke, {\it {Four-dimensional wall-crossing
  via three-dimensional field theory}},  {\em Commun.Math.Phys.} {\bf 299}
  (2010) 163--224, [\href{http://arxiv.org/abs/0807.4723}{{\tt
  arXiv:0807.4723}}].

\bibitem{Gaiotto:2009hg}
D.~Gaiotto, G.~W. Moore, and A.~Neitzke, {\it Wall-crossing, {H}itchin systems,
  and the {WKB} approximation},  {\em Adv. Math.} {\bf 234} (2013) 239--403,
  [\href{http://arxiv.org/abs/0907.3987}{{\tt arXiv:0907.3987}}].

\bibitem{Denef:2007vg}
F.~Denef and G.~W. Moore, {\it {Split states, entropy enigmas, holes and
  halos}},  {\em JHEP} {\bf 11} (2011) 129,
  [\href{http://arxiv.org/abs/hep-th/0702146}{{\tt hep-th/0702146}}].

\bibitem{Alim:2011ae}
M.~Alim, S.~Cecotti, C.~Cordova, S.~Espahbodi, A.~Rastogi, and C.~Vafa, {\it
  {BPS Quivers and Spectra of Complete N=2 Quantum Field Theories}},  {\em
  Commun. Math. Phys.} {\bf 323} (2013) 1185--1227,
  [\href{http://arxiv.org/abs/1109.4941}{{\tt arXiv:1109.4941}}].

\bibitem{Alim:2011kw}
M.~Alim, S.~Cecotti, C.~Cordova, S.~Espahbodi, A.~Rastogi, and C.~Vafa, {\it
  {$\mathcal{N} = 2$ quantum field theories and their BPS quivers}},  {\em Adv.
  Theor. Math. Phys.} {\bf 18} (2014), no.~1 27--127,
  [\href{http://arxiv.org/abs/1112.3984}{{\tt arXiv:1112.3984}}].

\bibitem{Manschot:2010qz}
J.~Manschot, B.~Pioline, and A.~Sen, {\it {Wall Crossing from Boltzmann Black
  Hole Halos}},  {\em JHEP} {\bf 07} (2011) 059,
  [\href{http://arxiv.org/abs/1011.1258}{{\tt arXiv:1011.1258}}].

\bibitem{Manschot:2012rx}
J.~Manschot, B.~Pioline, and A.~Sen, {\it {From Black Holes to Quivers}},  {\em
  JHEP} {\bf 11} (2012) 023, [\href{http://arxiv.org/abs/1207.2230}{{\tt
  arXiv:1207.2230}}].

\bibitem{Alexandrov:2008gh}
S.~Alexandrov, B.~Pioline, F.~Saueressig, and S.~Vandoren, {\it {D-instantons
  and twistors}},  {\em JHEP} {\bf 03} (2009) 044,
  [\href{http://arxiv.org/abs/0812.4219}{{\tt arXiv:0812.4219}}].

\bibitem{Jafferis:2008uf}
D.~L. Jafferis and G.~W. Moore, {\it {Wall crossing in local Calabi Yau
  manifolds}},  \href{http://arxiv.org/abs/0810.4909}{{\tt arXiv:0810.4909}}.

\bibitem{Szendroi:2007nu}
B.~Szendroi, {\it {Non-commutative Donaldson-Thomas theory and the conifold}},
  {\em Geom. Topol.} {\bf 12} (2008) 1171--1202,
  [\href{http://arxiv.org/abs/0705.3419}{{\tt arXiv:0705.3419}}].

\bibitem{Morrison:2011rz}
A.~Morrison, S.~Mozgovoy, K.~Nagao, and B.~Szendroi, {\it {Motivic
  Donaldson-Thomas invariants of the conifold and the refined topological
  vertex}},  \href{http://arxiv.org/abs/1107.5017}{{\tt arXiv:1107.5017}}.

\bibitem{Banerjee:2018syt}
S.~Banerjee, P.~Longhi, and M.~Romo, {\it Exploring 5d {BPS} spectra with
  exponential networks},  {\em Annales Henri Poincar{\'e}} (Oct, 2019)
  [\href{http://arxiv.org/abs/1811.02875}{{\tt arXiv:1811.02875}}].

\bibitem{Alday:2009fs}
L.~F. Alday, D.~Gaiotto, S.~Gukov, Y.~Tachikawa, and H.~Verlinde, {\it {Loop
  and surface operators in N=2 gauge theory and Liouville modular geometry}},
  {\em JHEP} {\bf 01} (2010) 113, [\href{http://arxiv.org/abs/0909.0945}{{\tt
  arXiv:0909.0945}}].

\bibitem{Gaiotto:2009fs}
D.~Gaiotto, {\it {Surface Operators in N = 2 4d Gauge Theories}},  {\em JHEP}
  {\bf 11} (2012) 090, [\href{http://arxiv.org/abs/0911.1316}{{\tt
  arXiv:0911.1316}}].

\bibitem{Gaiotto:2012rg}
D.~Gaiotto, G.~W. Moore, and A.~Neitzke, {\it {Spectral networks}},  {\em
  Annales Henri Poincare} {\bf 14} (2013) 1643--1731,
  [\href{http://arxiv.org/abs/1204.4824}{{\tt arXiv:1204.4824}}].

\bibitem{Dimofte:2010tz}
T.~Dimofte, S.~Gukov, and L.~Hollands, {\it {Vortex Counting and Lagrangian
  3-manifolds}},  {\em Lett. Math. Phys.} {\bf 98} (2011) 225--287,
  [\href{http://arxiv.org/abs/1006.0977}{{\tt arXiv:1006.0977}}].

\bibitem{Gaiotto:2011tf}
D.~Gaiotto, G.~W. Moore, and A.~Neitzke, {\it {Wall-Crossing in Coupled 2d-4d
  Systems}},  {\em JHEP} {\bf 12} (2012) 082,
  [\href{http://arxiv.org/abs/1103.2598}{{\tt arXiv:1103.2598}}].

\bibitem{Eager:2016yxd}
R.~Eager, S.~A. Selmani, and J.~Walcher, {\it {Exponential Networks and
  Representations of Quivers}},  {\em JHEP} {\bf 08} (2017) 063,
  [\href{http://arxiv.org/abs/1611.06177}{{\tt arXiv:1611.06177}}].

\bibitem{Mclean96deformationsof}
R.~C. Mclean, {\it Deformations of calibrated submanifolds},  {\em Commun.
  Analy. Geom} {\bf 6} (1996) 705--747.

\bibitem{Aganagic:2000gs}
M.~Aganagic and C.~Vafa, {\it {Mirror symmetry, D-branes and counting
  holomorphic discs}},  \href{http://arxiv.org/abs/hep-th/0012041}{{\tt
  hep-th/0012041}}.

\bibitem{Aganagic:2001nx}
M.~Aganagic, A.~Klemm, and C.~Vafa, {\it {Disk instantons, mirror symmetry and
  the duality web}},  {\em Z. Naturforsch.} {\bf A57} (2002) 1--28,
  [\href{http://arxiv.org/abs/hep-th/0105045}{{\tt hep-th/0105045}}].

\bibitem{Aganagic:2013jpa}
M.~Aganagic, T.~Ekholm, L.~Ng, and C.~Vafa, {\it {Topological Strings, D-Model,
  and Knot Contact Homology}},  {\em Adv. Theor. Math. Phys.} {\bf 18} (2014),
  no.~4 827--956, [\href{http://arxiv.org/abs/1304.5778}{{\tt
  arXiv:1304.5778}}].

\bibitem{Gabella:2017hpz}
M.~Gabella, P.~Longhi, C.~Y. Park, and M.~Yamazaki, {\it {BPS Graphs: From
  Spectral Networks to BPS Quivers}},  {\em JHEP} {\bf 07} (2017) 032,
  [\href{http://arxiv.org/abs/1704.04204}{{\tt arXiv:1704.04204}}].

\bibitem{Longhi:2016wtv}
P.~Longhi, {\it {Wall-Crossing Invariants from Spectral Networks}},  {\em
  Annales Henri Poincare} {\bf 19} (2018), no.~3 775--842,
  [\href{http://arxiv.org/abs/1611.00150}{{\tt arXiv:1611.00150}}].

\bibitem{Klemm:1996bj}
A.~Klemm, W.~Lerche, P.~Mayr, C.~Vafa, and N.~P. Warner, {\it {Selfdual strings
  and N=2 supersymmetric field theory}},  {\em Nucl. Phys.} {\bf B477} (1996)
  746--766, [\href{http://arxiv.org/abs/hep-th/9604034}{{\tt hep-th/9604034}}].

\bibitem{Gaiotto:2013sma}
D.~Gaiotto, S.~Gukov, and N.~Seiberg, {\it {Surface Defects and Resolvents}},
  {\em JHEP} {\bf 09} (2013) 070, [\href{http://arxiv.org/abs/1307.2578}{{\tt
  arXiv:1307.2578}}].

\bibitem{Klemm:2008yu}
A.~Klemm and P.~Sulkowski, {\it {Seiberg-Witten theory and matrix models}},
  {\em Nucl. Phys.} {\bf B819} (2009) 400--430,
  [\href{http://arxiv.org/abs/0810.4944}{{\tt arXiv:0810.4944}}].

\bibitem{Nekrasov:1996cz}
N.~Nekrasov, {\it {Five dimensional gauge theories and relativistic integrable
  systems}},  {\em Nucl. Phys.} {\bf B531} (1998) 323--344,
  [\href{http://arxiv.org/abs/hep-th/9609219}{{\tt hep-th/9609219}}].

\bibitem{Ashok:2017lko}
S.~K. Ashok, M.~Billo, E.~Dell'Aquila, M.~Frau, V.~Gupta, R.~R. John, and
  A.~Lerda, {\it {Surface operators, chiral rings and localization in $
  \mathcal{N} $ =2 gauge theories}},  {\em JHEP} {\bf 11} (2017) 137,
  [\href{http://arxiv.org/abs/1707.08922}{{\tt arXiv:1707.08922}}].

\bibitem{Ashok:2017bld}
S.~K. Ashok, M.~Billo, E.~Dell'Aquila, M.~Frau, V.~Gupta, R.~R. John, and
  A.~Lerda, {\it {Surface operators in 5d gauge theories and duality
  relations}},  {\em JHEP} {\bf 05} (2018) 046,
  [\href{http://arxiv.org/abs/1712.06946}{{\tt arXiv:1712.06946}}].

\bibitem{Longhi:2016bte}
P.~Longhi and C.~Y. Park, {\it {ADE Spectral Networks and Decoupling Limits of
  Surface Defects}},  {\em JHEP} {\bf 02} (2017) 011,
  [\href{http://arxiv.org/abs/1611.09409}{{\tt arXiv:1611.09409}}].

\bibitem{Cecotti:1992qh}
S.~Cecotti, P.~Fendley, K.~A. Intriligator, and C.~Vafa, {\it {A New
  supersymmetric index}},  {\em Nucl. Phys.} {\bf B386} (1992) 405--452,
  [\href{http://arxiv.org/abs/hep-th/9204102}{{\tt hep-th/9204102}}].

\bibitem{Cecotti:1991me}
S.~Cecotti and C.~Vafa, {\it {Topological antitopological fusion}},  {\em Nucl.
  Phys.} {\bf B367} (1991) 359--461.

\bibitem{Cecotti:1992rm}
S.~Cecotti and C.~Vafa, {\it {On classification of N=2 supersymmetric
  theories}},  {\em Commun. Math. Phys.} {\bf 158} (1993) 569--644,
  [\href{http://arxiv.org/abs/hep-th/9211097}{{\tt hep-th/9211097}}].

\bibitem{Cecotti:2010fi}
S.~Cecotti, A.~Neitzke, and C.~Vafa, {\it {R-Twisting and 4d/2d
  Correspondences}},  \href{http://arxiv.org/abs/1006.3435}{{\tt
  arXiv:1006.3435}}.

\bibitem{Cecotti:2013mba}
S.~Cecotti, D.~Gaiotto, and C.~Vafa, {\it {$tt^*$ geometry in 3 and 4
  dimensions}},  {\em JHEP} {\bf 05} (2014) 055,
  [\href{http://arxiv.org/abs/1312.1008}{{\tt arXiv:1312.1008}}].

\bibitem{Witten:1997sc}
E.~Witten, {\it {Solutions of four-dimensional field theories via M theory}},
  {\em Nucl. Phys.} {\bf B500} (1997) 3--42,
  [\href{http://arxiv.org/abs/hep-th/9703166}{{\tt hep-th/9703166}}].
  [,452(1997)].

\bibitem{Gaiotto:2009we}
D.~Gaiotto, {\it {$\mathcal{N}$=2 dualities}},  {\em JHEP} {\bf 1208} (2012)
  034, [\href{http://arxiv.org/abs/0904.2715}{{\tt arXiv:0904.2715}}].

\bibitem{Mikhailov:1997jv}
A.~Mikhailov, {\it {BPS states and minimal surfaces}},  {\em Nucl. Phys.} {\bf
  B533} (1998) 243--274, [\href{http://arxiv.org/abs/hep-th/9708068}{{\tt
  hep-th/9708068}}].

\bibitem{Longhi:2016rjt}
P.~Longhi and C.~Y. Park, {\it {ADE Spectral Networks}},  {\em JHEP} {\bf 08}
  (2016) 087, [\href{http://arxiv.org/abs/1601.02633}{{\tt arXiv:1601.02633}}].

\bibitem{Gopakumar:1998ii}
R.~Gopakumar and C.~Vafa, {\it {M theory and topological strings. 1.}},
  \href{http://arxiv.org/abs/hep-th/9809187}{{\tt hep-th/9809187}}.

\bibitem{Gopakumar:1998ki}
R.~Gopakumar and C.~Vafa, {\it {On the gauge theory / geometry
  correspondence}},  {\em Adv. Theor. Math. Phys.} {\bf 3} (1999) 1415--1443,
  [\href{http://arxiv.org/abs/hep-th/9811131}{{\tt hep-th/9811131}}]. [AMS/IP
  Stud. Adv. Math.23,45(2001)].

\bibitem{Katz:1999xq}
S.~H. Katz, A.~Klemm, and C.~Vafa, {\it {M theory, topological strings and
  spinning black holes}},  {\em Adv. Theor. Math. Phys.} {\bf 3} (1999)
  1445--1537, [\href{http://arxiv.org/abs/hep-th/9910181}{{\tt
  hep-th/9910181}}].

\bibitem{Hosono:2001gf}
S.~Hosono, M.-H. Saito, and A.~Takahashi, {\it {Relative Lefschetz action and
  BPS state counting}},  {\em Int. Math. Res. Not.} {\bf 15} (2001) 783--816,
  [\href{http://arxiv.org/abs/math/0105148}{{\tt math/0105148}}].

\bibitem{Maulik:2016rip}
D.~Maulik and Y.~Toda, {\it {Gopakumar-Vafa invariants via vanishing cycles}},
  \href{http://arxiv.org/abs/1610.07303}{{\tt arXiv:1610.07303}}.

\bibitem{Katz:2004js}
S.~H. Katz, {\it {Gromov-Witten, Gopakumar-Vafa, and Donaldson-Thomas
  invariants of Calabi-Yau threefolds}},  in {\em {Snowbird lectures on string
  theory. Proceedings, Joint Summer Research Conference, Snowbird, USA, June
  5-11, 2004}}, pp.~43--52, 2004.
\newblock \href{http://arxiv.org/abs/math/0408266}{{\tt math/0408266}}.

\bibitem{Witten:1993yc}
E.~Witten, {\it {Phases of N=2 theories in two-dimensions}},  {\em Nucl. Phys.}
  {\bf B403} (1993) 159--222, [\href{http://arxiv.org/abs/hep-th/9301042}{{\tt
  hep-th/9301042}}]. [AMS/IP Stud. Adv. Math.1,143(1996)].

\bibitem{Hori:2000kt}
K.~Hori and C.~Vafa, {\it {Mirror symmetry}},
  \href{http://arxiv.org/abs/hep-th/0002222}{{\tt hep-th/0002222}}.

\bibitem{2007arXiv0709.3079Y}
B.~{Young}, {\it {Computing a pyramid partition generating function with dimer
  shuffling}},  {\em arXiv e-prints} (Sep, 2007) arXiv:0709.3079,
  [\href{http://arxiv.org/abs/0709.3079}{{\tt arXiv:0709.3079}}].

\bibitem{Seidel:2000ia}
P.~Seidel and R.~P. Thomas, {\it {Braid group actions on derived categories of
  coherent sheaves}},  \href{http://arxiv.org/abs/math/0001043}{{\tt
  math/0001043}}.

\bibitem{Nishinaka:2010qk}
T.~Nishinaka and S.~Yamaguchi, {\it {Wall-crossing of D4-D2-D0 and flop of the
  conifold}},  {\em JHEP} {\bf 09} (2010) 026,
  [\href{http://arxiv.org/abs/1007.2731}{{\tt arXiv:1007.2731}}].

\bibitem{Aspinwall:2004bs}
P.~S. Aspinwall and S.~H. Katz, {\it {Computation of superpotentials for
  D-branes}},  {\em Commun. Math. Phys.} {\bf 264} (2006) 227--253,
  [\href{http://arxiv.org/abs/hep-th/0412209}{{\tt hep-th/0412209}}].

\bibitem{Iqbal:2007ii}
A.~Iqbal, C.~Kozcaz, and C.~Vafa, {\it {The Refined topological vertex}},  {\em
  JHEP} {\bf 10} (2009) 069, [\href{http://arxiv.org/abs/hep-th/0701156}{{\tt
  hep-th/0701156}}].

\bibitem{Bouchard:2011ya}
V.~Bouchard and P.~Sulkowski, {\it {Topological recursion and mirror curves}},
  {\em Adv. Theor. Math. Phys.} {\bf 16} (2012), no.~5 1443--1483,
  [\href{http://arxiv.org/abs/1105.2052}{{\tt arXiv:1105.2052}}].

\end{thebibliography}\endgroup
\bibliographystyle{JHEP}

\end{document}